\documentclass[aps,prl,groupedaddress,superscriptaddress,showpacs,reprint,amsmath,amssymb]{revtex4-2}
\usepackage{xr-hyper}
\usepackage{hyperref}
\usepackage{stackengine}
\usepackage{color}
\usepackage{subfigure}
\usepackage{lipsum}
\usepackage{longtable}

\usepackage{dcolumn}% Align table columns on decimal point
\usepackage{bm}% bold math

\usepackage[table]{xcolor}
\usepackage{amsmath,amssymb,graphicx,esint,color,setspace,diagbox}
\usepackage[margin=2.6 cm]{geometry}
\usepackage{url}

\makeatletter
\@ifundefined{date}{}{\date{}}

\newcommand{\de}{{\rm{d}}}
\newcommand{\pe}{p_{\text{edge}}}

\makeatletter
\newcommand*{\addFileDependency}[1]{% argument=file name and extension
  \typeout{(#1)}
  \@addtofilelist{#1}
  \IfFileExists{#1}{}{\typeout{No file #1.}}
}
\makeatother

\begin{document}

\preprint{APS/123-QED}

\title{Efficient population coding of sensory stimuli}

\author{Shuai Shao}
\affiliation{Computation in Neural Circuits Group, Max Planck Institute for Brain Research, Frankfurt, Germany}
\affiliation{Donders Institute and Faculty of Science, Radboud University, Nijmegen, Netherlands}

\author{Markus Meister}
\affiliation{Division of Biology and Biological Engineering, California Institute of Technology, Pasadena, CA, USA}

\author{Julijana Gjorgjieva}
\email[]{gjorgjieva@tum.de}
\affiliation{School of Life Sciences, Technical University of Munich, Freising, Germany}
\affiliation{Computation in Neural Circuits Group, Max Planck Institute for Brain Research, Frankfurt, Germany}

\date{\today}

\begin{abstract}
The efficient coding theory postulates that single cells in a neuronal population should be optimally configured to efficiently encode information about a stimulus subject to biophysical constraints. This poses the question of how multiple neurons that together represent a common stimulus should optimize their activation functions to provide the optimal stimulus encoding. Previous theoretical approaches have solved this problem with binary neurons that have a step activation function, and have assumed that spike generation is noisy and follows a Poisson process. Here we derive a general theory of optimal population coding with neuronal activation functions of any shape, different types of noise and heterogeneous firing rates of the neurons by maximizing the Shannon mutual information between a stimulus and the neuronal spiking output subject to a constrain on the maximal firing rate. We find that the optimal activation functions are discrete in the biological case of non-negligible noise and demonstrate that the information does not depend on how the population is divided into ON and OFF cells described by monotonically increasing vs.\ decreasing activation functions, respectively. However, the population with an equal number of ON and OFF cells has the lowest mean firing rate, and hence encodes the highest information per spike. These results are independent of the shape of the activation functions and the nature of the spiking noise. Finally, we derive a relationship for how these activation functions should be distributed in stimulus space as a function of the neurons' firing rates.
\\
\\
\end{abstract}

%\keywords{Suggested keywords}%Use showkeys class option if keyword
                              %display desired
\maketitle

%\tableofcontents

\section*{I. Introduction}
In many neuronal systems, sensory information is processed by multiple neurons in parallel, forming a population code. However, how a population of neurons works together to efficiently encode a sensory stimulus in the presence of different biological constraints is still an open question. Many experimental and theoretical studies have proposed that neuronal coding is optimal \cite{Laughlin2001, Schreiber2002, Atick1990, Laughlin1981, Hateren1992}. Determining optimality is typically considered in the context of various constrains provided by the biological system in question. These include various assumptions made about the structure of the neuronal population, the relationship between stimulus and neuronal firing, the source and magnitude of sensory noise, and different measures used to quantify coding efficiency. For example, a common way to describe the firing rate of a neuron as a function of the stimulus is through an activation function, which usually describes a nonlinear dependence determined by the various ion channels embedded in the neuron's membrane or elaborate dendrites morphologies \cite{Hodgkin1952, Weiler2022}. The activation functions of sensory neurons can be monotonically increasing or decreasing as a function of the stimulus, referred to as ON or OFF, respectively (Fig.~\ref{fig:model}A), although in some sensory systems ON-OFF cells with non-monotonic activation functions also exist \cite{Wightman1992, Harper2004}. ON and OFF cells are found in many sensory systems, including the retina where ON (OFF) ganglion cells code for increases (decreases) in visual stimulus intensity or contrast \cite{Berens2017, Ratliff2010} and the insect mechanosensory system where they code for increases and decreases in leg angle \cite{Mamiya2018}. In line with most optimal coding theories of neuronal populations, here we assume that multiple cells together encode a sensory stimulus more efficiently than single cells in the presence of sensory noise and biophysical constraints.

Populations of sensory neurons are typically affected by noise which can come from different sources including from the sensory environment and biophysical constraints. Assuming a description of neuronal firing by activation functions, noise can enter before or after the activation function, called input vs.\ output noise, respectively, and can have a different influence of stimulus coding \cite{Faisal2008, Roth2021}. Since neurons communicate via action potentials, theoretical studies of optimal coding have commonly assumed that individual neurons generate spike counts in fixed coding windows following Poisson statistics \cite{Shamai1990, Nikitin2009, Pitkow2012, Gjorgjieva2019}. Under conditions of low spike count intensity of the Poisson process, the optimal activation functions of single neurons can be proven to be discrete with a single step, i.e., binary \cite{Shamai1990, Nikitin2009}. However, when the spike count intensity increases, binary neurons are no longer optimal, but rather the number of steps in the activation function increases as a function of spike count intensity \cite{Nikitin2009}. Especially in biological systems, many of these assumptions need to be relaxed. First, activation functions in different sensory systems usually do not manifest as binary and may appear continuous due to the presence of noise \cite{Taberner2004, Kastner2015, Roth2021}. Neuronal spike counts can also be non-Poisson, for instance, in the retina \cite{Berry1997, Kara2000}. Therefore, it is an interesting question what optimal configuration of activation functions can be achieved in theoretical frameworks of efficient coding where spike counts follow statistics other than Poisson.

What quantity might neural populations optimize? Two measures have been commonly used \cite{seung_93, warland_97, Bethge2003, Pitkow2012, rieke_97, bialek_91}. The Shannon mutual information between the stimulus and neuronal responses does not assume how the information should be decoded downstream. Alternatively, the stimulus can be estimated using a decoder and the difference between the stimulus and the estimate can be minimized. These two measures can generate very different predictions about the optimal population coding strategy \cite{Gjorgjieva2014, Gjorgjieva2019}.

Here, we develop a general efficient coding theory based on a population coding model with multiple ON and OFF neurons that code for a scalar stimulus from a given distribution assuming any (monotonic) nonlinear activation function and any noise statistics. We use the Shannon mutual information between the stimulus and the neuronal spikes as a measure of coding efficiency, and discover that this measure is independent of how the population is divided into ON and OFF neurons. We also investigate how the optimal firing thresholds of ON and OFF neurons partition the stimulus space as a function of the maximal neuronal firing rates. When these firing rates are equal for all neurons, we find that the thresholds divide the stimulus distribution into surprisingly regular stimulus regions.

\section*{II. Theoretical framework \label{sec:framework}}
\begin{figure*}[htb]
\includegraphics[width=\textwidth]{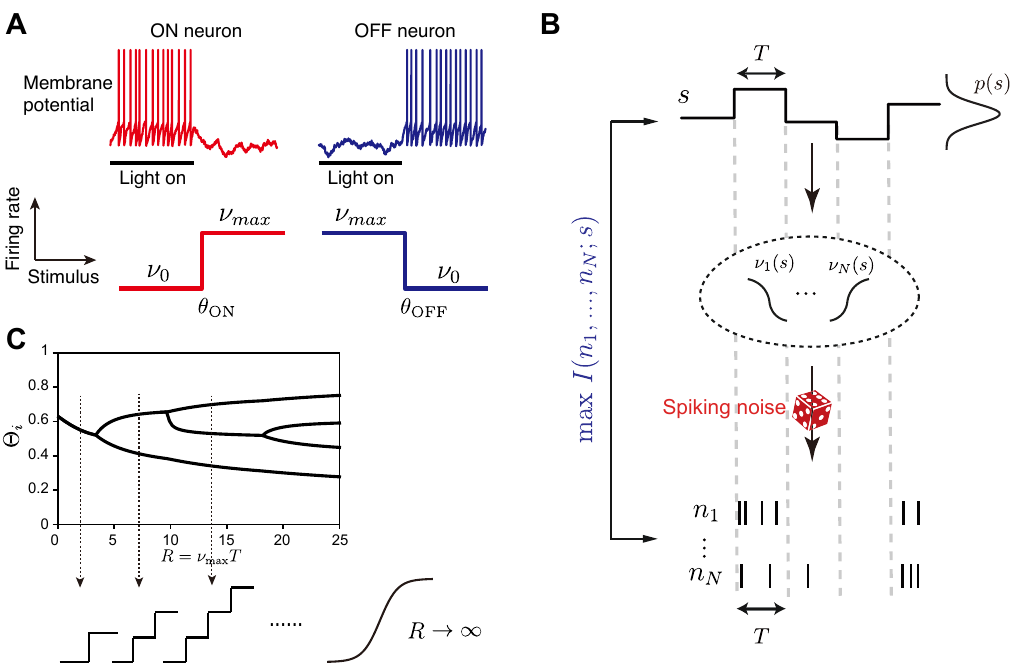}
\centering
\caption{\label{fig:model} \textbf{Efficient coding framework of a population of ON and OFF neurons.} \textbf{A.} A schematic of ON and OFF neurons. An ON neuron fires more frequently when the stimulus (which is light in this example) is high and fires at the spontaneous rate (here 0) when the stimulus is absent. The opposite is true for an OFF neuron. \textbf{B.} The population coding model. Sensory stimuli $s$, which are constant in the coding windows of size $T$, are drawn from a distribution $p(s)$. The stimuli are encoded by a population of neurons with firing rates $\nu_i(s)$, which fire noisy spike trains, $n_i$. The distribution of $n_i$ is given by the conditional probability $p(n_i|\nu_i(s))$, which denotes the spiking noise. The efficiency of the neuronal coding is quantified by the Shannon mutual information between the stimuli $s$ and the spike trains $n_i$, i.e., $I(n_1, ..., n_N; s)$.  \textbf{C.} The optimal activation function which maximizes the mutual information for a single ON neuron is discrete. Upper: The optimal thresholds of a single neuron that maximizes the Shannon mutual information. $\Theta_i$ denotes the cumulative probability of $s$ above a threshold. Lower: Schematics depicting that the number of steps of the optimal activation function increases with the product of the maximal firing rate and the coding window, i.e., $R=\nu_{\max}T$. For low $R$, the optimal activation function is binary and has one threshold ($i=1$). As $R$ increases, the optimal activation function become ternary ($i=2$), etc. The activation function becomes continuous in the limit of $R \rightarrow \infty$.}
\end{figure*}

We propose a theoretical framework of population coding with the following assumptions (Fig.~\ref{fig:model}A): 
\begin{itemize}
	\item[(i)] A population of ON and OFF neurons code for a one dimensional stimulus, with monotonically increasing and decreasing firing rates as a function of the stimulus (respectively), called activation functions;
	\item[(ii)] Each neuron in the population $i$ has a minimum (spontaneous) firing rate $\nu_0$ usually assumed to be 0, and a maximal firing rate constraint $\nu_{\max,i}$;
	\item[(iii)] The dynamic range of each neuron $i$, defined as the stimulus that leads to non-zero and non-maximal firing rate $\nu_i$ (with $\nu_0 < \nu_i < \nu_{\max,i}$), does not overlap with the dynamic range of other neurons;
	\item[(iv)] The dynamic ranges of OFF neurons are lower than those of ON neurons.
\end{itemize}

For the second assumption, we start with a simple case in which the maximal firing rates in a population are identical across the cells, i.e., $\nu_{\max,i}=\nu_{\max}$. Later in this paper (Section~III.D and III.H) we also consider neuron populations with heterogeneous $\nu_{\max,i}$. The assumption of zero spontaneous firing rate ensures analytical tractability. Our conclusions hold, at least in the case of binary activation functions for all cells with Poisson noise, even if this assumption is relaxed \cite{Gjorgjieva2019}.

We denote the sensory stimulus to be encoded by a population of $N$ cells as the scalar $s$ which is drawn from a distribution $p(s)$. We denote the activation function of each neuron as $\nu_i(s)$, where the subscript $i$ is the index of neurons in the population. We define ``the coding window" $T$ as the time period when the stimulus $s$ is constant (Fig.~\ref{fig:model}B). The coding window depends on the neuronal dynamics in the specific sensory population. For instance, in the mammalian retina, retinal ganglion cells have a coding window of 10 to 50~ms \cite{Pitkow2012, Pillow2008, Uzzell2004}. In the mouse auditory system, auditory nerve fibers, have a coding window of 50~ms \cite{Taberner2004, Roth2021}. Defining a coding window allows us to define the spike count $n_i$ for neuron $i$ within a coding window $T$ which has an expected value of $\nu_i(s) T$. Therefore, the stimulus $s$ is encoded by a vector of noisy spike counts $\vec n = \lbrace n_1, ..., n_N \rbrace$, which represents the population code. 

We consider a general noise model where the spike counts follow a probability distribution $p(\vec n|s)$, which only directly depends on the expected value $\vec \nu(s) T$. Since the firing rate vector $\vec \nu$ is a deterministic function of the stimulus $s$, and assuming the noise of different neurons is independent of each other, the probability distribution $p(\vec n|s)$ can also be written as a product of the spike count probability distribution of every neuron, i.e., $p(\vec n|\vec \nu(s)) = \prod_i p(n_i|\nu_i(s))$.  Because $\nu_i$ is the firing rate and $\nu_i T$ is the expected value of the spike count $n_i$ of neuron $i$, by definition, we have
\begin{eqnarray}
\sum_{n_i=0}^{+\infty} p(n_i| \nu_i) &= 1 
\label{eq:definition_L}\\
\sum_{n_i=0}^{+\infty} p(n_i| \nu_i)n_i &= \nu_i T.
\label{eq:definition_nu}
\end{eqnarray}
While the noise can follow any distribution, a special case commonly used in previous work is the Poisson noise where $p(\vec n|s) = p(\vec n|\vec \nu(s)) = \prod_i \frac{(\nu_i(s)T)^{n_i}}{n_i!} e^{-\nu_i(s)T}$.
We quantify the coding efficiency of this population code using the Shannon mutual information between the population spike count $\vec n$ and stimulus $s$:
\begin{equation}
\begin{aligned}
&I(s, \vec n) = \sum_{\vec n} \int \de s \, p(s)p(\vec n|s) \,\text{log}\frac{p(\vec n|s)}{P(\vec n)}
%\\
%&=\sum_{\vec n} \int \de s \, p(s)\prod_i L(n_i, \nu_i(s) T) \,\text{log}\frac{\prod_i L(n_i, \nu_i(s) T)}{P(\vec n)}
\label{eq:info}
\end{aligned}
\end{equation}
where 
\begin{equation}
P(\vec n) = \int \de s \, p(s)\prod_i p(n_i| \nu_i),
\label{eq:def_Pn}
\end{equation}
and $\sum_{\vec n} = \sum_{n_1=0}^{+\infty} ... \sum_{n_N=0}^{+\infty}$ denotes the sum over all possible spike counts of all the neurons.

Because the firing rates $\vec \nu$ depend deterministically on the stimulus $s$, the mutual information between $s$ and $\vec n$ is the same as the mutual information between $\vec \nu$ and $\vec n$ (proof in Supplemental Material 1), 
\begin{equation}
I(s, \vec n) = I(\vec \nu, \vec n) = \sum_{\vec n} \int_{\vec \nu} \text d^N \vec \nu \, p(\vec \nu)p(\vec n|\vec \nu) \, \text{log}\frac{p(\vec n|\vec \nu)}{P(\vec n)}.
\label{eq:info_rate}
\end{equation}

\section*{III. Results}
We seek to derive the optimal activation functions $\lbrace \nu_i(\cdot) \rbrace_i$ of an entire population of ON and OFF neurons, which maximize the mutual information $I(s, \vec n)$ (Eq.~\ref{eq:info_rate}), when the conditional probability $p(n_i|\nu_i)$ is given. We also aim to determine how this maximal mutual information depends on the ON-OFF composition of the neuronal population.

\subsection*{A. The optimal activation function for a single neuron is discrete
\label{sec:discrete_tuning_1}}

We first investigate a population with only a single neuron subject to the constraints from Section II. Previous studies have found that under these conditions and with Poisson-distributed spike counts, the optimal activation function for a single neuron should be discrete, with an increasing number of steps as a function of the product $R=\nu_{\max}T$, i.e., the maximum expected spike count \cite{Shamai1990, Nikitin2009} (Fig.~\ref{fig:model}C). In two steps, we generalize this result to any analytic conditional probability $p(n| \nu)$ (analytic in terms of $\nu$) using the fact that mutual information is convex in the input space \cite{Smith1971}.

In step 1, we prove that the mutual information $I(\nu, n)$ is distributed proportionally to the probability density $p(\nu)$ in the optimal configuration. Defining the ``density of mutual information" as
\begin{equation}
i(\nu) = \sum_{n} p(n|\nu)\text{log}\frac{p(n|\nu)}{P(n)}
\label{eq:info_density_rate}
\end{equation}
we can write
\begin{equation}
I(\nu, n) = \int_\nu \de \nu \, p(\nu)\, i(\nu).
\label{eq:info_rate_2}
\end{equation}
We can then prove that in the optimal case,
\begin{equation}
i(\nu) = I^{\max}  \text{ for all possible} \ \nu
\label{eq:equal_info_density}
\end{equation}
where $I^{\max}$ is the maximal mutual information (Supplemental Material 2). 

Then in step 2, we show that Eq.~\ref{eq:equal_info_density} cannot be true if the activation function $\nu(s)$ is continuous, therefore concluding that it must be discrete. To do this, we first redefine the activation function using a function $F_{\nu}$. For an ON neuron (the case for an OFF neuron follows similarly), we can write for any arbitrary firing rate $\tilde{\nu}$
\begin{equation}
    F_\nu(\tilde{\nu}) = \int_{-\infty}^{s_{\max}(\tilde{\nu})} \de s\, p(s). \label{eq:def_F_nu_new_ON}
\end{equation}
where $s_{\max}(\tilde{\nu})$ is defined as the highest $s$ that makes $\nu(s) \leq \tilde{\nu}$, i.e. $s_{\max} = \textbf{max} \lbrace s|\nu(s) \leq \tilde{\nu} \rbrace$. Because $\nu(s)$ is a monotonically increasing function of $s$, $s_{\max}(\tilde{\nu})$ is also monotonically increasing, making $F_\nu(\tilde{\nu})$ a monotonically increasing function of $\tilde{\nu}$. We can replace the variable in the integral of Eq.~\ref{eq:def_F_nu_new_ON}, leading to
\begin{equation}
    F_\nu(\tilde{\nu}) = \int_{\nu(s \rightarrow -\infty)}^{\nu(s = s_{\max}(\tilde{\nu}))} \de \nu\, p(\nu).
    \label{eq:F_nu_cdf_ON}
\end{equation}
Therefore, $F_{\nu}$ becomes the cumulative distribution function of the firing rate $\nu$:
\begin{equation}
    F_\nu(\tilde{\nu}) = \int_0^{\tilde{\nu}} \de \nu\, p(\nu).
    \label{eq:def_F_nu}
\end{equation}
Let $F_{\nu}^*$ denote the optimal activation function, which maximizes the mutual information, $I(\nu,n)$ (Eq.~\ref{eq:info_rate} and Eq.~\ref{eq:info_rate_2}). We explicitly include the dependence of the density of mutual information $i(\nu)$ (Eq.~\ref{eq:info_density_rate}) on the activation function $F_{\nu}$ by writing $i(\nu, F_{\nu})$ because $P(n)$ depends on $F_\nu$. Then, Eq.~\ref{eq:equal_info_density} can be rewritten as
\begin{equation}
i(\nu, F_{\nu}^*) = I(F_{\nu}^*) \ \ \ \text{for all } \nu \text{ in } E_{\nu}^*
\end{equation}
where $E_{\nu}^*$ is the set of points at which $F_{\nu}^*$ increases. 

From now on, we denote the conditional probability $p(n_i| \nu_i)$ by $L(n_i, \nu_i T)$, and call it the ``noise generation function". If we assume $L(n, \nu T)$ is analytic with respect to $\nu T$, then we can show that the optimal activation function has a finite number of steps, i.e., $E_{\nu}^*$ is a finite set of points. Note that because of Eq.~\ref{eq:def_F_nu}, $E_{\nu}$ is also the set of all possible firing rates, i.e., $E_{\nu} = \lbrace \nu|p(\nu) > 0 \rbrace$. If $E_{\nu}^*$ has a finite number of points, then the optimal $\nu(s)$ will have a finite number of steps. 

Let us first consider the case that $E_{\nu}^*$ is infinite. In the simplest case, if $F_{\nu}^*$ is continuous over the interval $[0, \nu_{\max}]$, then $E_{\nu}^* = [0, \nu_{\max}]$. As a result, $i(\nu, F_{\nu}^*) = \text{const}$ for any $\nu \in [0, \nu_{\max}]$. 

If $F_{\nu}^*$ is not continuous but $E_{\nu}^*$ has an infinite number of points (e.g.~$F_{\nu}^*$ is only continuous on a subinterval of $[0, \nu_{\max}]$), similar to previous work \cite{Smith1971,Shamai1990}, one can use the Bolzano Weierstrass theorem \cite{Bartle1964} to prove that $E_{\nu}^*$ has a limit point in $[0, \nu_{\max}]$. Then by the identity theorem for analytic functions \cite{Knopp1945}, if two analytic functions, in our case $i(\nu, F_{\nu}^*)$ and $I(F_{\nu}^*)$, have the same value on an infinite number of points and the limit of these points, then they are equal, i.e.~$i(\nu, F_{\nu}^*) = \text{const}$ for any $\nu \in [0, \nu_{\max}]$. In short, assuming $E_{\nu}^*$ has an infinite number of points also implies that $i(\nu, F_{\nu}^*)$ is a constant over the interval $[0, \nu_{\max}]$.

If $E_{\nu}^*$ is infinite, assuming optimal coding, based on Eq.~\ref{eq:equal_info_density}, we have
\begin{equation}
i(\nu) = \sum_{n=0}^{+\infty} L(n,\nu T)\log \frac{L(n, \nu T)}{P(n)} = I^{\max} = \text{const}.
\label{eq:info_density_const}
\end{equation}
Then the derivative with respect to $\nu T$ 
\begin{equation}
i'(\nu) = \sum_{n=0}^{+\infty} L'(n, \nu T)\log \frac{L(n, \nu T)}{P(n)} = 0
\end{equation}
where $L'(n, \nu T)$ denotes $\frac{\partial L(n, \nu T)}{\partial (\nu T)}$. Similarly, the second derivative
\begin{equation}
\begin{aligned}
&i''(\nu) = \\
&\sum_{n=0}^{+\infty} \left[L''(n, \nu T)\log \frac{L(n, \nu T)}{P(n)} + \frac{L'(n, \nu T)^2}{L(n, \nu T)} \right] = 0.\label{enq:2-nd_deriv}
\end{aligned}
\end{equation}
Using mathematical induction, one can prove that for any $m \in \mathbb{N^+}$, the $m^{th}$ derivative of $i(\nu)$ with respect to $\nu T$, $i^{(m)} (\nu)$, contains the term
\begin{equation}
\sum_{n=0}^{+\infty} L^{(m)} (n, \nu T)\log \frac{L(n, \nu T)}{P(n)}. \label{eqn:m-th_deriv}
\end{equation}
According to Eq.~\ref{eq:definition_nu}, $\sum_{n} L(n,\nu T)n = \nu T$, we have
\begin{equation}
L(0, 0) = 1, \ \ \ L(n\geq 1, 0) = 0.
\label{eq:L_boundary_condition}
\end{equation} 
Based on these two boundary conditions, $L(n, \nu T)$ can be written as a Maclaurin series
\begin{equation}
L(0, \nu T) = 1+\sum_{k=1}^{+\infty} a_{0k}(\nu T)^k,
\label{eq:maclaurin_series_0}
\end{equation}
\begin{equation}
L(n, \nu T) = \sum_{k=1}^{+\infty} a_{nk}(\nu T)^k.
\label{eq:maclaurin_series_n}
\end{equation}
Substituting these two series into the fractional or polynomial terms of the derivatives of the noise generation function $L(n, \nu T), L'(n, \nu T), ..., L^{(m-1)} (n, \nu T)$, and also in the terms $\sum_n L^{(m)} (n, \nu T)\log P(n)$ in the derivatives $i^{(m)}(\nu)$ (Eq.~\ref{eqn:m-th_deriv}), we find that they all become fractional or polynomial terms of $\nu T$ after doing the Maclaurin expansion with respect to $\nu T$ around 0. For example, in $i''(\nu)$ (Eq.~\ref{enq:2-nd_deriv}),
\begin{equation}
\begin{aligned}
&\frac{L'(n, \nu T)^2}{L(n, \nu T)} \\
&= \frac{(\sum_{k=1}^{+\infty} a_{nk}k(\nu T)^{k-1})^2}{\sum_{k=1}^{+\infty} a_{nk}(\nu T)^k}\\
&= \frac{(\sum_{k=1}^{+\infty} a_{nk}k(\nu T)^{k-1})^2}{a_{n1}\nu T}\left (1+\sum_{k=2}^{+\infty} \frac{a_{nk}}{a_{n1}}(\nu T)^{k-1}\right )^{-1}\\
& = \frac{a_{n1}}{\nu T} + 3a_{n2} +\left  (5a_{n3} + \frac{a_{n2}^2}{a_{n1}}\right )\nu T + ...\ \ \ (n \geq 1).
\end{aligned}
\end{equation}
The only exception is the term containing $\log (\nu T)$ apart from the polynomial terms:
\begin{widetext}
\begin{equation}
\begin{aligned}
&\sum_{n=0}^{+\infty} L^{(m)} (n, \nu T)\log L(n, \nu T) \\
%&= \sum_{k=m}^{+\infty}a_{0k}\frac{k!}{(k-m)!}(\nu T)^{k-m}\log \left [1+\sum_{k=1}^{+\infty} a_{0k}(\nu T)^k\right ] \\
%&+ \sum_{n=1}^{+\infty}\sum_{k=m}^{+\infty}a_{nk}\frac{k!}{(k-m)!}(\nu T)^{k-m}\log \sum_{k=1}^{+\infty} a_{nk}(\nu T)^k\\
&=\sum_{k=m}^{+\infty}a_{0k}\frac{k!}{(k-m)!}(\nu T)^{k-m}\log \left[1+\sum_{l=1}^{+\infty} a_{0l}(\nu T)^l\right ]  \\
&+ \sum_{n=1}^{+\infty}\left [\sum_{k=m}^{+\infty}a_{nk}\frac{k!}{(k-m)!}(\nu T)^{k-m} \right] \left[\log a_{n,j(n)} + j(n)\log (\nu T) + \log \left(1+\sum_{l>j(n)} \frac{a_{nl}}{a_{nj}}(\nu T)^{l-j(n)}\right)\right]
\end{aligned}
\label{eq:log_diverge}
\end{equation}
\end{widetext}
where $j(n)$ is the minimal index of $k$ that makes $a_{nk} > 0$ when $n$ is given. 
When $\nu T \rightarrow 0$, we can see that the first term in Eq.~\ref{eq:log_diverge} is finite. The second term can be expanded as the sum of polynomial terms and other terms proportional to $(\nu T)^{k-m} \log (\nu T)$, which converge to 0 if $k>m$. The only diverging term is $(\nu T)^{k-m} \log (\nu T)$ when $k=m$, which becomes $\log (\nu T)$. Hence, the second term diverges as
\begin{equation}
\sum_{n=1}^{+\infty}a_{nm}\,j(n)\log (\nu T)
\label{eq:sum_of_coefficient_log}
\end{equation}
while other terms of $i^{(m)}(\nu)$ either converge to a finite value or diverge even faster than $\log (\nu T)$, because they are either polynomial or fractional terms of $\nu T$. The sum of the coefficients  $a_{nm}$ of all the fractional terms with the same order should then be 0. If we could not find a relationship among $a_{nm}$ that make the sum 0, a paradox would arise completing the proof. In addition, the sum of the coefficients  $a_{nm}$ of $\log (\nu T)$ terms should also be 0, i.e.
\begin{equation}
\sum_{n=1}^{+\infty}a_{nm}\,j(n) = 0 \text{ for all } m \geq 1.
\label{eq:sum_of_coefficient_0}
\end{equation}
According to Eq.~\ref{eq:L_boundary_condition}, when $\nu T = 0$, $L(n \geq 1, \nu T)$ reaches its lower bound 0. Then the derivative $L'(n, 0)$, which equals to $a_{n1}$ (see Eq.~\ref{eq:maclaurin_series_n}), is positive or 0 for any $n \geq 1$, i.e.,
\begin{equation}
    a_{n1} \geq 0.
\end{equation}
Combining with Eq.~\ref{eq:sum_of_coefficient_0}, and noting that $j(n) > 0$, we have
\begin{equation}
    a_{n1} = 0 \text{ for all } n \geq 1.
\end{equation}
Similarly, based on $a_{n1} = 0$, we can derive $a_{n2} = 0$. This is because the second derivative $L''(n \geq 1, 0)$ also needs to be positive or 0, given that $L(n \geq 1, \nu T)$ is at its lower bound and its first derivative is 0. Continuing this process, we get
\begin{equation}
    a_{nm} = 0
\end{equation}
for all $n \geq 1$ and $m \geq 1$. Substituting into Eq.~\ref{eq:maclaurin_series_n}, we have
\begin{equation}
L(n, \nu T) = 0 \text{ for any } n \geq 1 \text{ and any } \nu,
\end{equation}
which leads to
\begin{equation}
L(0, \nu T) = 1 \text{ for any } \nu.
\label{eq:contradiction_single_neuron}
\end{equation}
This is in contradiction to Eq.~\ref{eq:definition_nu}, $\sum_{n} L(n,\nu T)n = \nu T$, since $\nu > 0$ means that the neuron fires and $L(0, \nu T)$ cannot be 1. Therefore, Eq.~\ref{eq:info_density_const} leads to a paradox, which indicates that the set of increasing points $E_\nu^*$ cannot be infinite.

In summary, this proves that a continuous activation function is inconsistent with Eq.~\ref{eq:equal_info_density}. This means that the optimal activation function for a single neuron must be discrete for any noise generation function.

\subsection*{B. The optimal activation functions for a population of neurons are discrete
\label{sec:discrete_tuning_N}}

Next, we investigate a population of $N$ neurons, made up of ON and OFF neurons that have monotonically increasing and decreasing activation functions as a function of the stimulus $s$, respectively. We continue to consider the same constraints of a maximal firing rate and zero spontaneous firing rate (Fig.~\ref{fig:model}A). Under these conditions, the optimal activation functions for all neurons in the population continues to be discrete for any analytic noise generation function $L(n_i, \nu_i T)$. 

We define the ``dynamic range" of a neuron to be the interval of $s$ that leads to unsaturated firing rates, i.e.~$\lbrace s|0 < \nu_i(s) < \nu_{\max}\rbrace$ for neuron $i$ (see Section II). For a discrete activation function, the dynamic range is the interval between the lowest and highest threshold. We assume that the dynamic ranges of any two neurons do not overlap and also assumed that any OFF neuron encodes smaller stimuli than any ON neuron (see Section II), which is consistent with experimental measurements \cite{Mamiya2018} and previous theoretical work \cite{Gjorgjieva2019}. 

We consider a mixed population of $m$ ON neurons and $N-m$ OFF neurons. To proceed, we label all ON neurons with decreasing indices ($m$ to 1) from low to high dynamic ranges, where the ON neuron with the highest dynamic range has index 1. Similarly, we label all OFF neurons with increasing indices ($m+1$ to $N$) from low to high dynamic ranges to ensure symmetry in our mathematical expressions (note this ordering is different from previous work \cite{Gjorgjieva2019} to ensure symmetry of the expressions). 

If one of the ON neurons $1, 2, ..., m$ fires, assuming that spontaneous firing rates are 0, we know that the stimulus $s$ is higher than, or is at least within the dynamic range of neuron $m$. Then we also know the firing rates of neurons $m+1, m+2, ..., N$, which means the spike counts of these neurons cannot give any new information about the stimulus $s$. Based on this, we can write the mutual information encoded by the mixture of $m$ ON neurons and $N-m$ OFF neurons as
\begin{equation}
\begin{aligned}
&I_N\left(F_1, ..., F_N\right) \\
&= I_m\left(F_1, ..., F_m\right) + Q_m I_{N-m}\left(F_{m+1}^{(m)}, ..., F_N^{(m)}\right).
\end{aligned}
\label{decompose_mutual_info_F}
\end{equation}
Here $F_i = F_{\nu_i}$ is defined in the same way as before (Eq.~\ref{eq:def_F_nu}), while $Q_m$ denotes the probability that none of the ON neurons $1, 2, ..., m$ fires. We additionally define the terms $F^{(m)}$ to denote the `revised' distribution functions under the condition that none of the neurons $1, 2, ..., m$ fires, i.e., given an arbitrary firing rate $\tilde{\nu}$,
\begin{equation}
    F_i^{(m)}(\tilde{\nu}) = \text{Prob}(\nu_i \leq \tilde{\nu}|n_1 = ... = n_m = 0).
\end{equation}
From Bayes rule, we can write
\begin{equation}
    F_i^{(m)}(\tilde{\nu}) = \frac{F_i(\tilde{\nu})\, \text{Prob}(n_1 = ... = n_m = 0|\nu_i \leq \tilde{\nu})}{Q_m}.
    \label{eq:F_i_bayesian}
\end{equation}
Here, $Q_m$ does not depend on $\tilde{\nu}$. Within the dynamic range of neuron $i$ (where $i > m$), the firing rate of neurons $1, ..., m$ are all 0, which means $\text{Prob}(n_1 = ... = n_m = 0|\nu_i \leq \tilde{\nu})$ also does not depend on $\tilde{\nu}$ in the dynamic range of $F_i$. Therefore, if $F_i$ is discrete, $F_i^{(m)}$ will also be discrete, and vice versa. This relationship also exists between $F_i$ and $F_i^{(j)}$ where $j$ is an arbitrary positive integer smaller than $i$.

Following a similar logic, we can also decompose the mutual information encoded by a population of $N$ neurons in Eq.~\ref{decompose_mutual_info_F} into $N$ single terms, each containing the mutual information encoded by one neuron, i.e.,
\begin{widetext}
\begin{equation}
I_N = I(F_1) + P_1(0)\left \{ I\left(F_2^{(1)}\right)+ P_2^{(1)}(0) \left[I\left(F_3^{(2)}\right) + ...+P_{N-1}^{(N-2)}(0)\,I\left(F_N^{(N-1)}\right)\right]\right \}
\label{recursive_calculation_F}
\end{equation}
\end{widetext}
where $P_i(0) = \int L(0, \nu_i T)\, \de F_i$ denotes the probability that neuron $i$ does not fire, i.e., $n_i = 0$. Furthermore, we have used $I\left(F_i^{(i-1)}\right)$ to denote the mutual information of neuron $i$ assuming that neurons $1, ..., i-1$ do not fire. Since $m$ does not explicitly appear in this equation, Eq.~\ref{recursive_calculation_F} applies to any mixed ON-OFF population, including homogeneous ON populations (where $m = N$) or homogeneous OFF populations (where $m = 0$).

We use mathematical induction to demonstrate that the optimal activation functions in a population are all discrete. Having already shown this for a single neuron, we assume it is true for a population of $N-1$ cells. Then we add an additional neuron and show the optimal activation functions of all $N$ neurons are discrete. Without loss of generality, we assume that the newly added neuron is an ON neuron with a highest dynamic range, labeled with 1, and the remaining $N-1$ neurons $2, ..., N$. The sum of all the terms multiplying $P_1(0)$ in Eq.~\ref{recursive_calculation_F} has the same mathematical form as $I_{N-1}$. As a result, the sum equals $I_{N-1}^{\max}$ when optimizing $F_2^{(1)}, ..., F_N^{(N-1)}$, allowing us to write
\begin{equation}
I_N = I(F_1) + P_1(0)I_{N-1}^{\max}.
\label{eq:IN_on_F1}
\end{equation}
Meanwhile, because we assumed that optimal activation functions are discrete in a population of $N-1$ neurons, the optimal $F_2^{(1)}, ..., F_N^{(N-1)}$ are all discrete. As we argued before, since $F_i$ and $F_i^{(j)}$ are either both discrete or both continuous, this means that $F_2, ..., F_N$ are all discrete. As before (Eq.~\ref{eq:info_density_rate}), we can also define the density of mutual information here as
\begin{equation}
\begin{aligned}
    \widetilde i(\nu_1) &= \sum_{n_1} p(n_1|\nu_1)\log \frac{p(n_1|\nu_1)}{P(n)}+ p(n_1=0|\nu_1)I_{N-1}^{\max} \\
    &= \sum_{n_1} L(n_1, \nu_1 T)\log \frac{L(n_1, \nu_1 T)}{P(n)}+ L(0, \nu_1 T)I_{N-1}^{\max}.
\end{aligned}
\label{eq:def_mi_density_N}
\end{equation}
Therefore, maximizing $I_N$ is equivalent to optimizing $F_1$ assuming optimal $F_2^{(1)}, ..., F_N^{(N-1)}$ as in Eq.~\ref{eq:IN_on_F1}. If the optimal $F_1$ is continuous, when $I_N$ is maximized we have
\begin{equation}
\widetilde i(\nu_1) =I_N^{\max},\,\, \nu_1 \in [0, \nu_{max}]
\label{eq:mutual_info_density_N}
\end{equation}
and this leads to (detailed proof in Supplemental Material 3)
\begin{equation}
L(n_1 = 0, \nu_1 T) = 1 \text{ for any } \nu_1.
\label{eq:contradiction_multiple_neurons}
\end{equation}
Similar to Eq.~\ref{eq:contradiction_single_neuron}, here Eq.~\ref{eq:contradiction_multiple_neurons} is also in contradiction to Eq.~\ref{eq:definition_nu}, $\sum_{n_1} L(n_1,\nu_1 T)n_1 = \nu_1 T$. Therefore, the optimal $F_1$ must be discrete and we have proved that all the $N$ optimal activation functions need to be discrete.

Hence, using mathematical induction, we have proved that all the neurons' optimal activation functions in a population of any number of neurons are discrete.

\subsection*{C. The optimal thresholds and the maximal mutual information for a population of binary neurons
\label{sec:binary}}
\begin{figure}[ht]
\includegraphics[width=0.48\textwidth]{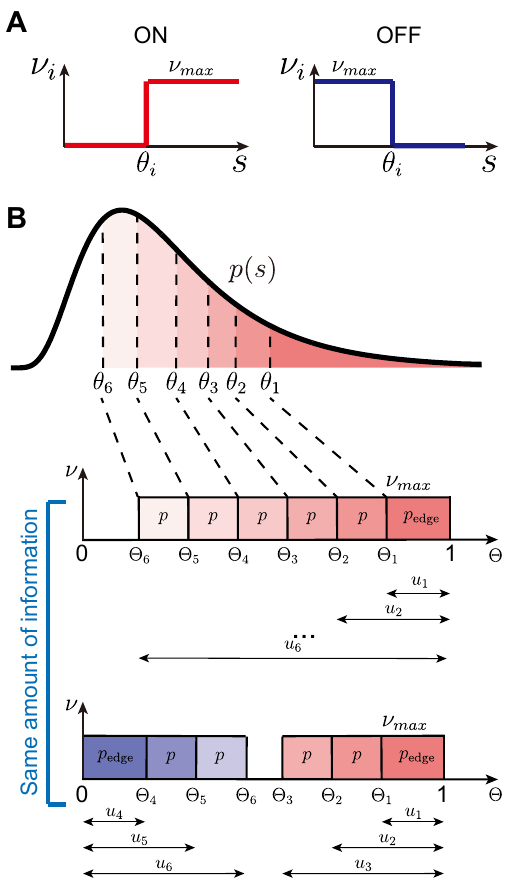}
\centering
\caption{\label{fig:binary} \textbf{Efficient population coding of binary neurons.} \textbf{A.} Activation functions of ON and OFF binary neurons. Each neuron has the same maximal firing rate $\nu_{\max}$, with an activation function described by a single threshold,  $\theta_i$. \textbf{B.} Optimal configurations of homogeneous populations with only ON neurons and mixed ON and OFF neurons. $\Theta_i$ denotes the cumulative probability of $s$ above threshold $\theta_i$. The optimal thresholds partition the cumulative stimulus space into regular intervals (Eqs.~\ref{eq:cumulative_theta}-\ref{eq:binary_OFFthresholds}). The optimized mutual information is independent of the ON-OFF mixture for any noise generation function (Eq.~\ref{eq:I_p_binary}).}
\end{figure}
Having shown that the optimal activation functions in population are discrete for any noise generation function, we first consider the simplest discrete activation function, which is binary, to derive the optimal thresholds and the maximal mutual information. As before, we study a combination of a total of $N$ neurons, $m$ ON and $N-m$ OFF neurons. Assuming that the spontaneous firing rate (the firing rate when the stimulus $s$ is subthreshold) is 0 (Fig.~\ref{fig:binary}A), only two parameters characterize the activation function of neuron $i$, $\nu_i(s)$: the threshold (denoted as $\theta_i$) and the maximal firing rate, which as before we assume is the same for all neurons ($\nu_{\max,i} = \nu_{\max}$ for all $i$). 

Because there is only one threshold for every neuron, the dynamic range of every neuron is compressed to a single point, the neuron's threshold. Labeling all the neurons as before
\begin{equation}
\theta_{m+1} < ... < \theta_N < \theta_m < ... < \theta_1,
\label{eq:thresh_index_binary}
\end{equation}
we note that there is only one noisy firing level at the maximum firing rate. The absence of noise in the zero firing state enables us to lump all firing states with nonzero spike count into one \cite{Gjorgjieva2019} (see Supplemental Material 5).

Because the optimal activation functions are discrete, following \cite{Gjorgjieva2019}, we can replace the firing thresholds with the intervals of stimulus space partitioned by those thresholds (Fig.~\ref{fig:binary}A) and optimize them instead of directly optimizing thresholds, i.e., we define
\begin{equation}
 u_i = \text{Prob}(\nu_i = \nu_{\max}) = 
\begin{cases}
\int_{\theta_i}^{+\infty} \de s \, p(s), \, \text{for ON}\\
\int_{-\infty}^{\theta_i} \de s \, p(s), \, \text{for OFF.}
\end{cases}
\label{eq:binary_u_def}
\end{equation}
Denoting 
\begin{equation}
\begin{aligned}
R &= \nu_{\max}T, \\
q &= L(0, R) = 1 - \sum_{n=1}^{+\infty} L(n, R),
\end{aligned}
\end{equation}
we extend the finding of \cite{Gjorgjieva2019} to any noise generation function that  the maximal mutual information is
\begin{equation}
I^{\max}_N = \text{log} \left(1 + N(1-q)q^{q/(1-q)} \right) = -\text {log} \left(P(\vec 0)\right)
\label{eqn:binary_info}
\end{equation}
where $P(\vec 0)$ is the probability that spike counts are all 0 (see Supplemental Material 5). The maximal information $I^{\max}_N$ is independent of the composition of ON neurons and OFF neurons and only depends on the total number of neurons $N$. Hence, we have generalized the previously termed ``Equal Coding Theorem" to other noise generation functions than Poisson \cite{Gjorgjieva2019}. In addition, comparing the maximal mutual information of a single neuron population, $I_1^{\max}$, and of an $N$-neuron population, $I_N^{\max}$, reveals that the maximum mutual information encoded by a population of neurons increases logarithmically with the number of neurons:
\begin{equation}
I_N^{\max} = \log \left[N(\exp({I_1^{\max}})-1)+1 \right].
\label{eq:IN_I1}
\end{equation}

Given this maximum mutual information, we next calculate the optimal threshold distribution of the population's binary activation functions. We can show (see Supplemental Material 5) that the optimal $\lbrace u_i \rbrace$ for the ON neurons are 
\begin{equation}
u_i = \frac{1+(i-1)(1-q)}{N(1-q)+q^{-q/(1-q)}}
\label{eqn:binary_thresh_on}
\end{equation}
and for the OFF neurons 
\begin{equation}
u_i = \frac{1+(m-i+1)(1-q)}{N(1-q)+q^{-q/(1-q)}}.
\label{eqn:binary_thresh_off}
\end{equation}
The terms $\{ u_i \}$ represent an arithmetic progression for any noise generation function $L$, whereby all the firing thresholds equally partition the probability space of stimuli similar to the case with Poisson noise \cite{Gjorgjieva2019}. If we define 
\begin{equation}
\begin{aligned}
&p_1 = u_1,~p_{m+1} = u_{m+1}\\
&p_i = u_i - u_{i-1}, i = 2, ..., m, m+2, ..., N
\end{aligned}
\label{eq:binary_p_def}
\end{equation}
as the probabilities of the stimulus intervals, i.e., the intervals of stimuli $s$ that lead to the same firing rates $\vec \nu$ (Fig.~\ref{fig:binary}B), we have
\begin{equation}
\begin{aligned}
&p_1 = p_{m+1} \overset{\text{def}}{=} p_{\text {edge}} \\
&p_2 = ... = p_m = p_{m+2} = ... = p_N \overset{\text{def}}{=} p \\
&p = (1-q)p_{\text {edge}}.
\end{aligned}
\label{eq:binary_p_result}
\end{equation}
This gives us the optimal thresholds in cumulative stimulus space (Fig.~\ref{fig:binary}B), 
\begin{equation}
\Theta_i=\int_{-\infty}^{\theta_i} \de s \, p(s) \label{eq:cumulative_theta}
\end{equation}
for the ON cells as:
\begin{equation}
\begin{aligned}
\Theta_1 &= 1- p_{\text{edge}}\\
\Theta_2 &= 1- p_{\text{edge}} - p\\
...\\
\Theta_m &= 1- p_{\text{edge}} - (m-1)p
\end{aligned}
\label{eq:binary_ONthresholds}
\end{equation}
and for the OFF cells as:
\begin{equation}
\begin{aligned}
\Theta_{m+1} &= p_{\text{edge}}\\
\Theta_{m+2} &= p_{\text{edge}}+ p\\
...\\
\Theta_N &= p_{\text{edge}}+ (N-m-1) p
\end{aligned}
\label{eq:binary_OFFthresholds}
\end{equation}
Given these optimal thresholds, we can combine this  with Eq.~\ref{eqn:binary_info} to find the expression for the optimal mutual information
\begin{equation}
I^{\max}_N = -\log (1-Np).
\label{eq:I_p_binary}
\end{equation}

Hence, we conclude that, for a mixed ON-OFF population with binary activation functions, the optimal thresholds and mutual information look exactly the same for any noise generation function as for Poisson \cite{Gjorgjieva2019}. Homogeneous populations with only ON or OFF neurons, and mixed ON-OFF populations with any ON-OFF mixture can encode the same amount of information.
\subsection*{D. The optimal thresholds for a population of binary neurons with heterogeneous maximal firing rates
\label{sec:binary_het}}
\begin{figure}[th!]
\includegraphics[width=0.48\textwidth]{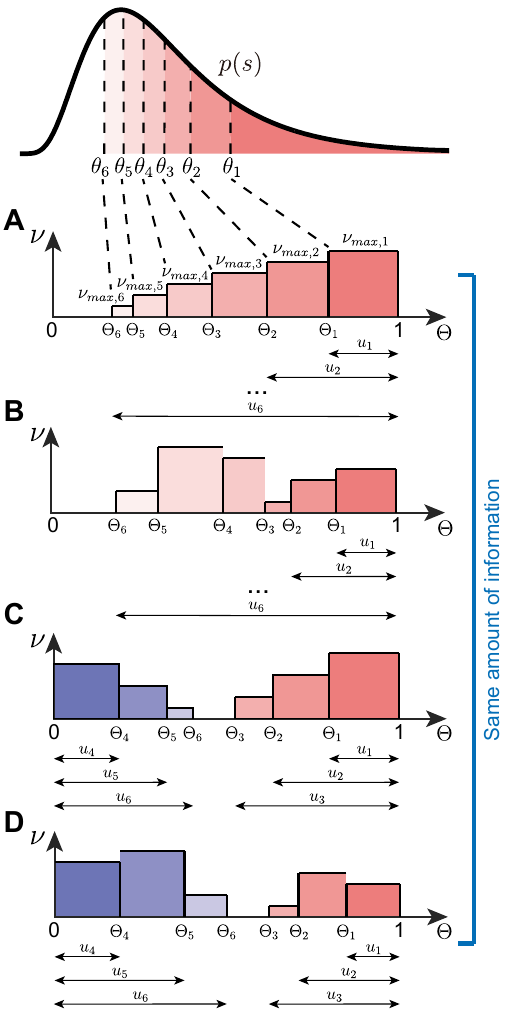}
\centering
\caption{\label{fig:binary_hetero} \textbf{Efficient population coding of binary neurons with heterogeneous maximal firing rates.} \textbf{A.} Optimal configurations of homogeneous populations with only ON neurons. $\Theta_i$ denotes the cumulative probability of $s$ above threshold $\theta_i$.  The optimal thresholds partition the cumulative stimulus space into intervals which increase with the maximal firing rate of the neurons within a population. \textbf{B.} Same as A but with maximal firing rates shuffled. \textbf{C.} Same as A but for mixed populations of ON and OFF neurons. \textbf{D.} Same as C but with maximal firing rates shuffled. All populations code the same amount of mutual information assuming the same distribution of the maximal firing rates  (Eq.~\ref{eq:I_max_binary_hetero}).}
\end{figure}

Different neurons might have different maximal firing rate constraints. For example, ON ganglion cells in the primate retina have higher firing rates than OFF ganglion cells \cite{Uzzell2004}. To explore the effect of these maximal firing rate differences on efficient coding, next we assume that the different neurons in the population might have different maximal firing rates, and consider a heterogeneous population of ON and OFF neurons.  In this case, we define $\nu_{\max, i}$ as the maximal firing rate of neuron $i$, 
\begin{equation}
\begin{aligned}
    R_i &= \nu_{\max,i}\,T, \\
    q_i &= L(0, R_i).
\end{aligned}
\end{equation}
Similar to Eq.~\ref{eq:binary_u_def}, we define $u_i$ as the probability that neurons $i$ fires at its maximal firing rate, i.e., $u_i = \text{Prob}(\nu_i = \nu_{\max,i})$. Then we can prove that the optimal thresholds are (see Supplemental Material 6, Eq. S6.9 and Eq. S6.10)
\begin{equation}
    u_i = \frac{q_i^{q_i/(1-q_i)}+\sum_{j=1}^{i-1}(1-q_j)\,q_j^{q_j/(1-q_j)}}{1+\sum_{j=1}^N(1-q_j)\,q_j^{q_j/(1-q_j)}}
\end{equation}
for ON neurons and
\begin{equation}
    u_i = \frac{q_i^{q_i/(1-q_i)}+\sum_{j=m+1}^{i-1}(1-q_j)\,q_j^{q_j/(1-q_j)}}{1+\sum_{j=1}^N(1-q_j)\,q_j^{q_j/(1-q_j)}}
\end{equation}
for OFF neurons. The maximal mutual information now becomes
\begin{equation}
    I_N = \log\left[1+\sum_{j=1}^N(1-q_j)\,q_j^{q_j/(1-q_j)} \right].
    \label{eq:I_max_binary_hetero}
\end{equation}
This result tells us that as long as the distribution of the maximal firing rates is the same (i.e., the same set of $\lbrace q_i \rbrace$), shuffling the thresholds within the ON and OFF subpopulations, replacing ON neurons with OFF, or replacing OFF neurons with ON, does not change the maximal mutual information (Fig.~\ref{fig:binary_hetero}). 
Similar to Eq.~\ref{eq:binary_p_def}, defining
\begin{equation}
\begin{aligned}
    &p_1 = u_1 \\
    &p_{m+1} = u_{m+1} \\
    &p_i = u_i - u_{i-1}, \quad i = 2, ..., m, m+2, ..., N,
\end{aligned}
\label{eq:p_def_hetero}
\end{equation}
we can derive the stimulus intervals partitioned by the thresholds as:
\begin{equation}
    \begin{aligned}
    p_1 &= \frac{q_1^{q_1/(1-q_1)}}{1+\sum_{j=1}^N(1-q_j)\,q_j^{q_j/(1-q_j)}} \\
    &= e^{-I_N}\,q_1^{q_1/(1-q_1)}\\
    p_{m+1} &= \frac{q_{m+1}^{q_{m+1}/(1-q_{m+1})}}{1+\sum_{j=1}^N(1-q_j)\,q_j^{q_j/(1-q_j)}}\\
    &= e^{-I_N}\,q_{m+1}^{q_{m+1}/(1-q_{m+1})}\\
    p_i &= \frac{q_i^{q_i/(1-q_i)} - q_{i-1}^{1/(1-q_{i-1})}}{1+\sum_{j=1}^N(1-q_j)\,q_j^{q_j/(1-q_j)}} \\
    &= e^{-I_N}\left[q_i^{q_i/(1-q_i)} - q_{i-1}^{1/(1-q_{i-1})}\right], \\ &i = 2, ..., m, m+2, ..., N.
    \end{aligned}
\label{eq:p_hetero}
\end{equation}
One can show that (see Supplemental Material 6)
\begin{equation}
    \begin{aligned}
    \frac{d}{dq}\left(q^{q/(1-q)}\right) < 0, \quad \quad \frac{d}{dq}\left(q^{1/(1-q)}\right) > 0.
    \end{aligned}
\end{equation}
When $q_i \rightarrow 0$ and $q_{i-1} \rightarrow 0$, $p_i$ converges to its maximum $e^{-I_N}$. On the other hand, when $q_i \rightarrow 1$ and $q_{i-1} \rightarrow 1$, $p_i$ converges to its minimum 0. Therefore, given a fixed amount of mutual information $I_N$, $p_1$ increases with $\nu_{\max,1}$, $p_{m+1}$ increases with $\nu_{\max,{m+1}}$, and other $p_i$ increases with both $\nu_{\max,i}$ and $\nu_{\max,i-1}$ (Fig.~\ref{fig:binary_hetero}). This dependency indicates that when the maximal firing rates are heterogeneous in a population, stimulus intervals corresponding to high firing neurons are wider, which can be understood as a result of noise reduction.

\subsection*{E. The optimal activation functions with increasing maximal firing rate constraint \label{sec:beyond_binary}}

\begin{figure}[ht]
\includegraphics[width=0.48\textwidth]{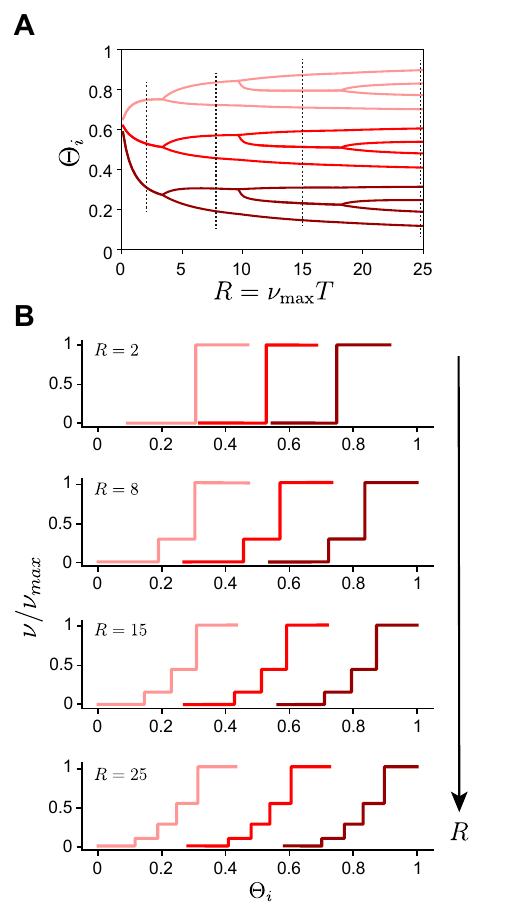}
\centering
\caption{\label{fig:discrete_act_func} \textbf{The optimal activation functions in a neuronal population are discrete with a number of steps (i.e., thresholds) increasing with the maximal firing rate.} \textbf{A.} The optimal thresholds in a population of 3 ON neurons with Poisson noise generation function. $\Theta_i$ denotes the cumulative probability of $s$ above a threshold. The activation functions acquire more discrete steps as the maximal firing rate $\nu_{\max}$ increases (assuming constant $T$). \textbf{B.} Optimal activation functions for the neurons in A. Note that all the activation functions have the same number of steps for a given noise level. The values of $R=\nu_{\max}T$ are indicated on the top left of each plot and shown as dashed lines in A.}
\end{figure}

For a single neuron, Nikitin et al., \cite{Nikitin2009} showed that the optimal activation function is discrete with an increasing number of steps as the maximal firing rate constraint increases (Fig.~\ref{fig:model}C, assuming the coding window $T$ is constant). Given our result that even in a population of neurons the optimal activation functions remain discrete with increasing maximal firing rate constraint (Section III.B), we next determine the number of steps in the activation function for each neuron.

Similar to a single neuron \cite{Nikitin2009}, the number of steps for all neurons in the population increases as a function of increasing maximal firing rate ($\nu_{\max,i}$), or equivalently decreasing level of noise ($R_i=\nu_{\max,i}T$). In a population of neurons with identical maximal firing rates ($\nu_{\max,i} = \nu_{\max}$ and hence $R_i = R$ for all $i$), all the activation functions have the same number of steps for a given noise level, such that, for example, the optimal activation function cannot be binary for one neuron and ternary for another neuron in the same population. To verify this, we perform extensive numerical calculations with multiple noise generation functions $L(n, \nu T)$ for small populations of neurons (Fig.~\ref{fig:discrete_act_func}, Fig. S1, also see Supplemental Material 4), although we lack an analytical proof that this is universally true.

\subsection*{F. The optimal thresholds and the maximal mutual information for a population of ternary neurons
\label{sec:ternary}}
\begin{figure}[th!]
\includegraphics[width=0.48\textwidth]{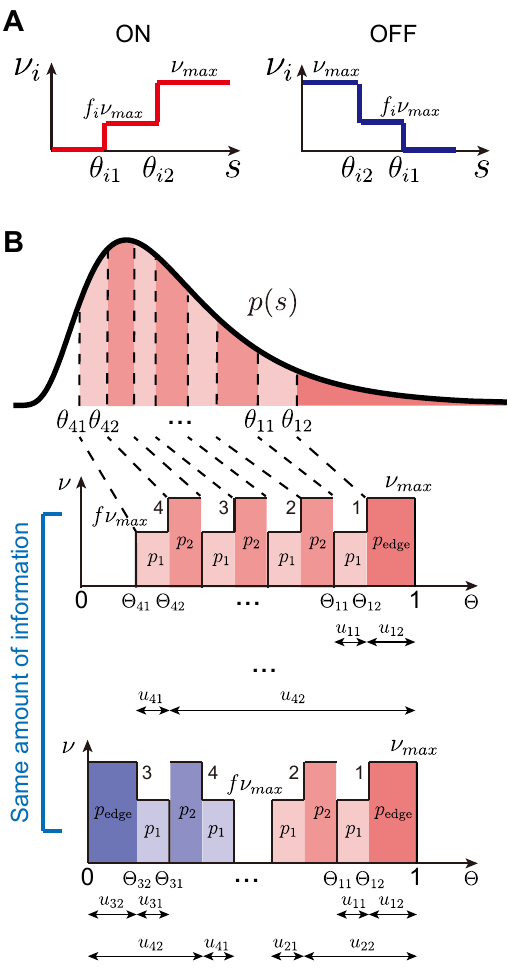}
\centering
\caption{\label{fig:ternary} \textbf{Efficient population coding of ternary neurons.} \textbf{A.} Activation functions of ON and OFF ternary neurons. Each neuron has the same maximal firing rate $\nu_{\max}$, with an activation function described by two thresholds, $\theta_{i,1}$ and $\theta_{i,2}$, and an intermediate firing rate, $f_i \nu_{\max}$. \textbf{B.} Optimal configurations of homogeneous populations with only ON neurons and mixed ON and OFF neurons. $\Theta_{ij}$ denotes the cumulative probability of $s$ above a threshold $\theta_{ij}$ for $j=\{1,2\}$. The optimal thresholds partition the cumulative stimulus space into regular intervals (Eqs.~\ref{eq:u1_u2_solution}-\ref{eq:ternary_p_results_ON_OFF}). The optimized mutual information is independent of the ON-OFF mixture for any noise generation function (Eq.~\ref{eq:I_p1_p2_ternary}).}
\end{figure}

Knowing that at some critical level of maximal firing rate the optimal activation functions of all neurons in the population increase the number of steps, we next generalize our results beyond binary neurons to activation functions with any number of steps and any noise generation function. For the time being, we assume identical maximal firing rates across all neurons in the population. We first start with ternary activation functions with three possible firing rate levels (Fig.~\ref{fig:ternary}A). In this case, one can no longer lump the firing states as we did previously for binary neurons because there is more than one noisy firing level in the activation function. We show that the ``Equal Coding Theorem'' remains valid in that the same maximal information is encoded by any mixture of ON and OFF neurons, and the thresholds of the ternary activation functions of each neuron divide the cumulative stimulus space equally (Fig.~\ref{fig:ternary}B).

Two thresholds, $\theta_{i1}$ and $\theta_{i2}$,  describe a ternary neuron $i$, separating the spontaneous firing rate $\nu = 0$, an intermediate firing rate $\nu = f_i\nu_{\max}$, and the maximal firing rate $\nu = \nu_{\max}$ (Fig.~\ref{fig:ternary}A). In a population of $m$ ON neurons and $N-m$ OFF neurons, we  assign all ON neuron indices, from the highest to the lowest threshold, and for the OFF neurons, from the lowest to the highest, as for the binary neurons  (Eq.~\ref{eq:thresh_index_binary} and Fig.~\ref{fig:binary}B). 

Because we assumed that the dynamic ranges of neurons do not overlap (see Section II), we have
\begin{equation}
\begin{aligned}
    &\theta_{m+1, 2} \leq \theta_{m+1, 1} < ... < \theta_{N, 2} \leq \theta_{N, 1} <\\
    &\quad \theta_{m, 1} \leq \theta_{m, 2} < ... < \theta_{11} \leq \theta_{12}.
    \label{eq:thresh_12_ON_OFF}
\end{aligned}
\end{equation}

We transform the firing thresholds $\lbrace \theta_{i1} \rbrace$ and $\lbrace \theta_{i2} \rbrace$ to intervals of stimulus $s$, as for the binary neurons in Eq.~\ref{eq:binary_u_def}, for the ON neurons,
\begin{equation}
u_{i1} = \int_{\theta_{i1}}^{\theta_{i2}}\de s \, p(s),\ \ u_{i2} = \int_{\theta_{i2}}^{+\infty}\de s\,  p(s)
\label{eq:ternary_u_def}
\end{equation}
and for the OFF neurons,
\begin{equation}
u_{i1} = \int_{\theta_{i2}}^{\theta_{i1}}\de s \, p(s),\ \ u_{i2} = \int_{-\infty}^{\theta_{i2}}\de s \, p(s).
\label{eq:u_ternary_off_neurons}
\end{equation}

We can still find direct relations between $(u_{i1}, u_{i2}, f_i)$ and $(u_{i+1,1}, u_{i+1,2}, f_{i+1})$ (Eq. S7.32, also see Supplemental Material 7.3). Finally, we can express the optimal thresholds and intermediate firing levels as
\begin{equation}
\begin{aligned}
u_{i1} &= \frac{u_1}{1 + (N-1)\left[u_1\left(1-q_1\right) + u_2\left(1-q_2\right) \right]} \\
u_{i2} &= \frac{u_2+(i-1)\left[u_1\left(1-q_1\right) + u_2\left(1-q_2\right) \right]}{1 + (N-1)\left[u_1\left(1-q_1\right) + u_2\left(1-q_2\right) \right]} \\
&\quad \quad \quad \quad\text{(ON neurons, } i = 1, ..., m \text{)} \\
u_{i2} &= \frac{u_2+(i-m-1)\left[u_1\left(1-q_1\right) + u_2\left(1-q_2\right) \right]}{1 + (N-1)\left[u_1\left(1-q_1\right) + u_2\left(1-q_2\right) \right]} \\
&\quad \quad \quad \quad\text{(OFF neurons, } i = m+1, ..., N \text{)} \\
f_i &= f,
\label{eq:u1_u2_solution}
\end{aligned}
\end{equation}
where $(u_1, u_2, f)$ is the optimal $(u_{11}, u_{12}, f_1)$ in a population of a single neuron ($N = 1$), $q_1 = L(0, fR)$ and $q_2 = L(0, R)$ with $R=\nu_{\max}T$.

Defining the cumulative stimulus intervals as Eq.~\ref{eq:binary_p_def},
\begin{equation}
\begin{aligned}
&p_{i1} = u_{i1}\\
&p_{12} = u_{12}, p_{m+1,2} = u_{m+1,2}\\
&p_{i2} = u_{i2} - u_{i-1, 2} - u_{i-1, 1}, \\ &i = 2, ..., m, m+2, ..., N
\end{aligned}
\label{eq:ternary_p_def}
\end{equation}
which are the probabilities of the stimulus intervals that have the same firing rates $\vec \nu$. With optimal $\lbrace u_{i1} \rbrace$, $\lbrace u_{i2} \rbrace$, and $\lbrace f_i \rbrace$ (Eq.~\ref{eq:u1_u2_solution}), these intervals can be expressed as
\begin{widetext}
\begin{equation}
\begin{aligned}
&p_{11} = p_{21} = ... = p_{N1} = \frac{u_1}{1 + (N-1)\left[u_1\left(1-q_1\right) + u_2\left(1-q_2\right) \right]} \overset{\text {def}}{=} p_1 \\
&p_{22} = ... = p_{m2} = p_{m+2, 2} = ... = p_{N2} = \frac{-u_1q_1 + u_2\left(1-q_2\right) }{1 + (N-1)\left[u_1\left(1-q_1\right) + u_2\left(1-q_2\right) \right]} \overset{\text {def}}{=} p_2 \\
&p_{12} = p_{m+1, 2} = \frac{u_2}{1 + (N-1)\left[u_1\left(1-q_1\right) + u_2\left(1-q_2\right) \right]} \overset{\text {def}}{=} \pe \\
&p_1 q_1 + p_2 = \pe\left(1-q_2 \right).
\end{aligned}
\label{eq:ternary_p_results_ON_OFF}
\end{equation}
\end{widetext}
This derives the optimal thresholds for a mixed population of ON and OFF cells (Fig.~\ref{fig:ternary}B).

Similar to binary neurons (Eq.~\ref{eqn:binary_info}, also see Supplemental Material 7.1), the maximal mutual information can be related to $P(\vec 0)$ as
\begin{equation}
I_N^{\max} = -\log P(\vec 0).
\end{equation}
This allows us to derive the relationship between the maximal mutual information and the stimulus interval $p$ as (see Supplemental Material 7.8)
\begin{equation}
I_N^{\max} = -\log \left[1 - N(p_1 + p_2)\right].
\label{eq:I_max_p1_p2}
\end{equation}
Using the optimal values of $p_1$ and $p_2$ (Eq. S7.57 and Eq.~\ref{eq:ternary_p_results_ON_OFF}), we have
\begin{equation}
I_N^{\max} = -\log \frac{1 - u_1\left(1-q_1\right) - u_2\left(1-q_2\right)}{1 + (N-1)\left[u_1\left(1-q_1\right) + u_2\left(1-q_2\right) \right]}.
\label{eq:I_p1_p2_ternary}
\end{equation}
We can then write the maximal mutual information of an $N$-neuron ternary population as a function of the maximal mutual information of a single neuron population:
\begin{equation}
I_N^{\max} = \log \left[N(\exp({I_1^{\max}})-1)+1 \right],
\label{eq:IN_I1_ternary}
\end{equation}
similar to the case with binary neurons (Eq.~\ref{eq:IN_I1}).

\subsection*{G. The optimal thresholds and the maximal mutual information for a population of neurons with any shapes of activation functions
\label{sec:M_ary}}

\begin{figure*}[th!]
\includegraphics[width=17.2cm]{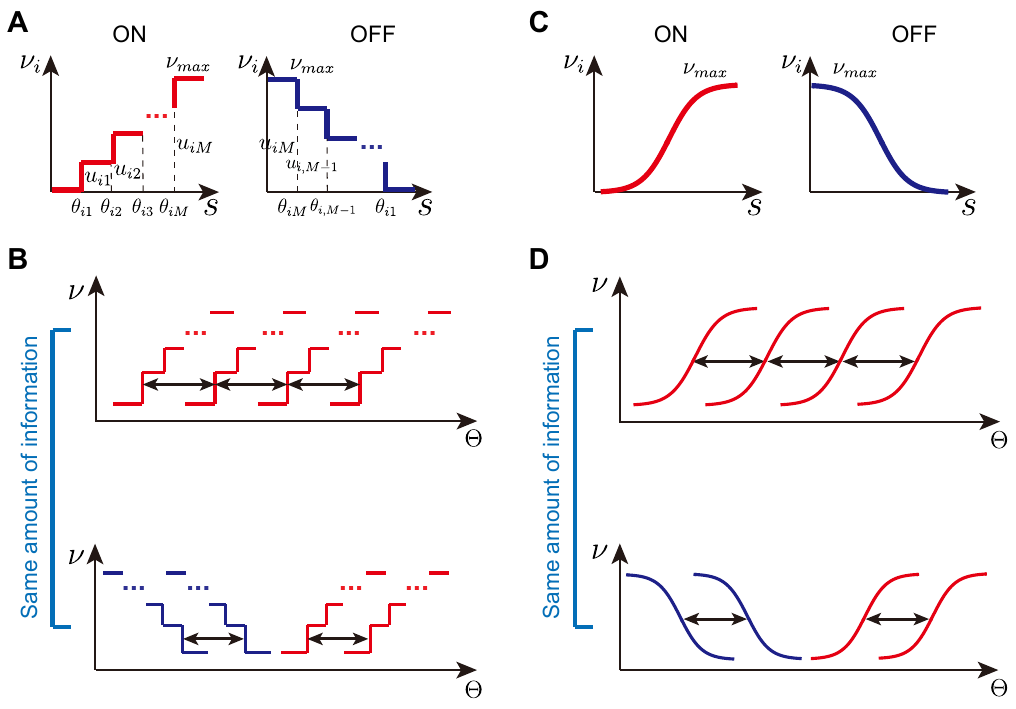}
\centering
\caption{\label{fig:any_tuning_curves} \textbf{A.} \textbf{Efficient population coding of neurons with any activation function.} Activation functions of ON and OFF $(M+1)$-ary neurons. Each neuron has the same maximal firing rate $\nu_{\max}$, and $M+1$ possible firing rates. \textbf{B.} Optimal configurations of homogeneous ON population and ON-OFF mixtures of $(M+1)$-ary neurons. Black arrows indicate that the distances between each two adjacent activation functions are the same in the probability space of the stimuli $s$. \textbf{C.} As $M \rightarrow +\infty$, the activation functions of $(M+1)$-ary neurons become continuous. \textbf{D.} Correspondingly, the optimal configuration of activation functions also transforms into the continuous limit as $M \rightarrow +\infty$.}
\end{figure*}

After generalizing the optimal thresholds in our efficient population coding framework from binary to ternary neurons, it is now straightforward to generalize them to activation functions with any number of steps. Activation functions with more than three steps can be represented with multiple $p_{ij}$ and $f_i$. For example, an $(M+1)$-ary neuron $i$ can be described with $M$
thresholds, $\theta_{i1}, ..., \theta_{iM}$,  separating the spontaneous firing rate $\nu = 0$, intermediate firing rate $\nu = f_{i1}\nu_{\max}, ..., f_{i,M-1}\nu_{\max}$, and the maximal firing rate $\nu = \nu_{\max}$ (Fig.~\ref{fig:any_tuning_curves}A). For simplicity, we also define $f_{iM} = 1$ for any index $i$.

Defining the cumulative stimulus thresholds similarly to binary neurons (Eq.~\ref{eq:binary_u_def} and Eq.~\ref{eq:binary_p_def}) and ternary neurons (Eq.~\ref{eq:ternary_u_def} and Eq.~\ref{eq:ternary_p_def}), we have for ON neurons
\begin{equation}
\begin{aligned}
&u_{ik} = \int_{\theta_{ik}}^{\theta_{i,k+1}}\de s \, p(s), k = 1, ..., M-1 \\
&u_{iM} = \int_{\theta_{iM}}^{+\infty}\de s\,  p(s)
\end{aligned}
\label{eq:m+1_ary_u_def}
\end{equation}
and for OFF neurons
\begin{equation}
\begin{aligned}
&u_{ik} = \int_{\theta_{i,k+1}}^{\theta_{ik}}\de s \, p(s), k = 1, ..., M-1 \\
&u_{iM} = \int_{-\infty}^{\theta_{iM}}\de s\,  p(s).
\end{aligned}
\label{eq:M+1_ary_u_def}
\end{equation}
Then the cumulative stimulus intervals can be written as
\begin{equation}
\begin{aligned}
&p_{ik} = u_{ik},~ i = 1, ..., N;~ k = 1, ..., M-1 \\
&p_{1M} = u_{1M}, ~p_{m+1,M} = u_{m+1,M} \\
&p_{iM} = u_{iM} - \sum_{k=1}^M u_{i-1, k},~i = 2, ..., m, m+2, ..., N 
\end{aligned}
\label{eq:M+1_ary_p_def}
\end{equation}
and based on this we can calculate the mutual information and the optimal thresholds for discrete activation functions with any number of steps.

The calculations with $(M+1)$-ary neurons follow very similarly to the calculations with ternary neurons (see Supplemental Material 8). In summary, for a population of $m$ ON and $N-m$ OFF $(M+1)$-ary neurons, when the mutual information $I(s, \vec n)$ is maximized, we have
\begin{equation}
\begin{aligned}
&p_{1k} = p_{2k} = ... = p_{Nk} \overset{\text{def}}{=} p_k,~k = 1,  2, ..., M-1 \\
&p_{2M} = ... = p_{mM} = p_{m+2, M} = ... = p_{NM} \overset{\text{def}}{=} p_M \\
&p_{1M} = p_{m+1, M} \overset{\text{def}}{=} \pe \\
&f_{1k} = f_{2k} = ... = f_{Nk} \overset{\text{def}}{=} f_k,~k = 1,  2, ..., M-1 \\
&\sum_{k=1}^{M-1}p_k L(0,f_kR) + p_M = \pe\left(1-L(0,R)\right).
\end{aligned}
\label{eq:M+1_ary_pk}
\end{equation}
This derives the optimal thresholds for populations of $(M+1)$-ary neurons (Fig.~\ref{fig:any_tuning_curves}B). 

Also same as before (Eq. S7.21, Eq.~\ref{eq:I_max_p1_p2}, Eq.~\ref{eq:IN_I1}, and Eq.~\ref{eq:IN_I1_ternary}), we still have
\begin{equation}
I^{\max}_N = -\log P(\vec 0),
\end{equation}
\begin{equation}
I^{\max}_N = -\log \left[1 - N\sum_{k=1}^M p_k\right],
\label{eq:I_p_M_ary}
\end{equation}
and
\begin{equation}
I^{\max}_N = \log \left[N(\exp(I^{\max}_1) - 1) + 1\right].
\label{eq:IN_I1_M_ary}
\end{equation}
We find that the maximal mutual information increases logarithmically with the number of neurons, independent of the specific shape of the activation functions and the type of noise.

Even letting $M \rightarrow \infty$, which corresponds to the limiting case in which the activation functions are continuous (Fig.~\ref{fig:any_tuning_curves}C) the optimal activation functions have the same shape on the probability space of the stimuli $s$, and the displacement from one to the next is a constant (Fig.~\ref{fig:any_tuning_curves}D).

\subsection*{H. The optimal thresholds and the maximal mutual information for a population of neurons with any shapes of activation functions and heterogeneous maximal firing rates
\label{sec:M_ary_het}}

The optimal thresholds of ternary neurons or any shapes of activation functions that we derived for identical maximal firing rates for all cells can also be generalized to heterogeneous maximal firing rates which are different across the different cells. We first start with ternary neurons. Similarly as before (Eqs.~\ref{eq:thresh_12_ON_OFF}-\ref{eq:u_ternary_off_neurons}), for every neuron $i$ in a population, we define an intermediate firing rate $f_i\,\nu_{\max,i}$, the cumulative stimulus intervals
\begin{equation}
\begin{aligned}
    u_{i1} &= \text{Prob}\,(\nu_i = f_i\,\nu_{\max,i}) \\
    u_{i2} &= \text{Prob}\,(\nu_i = \nu_{\max,i}),
    \label{eq:u1_u2_def_ternary_hetero}
\end{aligned}
\end{equation}
and
\begin{equation}
\begin{aligned}
    q_{i1} &= L(0, f_iR_i) \\
    q_{i2} &= L(0, R_i).
\end{aligned}
\end{equation}
Instead of searching for relations between $(u_{i1}, u_{i2}, f_i)$ and $(u_{i+1,1}, u_{i+1,2}, f_{i+1})$ (Eq. S7.32), here we need to derive direct relations between neuron $i$ in a population and a population of a single neuron with the same maximal firing rate (also $\nu_{\max, i}$). Therefore, for such a population of a single neuron, given its maximal firing rate $\nu_{\max}$, we denote $R = \nu_{\max}T$, and then
\begin{itemize}
    \item $u_1^*(R)$ as its optimal $u_1$
    \item $u_2^*(R)$ as its optimal $u_2$
    \item $f^*(R)$ as its optimal $f$
    \item $q_1^*(R) = L(0, f^*(R)R)$ as the probability $p(n=0|\nu=f^*(R)R)$ after optimization
    \item $q_2^*(R) = L(0, R)$ as the probability $p(n=0|\nu=R)$ after optimization
    \item $P^*(0,R) = 1-u_1^*(R)[1-q_1^*(R)] - u_2^*(R)[1-q_2^*(R)]$ as the probability of observing no spike after optimization.
\end{itemize}

With these new definitions, we can derive the optimal thresholds and intermediate firing rates as (also see Supplemental Material 9)
\begin{equation}
    \begin{aligned}
        u_{i1} &= \frac{u_1^*(R_i)}{P^*(0, R_i)}e^{-I_N} \, \\
        u_{i2} &= \left[\frac{u_2^*(R_i)}{P^*(0, R_i)}-\sum_{j=i}^N \frac{1}{P^*(0, R_j)}+(N-i)\right]e^{-I_N} + 1 \\
        &\quad \quad \quad \quad\text{(ON neurons, } i = 1, ..., m \text{)}\\
        u_{i2} &= \Bigg[\frac{u_2^*(R_i)}{P^*(0, R_i)}-\sum_{j=i}^N \frac{1}{P^*(0, R_j)} \\
        &\quad \quad-\sum_{j=1}^m \frac{1}{P^*(0, R_j)}+(N-i+m)\Bigg]e^{-I_N} + 1 \\
        &\quad \quad \quad \quad\text{(OFF neurons, } i = m+1, ..., N \text{)}\\
        f_i &= f^*(R_i).
    \end{aligned}
\end{equation}
and the maximal mutual information as
\begin{equation}
    I_N = \log\left[\sum_{j=i}^N \frac{1}{P^*(0, R_j)}-(N-1)\right].
    \label{eq:I_N_ternary_hetero_nu}
\end{equation}
Note that similarly to the case of binary neurons with heterogeneous maximal firing rates (Eq.~\ref{eq:I_max_binary_hetero}, Fig.~\ref{fig:binary_hetero}), this equation indicates that the maximal mutual information is the same as long as all the distribution of maximal firing rates is the same, independent of how the population is mixed, any threshold shuffling within the ON and OFF subpopulations and any replacing of ON with OFF neurons and vice versa.

Defining $p_{i1} = u_{i1}, p_{12} = u_{12}, p_{i2} = u_{i-1,2} - u_{i-1,1} - u_{i2}\,\,(i>1)$, we can express the stimulus intervals as
\begin{widetext}
\begin{equation}
    \begin{aligned}
        p_{i1} &= \frac{u_1^*(R_i)}{P^*(0, R_i)}e^{-I_N} \, \\
        p_{i2} &= \left[\frac{u_2^*(R_i)-u_1^*(R_i)}{P^*(0, R_i)}-\frac{u_2^*(R_{i-1})\,q_2^*(R_{i-1})-u_1^*(R_{i-1})\left(1-q_1^*(R_{i-1})\right)}{P^*(0, R_{i-1})}\right]e^{-I_N}\\
        &\quad \quad(i = 2, ..., m, m+2, ..., N)\\
        p_{12} &= \left[\frac{u_2^*(R_1)}{P^*(0, R_1)}-\sum_{j=1}^N \frac{1}{P^*(0, R_j)}+(N-1)\right]e^{-I_N} + 1\\
        p_{m+1,2} &= \left[\frac{u_2^*(R_{m+1})}{P^*(0, R_{m+1})}-\sum_{j=1}^N \frac{1}{P^*(0, R_j)}+(N-1)\right]e^{-I_N} + 1\\
        f_i &= f^*(R_i).
    \end{aligned}
    \label{eq:optimal_p_ternary_hetero}
\end{equation}
\end{widetext}

\begin{figure*}
    \includegraphics[width=0.95\textwidth]{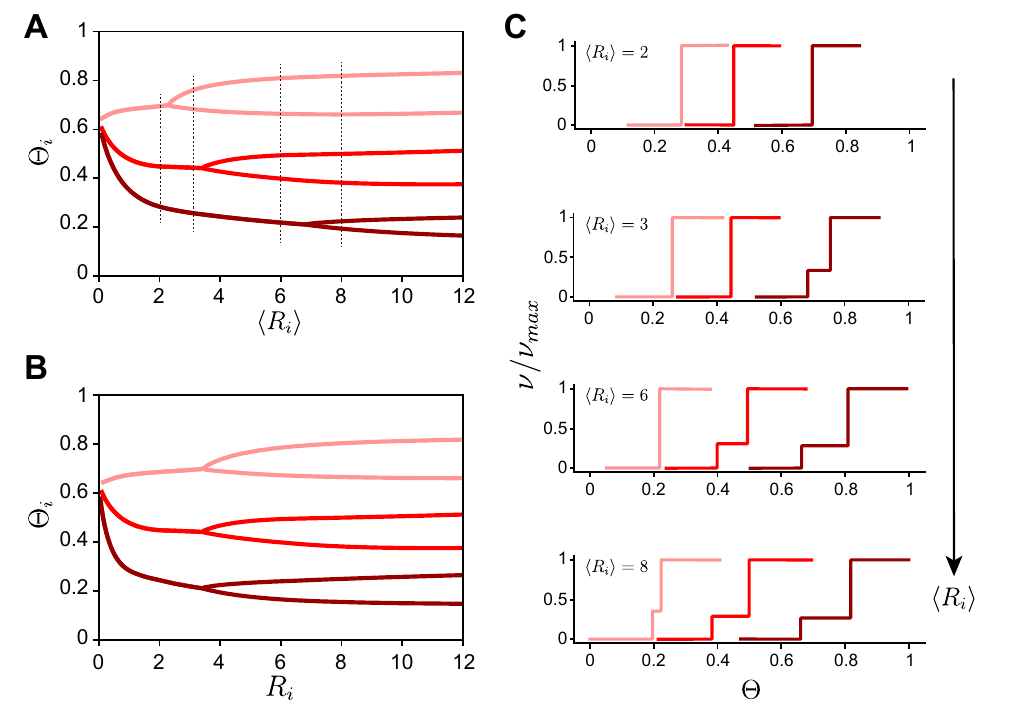}
    \centering
    \caption{\label{fig:bifurcation_hetero} \textbf{The optimal activation functions in a neuronal population with heterogeneous maximal firing rate.} The population consists of $N=3$ ternary ON neurons, with $R_1:R_2:R_3 = 1:2:3$.  
    \textbf{A.} Optimal thresholds as a function of $\langle R_i \rangle = (R_1+R_2+R_3)/3$. 
    \textbf{B.} Optimal thresholds as a function of $R_i$ of individual neurons.
    \textbf{C.} Optimal activation functions at different values of $\langle R_i \rangle$ (dashed lines in A). Note that the activation functions can have different numbers of steps in the same population. $\langle R_i \rangle$ values are indicated on the top left corner of each plot.}
\end{figure*}
Here, $f_i = f^*(R_i)$ indicates that in a population, the optimal activation functions of different neurons may consist of different numbers of steps depending on the maximal firing rate constraint of those neurons (Fig.~\ref{fig:bifurcation_hetero}). This result is unique to the case of heterogeneous maximal firing rates in the population, compared to the case of identical maximal firing rates for all cells where all the optimal activation functions in the population have same number of steps (Fig.~\ref{fig:discrete_act_func}). The bifurcations where the optimal activation functions acquire more steps are now neuron-specific (Fig.~\ref{fig:bifurcation_hetero}), only depending on the maximal firing rate of single neurons but not other neurons in the same population. This pattern of bifurcation  differs from the case of identical maximal firing rates for all cells where the bifurcations occur at the same maximal firing rates for the whole population (Fig.~\ref{fig:discrete_act_func}).

%% this paragraph can be deleted
Unlike binary neurons (Eq.~\ref{eq:p_hetero}), here we cannot analytically find how the stimulus intervals $\lbrace p_{i1}, p_{i2}\rbrace$ depend on the maximal firing rates $\nu_{\max,i}$, or equivalently, the maximum spike counts $\lbrace R_i \rbrace$. To obtain some understanding of the optimal thresholds in Eq.~\ref{eq:optimal_p_ternary_hetero}, we performed numerical calculations for Poisson noise (see Supplemental Material 9, Fig. S2, and Fig. S3). 

%leave the details out.... add them maybe to supplementary
%

Similarly to ternary neurons (also see Supplemental Material 9), we can also derive the optimal thresholds and intermediate firing rates for any activation function (with any number of steps) as
\begin{equation}
    \begin{aligned}
        u_{ik} &= \frac{u_k^*(R_i)}{P^*(0, R_i)}e^{-I_N} \quad k = 1, ..., M-1 \\
        u_{iM} &= \left[\frac{u_M^*(R_i)}{P^*(0, R_i)}-\sum_{j=i}^N \frac{1}{P^*(0, R_j)}+(N-i)\right]e^{-I_N} + 1 \\
        &\quad \quad \quad \quad\text{(ON neurons, } i = 1, ..., m \text{)}\\
        u_{iM} &= \Bigg[\frac{u_M^*(R_i)}{P^*(0, R_i)}-\sum_{j=i}^N \frac{1}{P^*(0, R_j)} \\
        &\quad \quad-\sum_{j=1}^m \frac{1}{P^*(0, R_j)}+(N-i+m)\Bigg]e^{-I_N} + 1 \\
        &\quad \quad \quad \quad\text{(OFF neurons, } i = m+1, ..., N \text{)}\\
        f_{ik} &= f_k^*(R_i), \quad k = 1, ..., M-1.
    \end{aligned}
\end{equation}
The maximal mutual information can be expressed in the same way as for ternary neurons (Eq.~\ref{eq:I_N_ternary_hetero_nu}):
\begin{equation}
    I_N = \log\left[\sum_{j=i}^N \frac{1}{P^*(0, R_j)}-(N-1)\right],
\end{equation}
which is again independent of the ON-OFF mixture and the shuffling within the ON or OFF subpopulations.
Finally, the optimal stimulus intervals are given by
\begin{widetext}
\begin{equation}
    \begin{aligned}
        p_{ik} &= \frac{u_1^*(R_i)}{P^*(0, R_i)}e^{-I_N} \quad k = 1, ..., M-1 \\
        p_{iM} &= \left[\frac{u_M^*(R_i)-\sum_{k=1}^{M-1}u_k^*(R_i)}{P^*(0, R_i)}-\frac{u_M^*(R_{i-1})\,q_M^*(R_{i-1})-\sum_{k=1}^{M-1}u_k^*(R_{i-1})\left(1-q_k^*(R_{i-1})\right)}{P^*(0, R_{i-1})}\right]e^{-I_N} \\
        &\quad \quad(i = 2, ..., m, m+2, ..., N)\\
        p_{1M} &= \left[\frac{u_M^*(R_1)}{P^*(0, R_1)}-\sum_{j=1}^N \frac{1}{P^*(0, R_j)}+(N-1)\right]e^{-I_N} + 1\\
        p_{m+1,M} &= \left[\frac{u_2^*(R_{m+1})}{P^*(0, R_{m+1})}-\sum_{j=1}^N \frac{1}{P^*(0, R_j)}+(N-1)\right]e^{-I_N} + 1\\
        f_{ik} &= f_k^*(R_i), \quad k = 1, ..., M-1.
    \end{aligned}
\end{equation}
\end{widetext}

In sum, we have generalized the efficient coding framework to any spiking noise, any shapes of activation functions, and any distribution of maximal firing rates. Although the solution becomes more complicated as we relax the assumptions, we can analytically link the optimal thresholds and optimal intermediate firing rates to those of a single neuron. Thus, the dependence of the optimal activation functions on the maximal firing rates, or equivalently the noise levels, can be interpreted intuitively. In addition, for any spiking noise, shapes of activation functions, and distribution of maximal firing rates, the maximal mutual information is always independent of the mixture of ON and OFF neurons, and the reordering of firing thresholds within ON or OFF subpopulations.

\subsection*{I. Information per spike is the highest for equal ON-OFF mixtures
\label{sec:info_spike}}

\begin{figure*}[htb]
\includegraphics[width=0.95\textwidth]{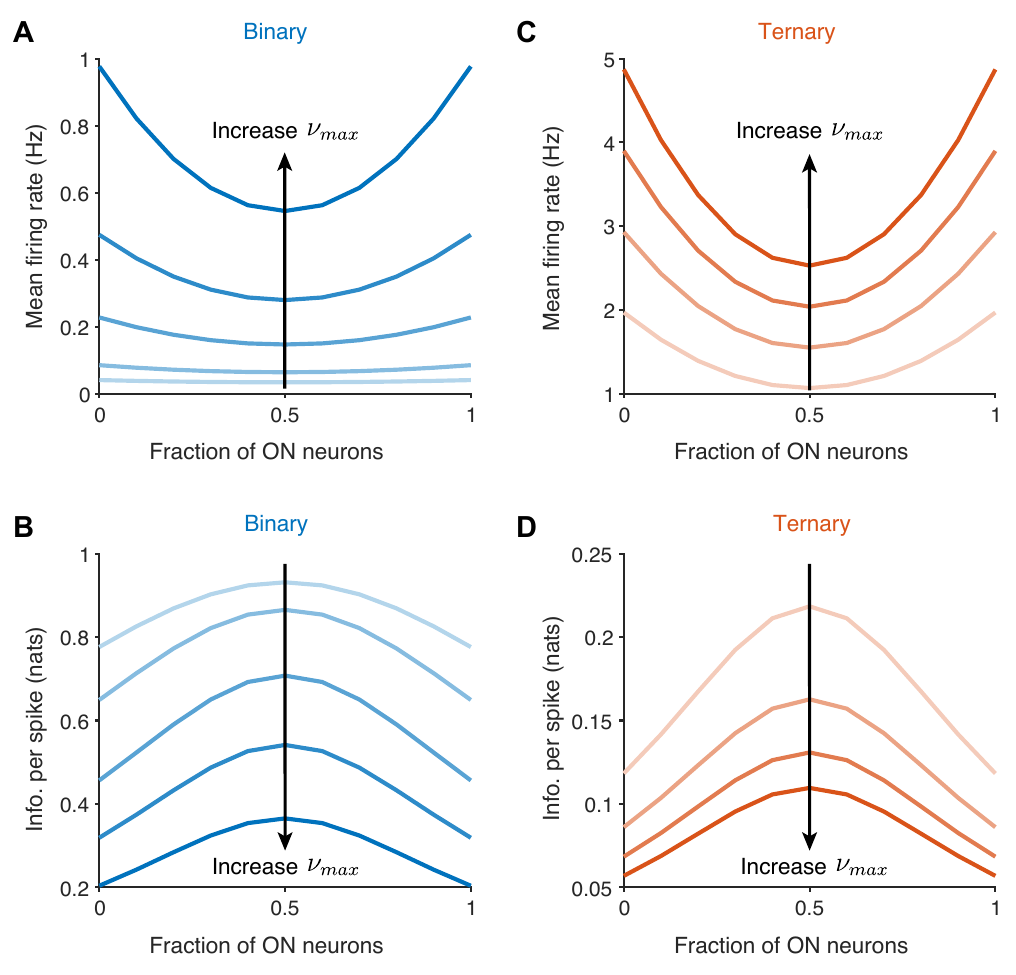}
\centering
\caption{\label{fig:info_per_spike} \textbf{Information per spike is the highest for equal ON-OFF mixtures.} \textbf{A.} The total mean firing rate $\bar \nu$ of neurons in populations with different fractions of ON neurons ($m/N$).  \textbf{B.} Information per spike ($I/(\bar \nu T)$) of neuron populations of different ON neuron ratios ($m/N$). In A and B, $T = 1\,s$ and different line colors correspond to $\nu_{\max} = 0.1, \, 0.2, \, 0.5, \, 1, \, 2$.  \textbf{C.} Same as A but for ternary neurons. \textbf{D.} Same as B but for ternary neurons. In C and D, $\nu_{\max} = 4, \, 6, \, 8, \, 10$. Population size is $N = 10$ for all the panels.}
\end{figure*}

While the total mutual information about a stimulus encoded by any mixture of ON and OFF cells is identical, the different ON and OFF populations have different total metabolic costs, measured in terms of the total mean firing rate of the population. Using the derived optimal thresholds, we can calculate the total population mean firing rate for  different ON and OFF mixtures. For a population of binary neurons, same as \cite{Gjorgjieva2019}, the mean firing rate $\bar \nu$ can be calculated as
\begin{equation}
    \bar \nu = \left[\pe + \frac{N-1}{2} p + \frac{m}{N} \left(m -N\right)p \right]\nu_{\max},
    \label{eq:avg_fr_binary}
\end{equation}
which reaches its maximum when $m = 0$ or $m = N$ (a homogeneous ON or homogeneous OFF population), and reaches its minimum when $m \rightarrow N/2$ (an equally mixed ON-OFF population, Fig.~\ref{fig:info_per_spike}A). 

Since the maximal mutual information is the same for homogeneous populations (all ON or all OFF) and all ON-OFF mixtures, the information per spike is the highest for equal ON-OFF mixtures and the lowest for homogeneous populations with only ON or OFF neurons (Fig.~\ref{fig:info_per_spike}B).

With ternary neurons, the average firing rate $\bar \nu$ in the population with optimal thresholds becomes (see Supplemental Material 7.7):
\begin{equation}
\begin{aligned}
    \bar \nu &= \Big[ \pe + p_1 f + \frac{N-1}{2}\left(p_1 + p_2\right) + \\
    & \quad \quad \frac{m}{N}\left(m-N\right)\left(p_1+p_2 \right) \Big] \nu_{\max}.
\end{aligned}
\label{eq:avg_fr_ternary}
\end{equation}

%Again, $\bar \nu$ reaches its maximum when $m = 0$ or $m = N$ (a homogeneous ON or homogeneous OFF population), and reaches its minimum when $m \rightarrow N/2$ (an equally mixed ON-OFF population, Fig.~\ref{fig:info_per_spike}C). As a result, $m \rightarrow N/2$ generates the highest information per spike (Fig.~\ref{fig:info_per_spike}D).

And generalizing this to $(M+1)$-ary neurons, the mean firing rate of the population $\bar \nu$ becomes 
\begin{equation}
\begin{aligned}
    \bar \nu &= \Big[ \pe + \frac{N-1}{2}\sum_{k = 1}^M p_k \\
    &+ \frac{m}{N}\left(m-N\right)\sum_{k = 1}^M p_k + \sum_{k=1}^{M-1} p_k f_k \Big] \nu_{\max}.
\end{aligned}
\label{eq:avg_fr_M+1_ary}
\end{equation}
As in the population with binary neurons, also here the highest mean firing rate is achieved for the purely ON or OFF populations with $m = 0$ or $m = N$ and the minimum is reached when $m \rightarrow N/2$ (also see Supplemental Material 8.3). Thus, similar to binary or ternary neurons, the information per spike is the highest in the equally mixed ON and OFF population, i.e., when $m \rightarrow N/2$.

When the maximal firing rates are heterogeneous, the average firing rate $\bar \nu$ does not only depend on the ratio of ON vs.~OFF neurons, but also depends on the order of neurons, i.e.~the shuffling of thresholds in the population. We can demonstrate that for binary neurons (see Supplemental Material 6.2), (1) homogeneous ON or OFF populations with only ON or OFF cells generate the highest $\bar \nu$. (2) To reach the lowest $\bar \nu$, neurons with higher maximal firing rates should have smaller probability to fire (smaller $u_i$). Mathematically, given the predefined order of thresholds (Eq.~\ref{eq:thresh_index_binary}), that is,
\begin{equation}
\nonumber
\theta_{m+1} < ... < \theta_N < \theta_m < ... < \theta_1,
\end{equation}
there should be 
\begin{equation}
\begin{aligned}
    &\nu_{\max, 1} \geq ... \geq \nu_{\max, m}\\
    &\nu_{\max, m+1} \geq ... \geq \nu_{\max, N}.
    \label{eq:binary_hetero_lowest_nu_1}
\end{aligned}
\end{equation}
As an intuitive example, in Fig.~\ref{fig:binary_hetero}, $\bar \nu$ is lower in panel C than in D, and also lower in A compared to B.
(3) When the lowest $\bar \nu$ is reached, the number of ON neurons $m$ should satisfy
\begin{equation}
    \frac{\sum_{i = m+1}^N \log q_i}{\sum_{i=1}^{m-1}(1-q_i)q_i^{q_i/(1-q_i)}} \simeq \frac{\log q_m}{(1-q_m)q_m^{q_m/(1-q_m)}}.
    \label{eq:binary_hetero_lowest_nu_2}
\end{equation}

Point (3) can be understood as the balance between the activity of ON and OFF neurons. $|\log q_i|$ and $(1-q_i)q_i^{q_i/(1-q_i)}$ both decrease with $q_i$, and thus, increase with $R_i$. Therefore, the denominator of the left side of Eq.~\ref{eq:binary_hetero_lowest_nu_2}, $\sum_{i=1}^{m-1}(1-q_i)q_i^{q_i/(1-q_i)}$, is approximately proportional to the total firing rate of ON neurons, that is $\sum_{i=1}^m \nu_{\max,i}$. Similarly, the numerator $\sum_{i = m+1}^N \log q_i$ is approximately proportional to the total firing rate of OFF neurons, i.e., $\sum_{i=m+1}^N \nu_{\max,i}$. Therefore, Eq.~\ref{eq:binary_hetero_lowest_nu_2} provides  the optimal firing rate ratio of ON vs.~OFF neurons  to minimize the mean firing rate of the mixed population. As a special case, when the maximal firing rates are identical for all cells, i.e.~$q_i = q$, Eq.~\ref{eq:binary_hetero_lowest_nu_2} degenerates into $m \simeq (N+1)/2$, which is consistent with the optimal $m$ in Eq.~\ref{eq:avg_fr_binary}-\ref{eq:avg_fr_M+1_ary} above (also see Fig.~\ref{fig:info_per_spike}).

In sum, for any shape of activation functions, although the maximal mutual information is independent of the numbers of ON and OFF cells, equal ON-OFF mixtures are more efficient in that their mean firing rate is the lowest, hence producing the highest information per spike. In the case of heterogeneous maximal firing rates, it is more efficient if neurons with higher maximal firing rates fire with lower probabilities. Intuitively, the mean firing rate of a neuron population can be related to the overlap of the neurons' high firing states -- when the overlap is bigger, the neurons fire simultaneously, increasing the metabolic cost. 

\section*{IV. Discussion}
Many neural systems process information using multiple neurons in parallel to encode a sensory stimulus in the presence of different biological constraints. In this work, we have developed an efficient coding theory that maximizes the Shannon mutual information between stimuli and neuronal spikes subject to neuronal noise. We considered several aspects of the neuronal populations to make them more biologically realistic, including ON and OFF neurons, different shapes of neuronal activation functions and a heterogeneity of maximal firing rates, different statistics of noise in the spike generation mechanism, and different numbers of neurons. 

We assumed that each neuron has a zero spontaneous firing rate and a fixed maximal firing rate, which can be interpreted as the inverse of the noise level. Generalizing previous results beyond Poisson spiking to any noise statistics, we first showed that the optimal activation function of a single neuron that maximizes the mutual information must be discrete. Additionally, we demonstrated that also in a population of any number of neurons and any noise statistics,  the optimal activation functions are all discrete. Intuitively, discrete activation functions could be advantageous because they create distinct firing levels that are more distinguishable in the presence of spiking noise, compared to continuous activation functions. As the maximal firing rate increases, the different firing levels become easier to separate, making room for more steps in the activation functions. Interestingly, for a population of fixed size with identical maximal firing rates for all cells, we found that the number of steps in the population increases with increasing the maximal firing rate constraint (or equivalently, decreasing noise level) simultaneously for all the neurons. This implies that at any noise level, all neurons have the same activation function shape, i.e.,~an activation function that consists of the same number of steps. When the maximal firing rates of the different cells are heterogeneous, the number of steps in the optimal activation function of each neuron depends on the maximal firing rate of that neuron only. Therefore, neurons in the same population can have different activation function shape.

A second aspect of optimal coding pertains to the optimal activation functions distribution of all neurons in stimulus space.  Remarkably, when the maximal firing rates of all cells are the same, the optimal activation functions divide the probability space of the stimulus into equal regions, hence implementing a coding strategy which emphasizes stimuli that occur with higher probability. While this result has been long known in the limit of low noise (high maximal firing rate) as `histogram equalization' \cite{Laughlin1981}, we show that it holds for any noise generation function and any amount of noise. When the maximal firing rates of the different cells are heterogeneous, the organization of thresholds is no longer regular but rather results in smaller (larger) stimulus regions when the firing rates are low (high) to compensate for the higher (lower) noise.

Finally, by considering populations with different proportions of ON and OFF neurons, we demonstrated that the maximal information is independent of the ON-OFF mixture. ON and OFF splitting of a sensory signals has been experimentally observed in various sensory systems, including the mammalian retina \cite{Ratliff2010, Berens2017}, the medulla of \textit{Drosophila} visual system \cite{Clark2011, Ketkar2022}, and the mechanosensory neurons in the legs of the adult \textit{Drosophila} \cite{Mamiya2018}. 
Our theory applies only to populations which code for the same one dimensional stimulus variable using just the spike count. However, in the vertebrate retina, diverse types of retinal ganglion cells represent a visual stimulus, and differ in their spatial and temporal processing characteristics \cite{sanes_15, Baden2016}. Nonetheless, certain types of ganglion cells exist both as ON and OFF, with otherwise similar spatial location and temporal processing features, and hence are consistent with the assumptions in our theory. Considering a theoretical framework for the encoding of a scalar stimulus which has no spatial or temporal correlations is clearly a limitation of our model,  since real-world stimuli possess rich temporal and spatial patterns \cite{Ratliff2010, Kayser2003, Dong1995, Simmons2013}. Several frameworks have begun exploring the impact of these complex sensory stimuli on efficient coding \cite{Park2013, Ganguli2014, Tesileanu2020}.  For example, mainly using the mammalian retina as a model, the optimal numbers of ON and OFF retinal ganglion cells were derived as a function of the spatial contrast statistics in these natural scenes \cite{Ratliff2010}. Other studies have built more extensive cascade models consisting of receptive fields and nonlinearities, and predicted the structure of retinal mosaics (spatial grids of cells that encode different sensory information) \cite{Karklin2011, Doi2012} and inter-mosaic relationships \cite{Jun2022, Roy2021} by considering the spatial structures of natural scenes. To account for the temporally rich dynamics, some models have extended similar cascade models to the temporal domain and predicted the spatial and temporal integration properties of retinal mosaics when training the models on natural videos \cite{Ocko2018,  Sederberg2018, Jun2022} or studied adaptation to a changing environment \cite{Mlynarski2018, Mlynarski2021}. While these are exciting recent developments that capitalize on new technologies to simultaneously record many cells and evaluate the theoretical predictions, they are often focused on particular systems, such as the vertebrate retina. Considering its applicability to the encoding of a one-dimensional static stimulus, our framework allows analytical tractability, and generalization across different dimensions, including noise functions, firing rate distributions, and activation functions, as well as being agnostic to the neural populations being modeled. Our model would need to be revised to account for the encoding of high dimensional spatial and temporal stimuli.

In addition to sensory stimuli, neural activity also exhibits a high degree of structure, for example, firing with specific temporal dynamics (sustained or transient \cite{Margolis2007, Berens2017}). Structured neural activity significantly impacts neural coding and could alter the conclusions of specific efficient coding theoretical results. For instance, noise correlations can significantly affect  coding performance \cite{Averbeck2006, Panzeri2022}. Bursting, another common and functional type of neural activity other than spiking, also has profound coding implications \cite{Zeldenrust2018}. Burst coding can be more efficient \cite{Park2019}, more biological realistic \cite{Williams2021}, and more multiplexed \cite{Naud2018} than rate coding.

The equality of mutual information across different ON-OFF mixtures predicted by our theoretical framework seems to be inconsistent with experimental data in the many sensory systems where different numbers of ON and OFF neurons have been observed (e.g., primate somatosensory cortex, fly visual system, and mammalian olfactory system, summarized in \cite{Gjorgjieva2019}). Nonetheless, in our theoretical results, despite the equality of total information the overall cost of spiking across the different ON-OFF mixtures differs. Hence, populations with equal ON-OFF mixtures have the lowest spiking cost, while homogeneous populations with only ON or OFF neurons have the highest cost. This implies despite the equality of information across the different ON-OFF mixtures, the information per spike is the highest for the equal ON-OFF mixture, and decreases monotonically when a single subpopulation (ON or OFF) begins to dominate the mixture. Taking into account the metabolic cost could be one answer to explaining different numbers of ON and OFF cells across different sensory systems. Other possibilities include different measures of coding efficiency than mutual information that might be used by the different sensory systems \cite{Bethge2003, Harper2004}, different sources of noise (input vs.~output \cite{Roth2021, Kastner2011, Brinkman2016}), as well as including spatio-temporal statistics in natural images. 

Although we considered different noise statistics, we assumed that the noise was inherent in the spike generation mechanism. Noise can enter in different places along the encoding pathway, for example, in the activation functions of the neurons \cite{Faisal2008}, and have considerable effects on neuronal coding \cite{Rieke1995, Chichilnisky2005, Brinkman2016, Roth2021}. Furthermore, we only focused on stimulus encoding without discussing how the information can be further decoded from the neuronal spikes. The decoding of information could be another key aspect of efficient neuronal coding \cite{Park2013, Gjorgjieva2014, Wang2016, Gjorgjieva2019}. Previous work has shown that the measure of efficiency can significantly influence conclusions about the ON-OFF composition of the population and the activation function distribution \cite{Kastner2015, Brinkman2016, Gjorgjieva2019}. It is possible that different sensory systems have evolved optimal coding strategies by maximizing different objective functions, or may not even be optimal at all. Our work provides a general theoretical framework that in principle can be applied to various sensory systems without the nuances of each system.  

%\begin{acknowledgments}
\section*{Acknowledgements}
This work is supported by the funding from the Max Planck Society (SS and JG), the Technical University of Munich (JG) and NIH Grant R01 NS111477 (MM). We also would like thank the ‘Computation in Neural Circuits’ group for suggestions on this paper, especially our colleagues Claudia Cusseddu and Judith Parkinson-Schwarz for providing feedback on the manuscript.
%\end{acknowledgments}

%\bibliographystyle{ieeetr}
%\bibliography{reference}

\section{Appendix}
\begin{longtable}[h]{|l|p{5cm}|}% @{\extracolsep{\fill}}
\caption{\label{tab:parameters}Summary of the parameters in this paper}\\
\hline
$\textbf{Variables}$ & $\textbf{Definitions}$ \\ \hline
$N$ & Number of neurons \\ \hline
$s$ & The scalar stimulus \\ \hline
$p(s)$ & The stimulus distribution  \\ \hline
$\nu_{\max}$ & The maximal firing rate constraint \\ \hline
$\nu_i(s)$ & The activation function of neuron $i$ \\ \hline
$T$ & The coding window within which  the spike count is computed \\ \hline
$R$ & $\nu_{\max}T$, the expected value of spike count when a neuron fires at the maximal firing rate $\nu_{\max}$ \\ \hline
$n_i$ & The spike count of neuron $i$ \\ \hline
$\nu_i T$ & The expected value of neuron $i$'s spike count \\ \hline
$L(n_i, \nu_i T)$ &  The noise generation function, equivalent to $p(n_i|\nu_i)$ \\ \hline
$I(s,n)$ & Shannon mutual information between stimulus $s$ and spike count $n$ encoded by the neuronal population \\ \hline
$i(\cdot)$ & Density of mutual information \\ \hline
%$g(\cdot)$ & The mutual information encoded by one neuron, either independent or in a population \\ \hline
$F_\nu$ & The cumulative distribution function of the firing rate $\nu$ \\ \hline
%$E_\nu$ & The set of points at which $F_\nu$ increases \\ \hline
$m$ & The number of ON neurons \\ \hline
$Q_m$ & The probability that none of the ON neurons $1,2, ..., m$ fires \\ \hline
$F_i^{(j)}$ & cumulative distribution function of $\nu_i$, given that none of the neurons $1, 2, ..., j$ fires \\ \hline
$\theta_{ik}$ & The $k^{th}$ firing threshold of neuron $i$ whose activation function is discrete \\ \hline
$p_{ik}$ & The $k^{th}$ stimulus interval of neuron $i$ \\ \hline
$u_{ik}$ & The $k^{th}$ cumulative stimulus interval of neuron $i$ \\ \hline
$f_{ik}$ & The ratio of the $k^{th}$ intermediate firing rate \\ \hline
$P_i(0)$ & The probability that neuron $i$ does not fire \\ \hline
$u_{ik}^{(j,N)}$ & In a group of $N$ neurons, the $k^{th}$ cumulative stimulus interval of neuron $i$, given that none of the neurons $1, ..., j \, (j < i)$ fires \\ \hline
$P_i^{(k,N)}(0)$ & In a group of $N$ neurons, the $k^{th}$ cumulative stimulus interval of neuron $i$, given that none of the neurons $1, ..., j \, (j < i)$ fires \\ \hline
$q$ & $L(0, \nu T)$, the probability that the spike count is 0 given a nonzero firing rate \\ \hline
%$\pe$ & The highest or the lowest stimulus interval \\ \hline
\end{longtable}

\end{document}

\noindent \textbf{\Large{Supplemental Material}}\\
\\
\noindent \textbf{\Large{Efficient population coding of sensory stimuli}}\\

\noindent Shuai Shao$^{1,2}$, Markus Meister$^{2}$, and Julijana Gjorgjieva$^{1,2}$
\newline

\noindent
1. Computation in Neural Circuits Group, Max Planck Institute for Brain Research, Frankfurt, Germany

\noindent
2. Donders Institute and Faculty of Science, Radboud University, Nijmegen, Netherlands

\noindent
3. Division of Biology and Biological Engineering, California Institute of Technology, Pasadena, CA, USA

\noindent
4. School of Life Sciences, Technical University of Munich, Freising, Germany
\newline

\noindent Contact: gjorgjieva@tum.de

%\section{$I(s, \vec n) = I(\vec \nu, \vec n)$\label{sec:info_s_nu}}
\section{The mutual information between stimulus and spikes equals the mutual information between firing rates and spikes\label{sec:info_s_nu}}

\noindent In this section we prove the argument in the main text that the mutual information between the stimuli $s$ and the spike counts $\vec n$, equals the mutual information between the firing rates $\vec \nu$ and the spike counts $\vec n$, i.e., $I(s, \vec n) = I(\vec \nu, \vec n)$ (Eq. 5). This was also shown in previous literature \cite{Nikitin2009} but limited to a single neuron (i.e., $N = 1$, when $\vec n$ and $\vec \nu$ are scalars).
\\
\\
Since the spike counts of different neurons are independent of each other, we can write
\begin{equation}
p(\vec n|s) = \prod_i p(n_i|s).
\label{eq:indepedent_neurons}
\end{equation}
Inserting it into the formula of the Mutual Information (Eq. 3), we have
\begin{equation}
\begin{aligned}
I(s, \vec n) &= \sum_{\vec n} \int \de s \, p(s)\, p(\vec n|s)\, \text{log}\,\frac{p(\vec n|s)}{P(\vec n)}\\
&= \sum_{\vec n} \int \de s \, p(s)\, p(\vec n|s)\, \text{log}\, p(\vec n|s) - \sum_{\vec n}P(\vec n)\,\text{log}\, P(\vec n) \\
& = \sum_{\vec n} \sum_i \int \de s \, p(s) \, \text{log}\, p(n_i|s) \prod_k p(n_k|s) - \sum_{\vec n}P(\vec n)\,\text{log}\, P(\vec n) \\
& = \sum_i  \sum_{\vec n} \int \de s \, p(s) \, \text{log}\, p(n_i|s) \prod_k p(n_k|s) - \sum_{\vec n}P(\vec n)\,\text{log}\, P(\vec n). 
\end{aligned}
\end{equation}
Note that $\prod_k p(n_k|s) = p(n_i|s)\prod_{k \neq i} p(n_k|s)$, summing over all $k \neq i$, we have
\begin{equation}
\begin{aligned}
I(s, \vec n) & = \sum_i  \sum_{n_i} \int \de s \, p(s)\, p(n_i|s) \,  \text{log}\, p(n_i|s)  - \sum_{\vec n}P(\vec n)\, \text{log}\, P(\vec n) \\
& = \sum_i  \sum_{n_i} \int \text{d}\nu_i \, p(\nu_i)\,p(n_i|\nu_i) \,  \text{log}\, p(n_i|\nu_i)  - \sum_{\vec n}P(\vec n)\, \text{log}\, P(\vec n). 
\end{aligned}
\label{eq:transform_to_info_rate}
\end{equation}
Denoting by ${\vec {\nu}}^{(i)} = (\nu_1, ..., \nu_{i-1}, \nu_{i+1}, ..., \nu_N)$ the vector of all the $\nu$ except for $\nu_i$, we have
\begin{equation}
 \int \text d^{N-1}{\vec {\nu}}^{(i)} \,  p({\vec {\nu}}^{(i)}|\nu_i) = 1.
\end{equation}
Therefore, equation (Eq.~\ref{eq:transform_to_info_rate}) becomes
\begin{equation}
\begin{aligned}
I(s, \vec n) &= \sum_i  \sum_{n_i}\int \text d^{N-1 }{\vec {\nu}}^{(i)} \,p({\vec {\nu}}^{(i)}|\nu_i)\int \text{d}\nu_i \, p(\nu_i)\,p(n_i|\nu_i) \,  \text{log}\, p(n_i|\nu_i)   - \sum_{\vec n}P(\vec n)\,\text{log}\,P(\vec n) \\
&= \sum_i  \sum_{n_i} \int \text{d}^N \vec \nu \,  p(\vec \nu)\, p(n_i|\nu_i) \,  \text{log}\, p(n_i|\nu_i)   - \sum_{\vec n}P(\vec n)\,\text{log}\, P(\vec n) \\
& = \sum_i  \sum_{n_i} \int \text{d}^N \vec \nu \,  p(\vec \nu) \,p(n_i|\vec \nu) \,  \text{log}\, p(n_i|\vec \nu)   - \sum_{\vec n}P(\vec n)\,\text{log}\, P(\vec n)  \\
& = \sum_i  \sum_{\vec n} \int \text{d}^N \vec \nu \,  p(\vec \nu)  \,  \text{log}\, p(n_i|\vec \nu) \prod_k p(n_k|\vec \nu)   - \sum_{\vec n}P(\vec n)\,\text{log}\, P(\vec n).  
\end{aligned}
\end{equation}
Similar to Eq.~\ref{eq:indepedent_neurons}, we also have
\begin{equation}
p(\vec n|\vec \nu) = \prod_i p(n_i|\vec \nu)
\end{equation}
which leads to
\begin{equation}
\begin{aligned}
I(s, \vec n) &=  \sum_i  \sum_{\vec n} \int \text{d}^N \vec \nu \,  p(\vec \nu)\, p(\vec n|\vec \nu) \,  \text{log}\, p(n_i|\vec \nu)   - \sum_{\vec n}P(\vec n)\, \text{log}\, P(\vec n) \\
&=   \sum_{\vec n}  \int \text{d}^N \vec \nu \,  p(\vec \nu) \,p(\vec n|\vec \nu) \,  \sum_i \text{log}\, p(n_i|\vec \nu)   - \sum_{\vec n}P(\vec n)\,\text{log}\, P(\vec n) \\
& = \sum_{\vec n} \int \text{d}^N \vec \nu \,  p(\vec \nu)\, p(\vec n|\vec \nu) \,  \text{log}\, p(\vec n|\vec \nu)   - \sum_{\vec n}P(\vec n)\, \text{log}\, P(\vec n) \\
& =  \sum_{\vec n} \int_{\vec \nu} \text d^N \vec \nu \, p(\vec \nu)\, p(\vec n|\vec \nu) \, \text{log}\frac{p(\vec n|\vec \nu)}{P(\vec n)} \\
& = I(\vec \nu, \vec n).
\end{aligned}
\end{equation}

\section{Density of the mutual information for a single neuron is constant when optimized\label{sec:density_info}}

This section works as a first step to prove that the optimal activation function for a single neuron is discrete. We limit our discussion to a single neuron. Without loss of generality, we only consider an ON neuron with an activation function where the firing rate increases with stimulus intensity. Studying an OFF neuron is entirely symmetric. As in the main text, the maximal firing rate of this neuron is constrained to $\nu_{max}$ and the spontaneous firing rate is denoted by $\nu_0$.

\subsection{A neuron with a discrete activation function}

\setcounter{equation}{0}

%A neuron with a discrete activation function has a finite number of firing rates. We use the variable $\nu$ to denote the firing rate. So the activation function can be formulated as a function of the stimuli $s$, i.e. $\nu = \nu(s)$. We also denote the lowest firing rate as $\nu_0$, which we call the spontaneous firing rate of the neuron, and $\nu$ can be some discrete values up to the maximal firing rate, which is fixed at $\nu_{\max}$. In the main text and starting from the next chapter, $\nu_0 = 0$, but this constraint is not needed here.
%\\
%\\ 
%In the main text, we defined the noise generation function $L(\cdot)$ as
%
%\begin{equation}
%L(n, \nu T) = p(n|\nu)
%\end{equation}
%
%where $p(n|\nu)$ denotes the conditional probability of getting $n$ spikes in a time window $T$ when the neuron's firing rate is $\nu$. The expected value of $n$ is then $\nu T$, and we assume it is the only parameter of the noise generation function $L$. 
First, we consider a neuron with a discrete activation function. In this case, the firing rate can only be some discrete values between $\nu_0$ and $\nu_{\max}$. Therefore, we denote the probability that the firing rate is $\nu$ by $p_\nu$, instead of $p(\nu)$ that we commonly write. The mutual information is then
%also denote with $p_\nu$ the probability that the firing rate is $\nu$, and write
\begin{equation}
I(s, n) = I(\nu, n) = \sum_n \sum_\nu \, p_{\nu}\,p(n|\nu) \, \text{log}\frac{p(n|\nu)}{P(n)}.
\label{eq:info_rate_one_neuron}
\end{equation}
%
%\begin{equation}
%H(\vec n|s) = H(\vec n|\nu) = -\sum_n \sum_\nu p_\nu\, p(n|\nu)\log p(n|\nu) = -\langle \log p(n|\nu) \rangle_{n,\nu},
%\end{equation}
%
%in which
%
%\begin{equation}
%\langle \log p(n|\nu)\rangle_n = \sum_{n=0}^{+\infty}L(n,\nu T)\log L(n,\nu T).
%\end{equation}
%
We can define the entropy of the spike count at a given firing rate as
\begin{equation}
h(\nu) = -\sum_{n=0}^{+\infty}p(n|\nu)\log p(n|\nu)
\end{equation}
%
%and then we can write
%
%\begin{equation}
%H(n|s) = H(n|\nu) = \sum_\nu p_\nu \, h(\nu).
%\end{equation}
%
%The entropy of $n$ can be expressed as
%
%\begin{equation}
%H(n)=-\sum_nP(n)\log P(n)=-\sum_n \sum_\nu p_\nu \, p(n|\nu) \log P(n),
%\end{equation}
%
%which gives
%
and note that
\begin{equation}
P(n) = \sum_\nu p_\nu \, p(n| \nu),
\end{equation}
so that we have
\begin{equation}
I(\nu, n) = -\sum_\nu p_\nu \, h(\nu) -\sum_\nu p_\nu \sum_n p(n|\nu)\log P(n).
\end{equation}
Since $p_\nu$ are probabilities, we have the constraint that $\sum_\nu p_\nu = 1$, hence to optimize the objective function we include a Lagrange multiplier,
\begin{equation}
\widetilde I = I(\nu, n) + \lambda (\sum_\nu p_\nu -1).
\end{equation}
Assuming optimality,
\begin{equation}
\partial_{p_\nu} \widetilde I = -h(\nu) - \sum_n p(n|\nu) \log P(n)-\sum_n p(n|\nu) + \lambda = 0.
\end{equation}
Absorbing $-\sum_n p(n|\nu) = -1$ into $\lambda$, i.e.~$\lambda \rightarrow \lambda-1$, we have
\begin{equation}
-h(\nu)-\sum_n p(n|\nu) \log P(n) + \lambda = 0.
\label{eq:optimize_I_lambda}
\end{equation}
Multiplying both sides by $p_\nu$ and summing over $\nu$, we have
\begin{equation}
I(s,n)+\lambda = 0.
\end{equation}
We define
\begin{equation}
i(\nu) = \sum_{n} p(n|\nu)\, \text{log}\frac{p(n|\nu)}{P(n)}.
\end{equation}
Multiplying this equation with $p_\nu$ and summing over $\nu$, we have
\begin{equation}
%I = \int_s p(s)i(s) = \sum_\nu p_\nu i(\nu)
I = \sum_\nu p_\nu i(\nu).
\end{equation}
Therefore, we call $i(\nu)$ ``the density of mutual information", which is also defined by Eq. 6 in the main text. According to Eq.~\ref{eq:optimize_I_lambda}, we can write
\begin{equation}
I(\nu,n) = -\lambda = -\sum_n p(n|\nu)\log P(n) - h(\nu) = \sum_n p(n|\nu)\log \frac{p(n|\nu)}{P(n)} = i(\nu).
\label{eq:equal_info_density_supp}
\end{equation}
%
%\begin{equation}
%I(s,n) = \sum_n p(n|\nu)\log \frac{p(n|\nu)}{P(n)} = \sum_n p(n|s)\log \frac{p(n|s)}{P(n)} = i(s)
%\end{equation}
%
This means when the mutual information is optimized, $i(\nu)$ is a constant for all possible $\nu$. The convexity of the mutual information ensures that the optimal solution is unique. As a special case, if the spontaneous rate $\nu_0 = 0$, according to Eq. 1 and Eq. 2, we have $p(n = 0|\nu = 0) = 1$ and $p(n \neq 0|\nu = 0) = 0$. As a result,
\begin{equation}
I^{\max} = i(\nu = 0) = -\log P(0).
\label{eq:I_log_P0}
\end{equation}
Also, Eq.~\ref{eq:equal_info_density_supp} means the mutual information $I(\nu, n)$ is distributed proportionally to the probabilities $p_\nu$ when it is maximized. In addition, one can also define
\begin{equation}
i_s(s) = \sum_{n} p(n|s)\, \text{log}\frac{p(n|s)}{P(n)},
% = \sum_{n} L\left(n, \nu(s) T\right)\text{log}\frac{L\left(n, \nu(s) T\right)}{P(n)},
\label{eq:i_s_defn}
\end{equation}
then we have
\begin{equation}
I = \int \de s \, p(s)\, i_s(s)
\end{equation}
and
\begin{equation}
i(\nu) = i_s(s).
\end{equation}
Therefore, the maximal mutual information $I(s, n)$ will be distributed proportionally to the probability density of the stimulus $s$, denoted by $p(s)$ in the main text and Fig. 1B. The density function $i_s(s)$ is also a constant over the space of stimulus $s$.
%{\jg{Therefore,}}\com{Is it ok to use therefore here?} the density of mutual information ($i(s)$ or $i(\nu)$) is a constant over the entire input space ($s$ or $\nu$) in the optimal condition.\com{OK yes let us discuss about this. I get that that information is distributed proportionally. But I don't get why it is a \textit{constant over the entire input space}.} 
For example, if we have a ternary activation function with three possible firing rates 0, $\nu_{max}/2$, and $\nu_{max}$, and the stimulus $s$ follows a standard normal distribution, the input space in terms of $\nu$ is $\{ 0, \nu_{max}/2, \nu_{max} \}$, so we have $i(\nu = 0) = i(\nu = \nu_{max}/2) = i(\nu = \nu_{max})$. Similarly, the input space of $s$ is then the set of all real numbers $\mathbb{R}$, and we have $i_s(s) = \text{const}, s \in \mathbb{R}$.

%%%%%%%%%%%%%%%%%%%%%%%%%%%%%
\subsection{A neuron with a continuous activation function}

We assume that the neuron has a continuous and smooth (analytic) activation function, with the lowest rate (i.e., the spontaneous firing rate) $\nu_0$ and the maximal firing rate $\nu_{max}$.
%with the expected value of the spike count equal to $R$ when $\nu = \nu_{\max}$, and the expected value $r_0$ when $\nu = \nu_0$. {\jg{Note that given an appropriate coding window $T$ in which spike counts can be computed, $R =  \nu_{\max} T$ and $r_0 = \nu_0 T$.}}  
Then, the mutual information can be written as:
\begin{equation}
I = \sum_{n=0}^{+\infty}\int_{\nu_0}^{\nu_{max}} \de s\, p(s)\, p(n|s)\log \frac{p(n|s)}{P(n)} =  \sum_{n=0}^{+\infty}\int_{\nu_0}^{\nu_{max}} \de \nu\, p(\nu)\, p(n|\nu)\log \frac{p(n|\nu)}{P(n)}.
\end{equation}
Define $\widetilde I = I+\lambda \left(\int_{\nu_0}^{\nu_{max}} p(\nu)\text d\nu - 1 \right)$, then
\begin{equation}
\widetilde I = -\sum_{n=0}^{+\infty}P(n)\log P(n)+\sum_{n=0}^{+\infty}\int_{\nu_0}^{\nu_{max}} \de \nu\, p(n|\nu)\, p(\nu)\log p(n|\nu) + \lambda \left(\int_{\nu_0}^{\nu_{max}} \de \nu\,p(\nu) -1\right ).
\end{equation}
When optimized,
\begin{equation}
\delta \widetilde I = -\sum_{n=0}^{+\infty}\left (\log \, P(n)+1 \right)\, \delta P(n)-\int_{\nu_0}^{\nu_{max}} \de \nu\,h(\nu)\, \delta p(\nu) + \lambda \int_{\nu_0}^{\nu_{max}} \de \nu\,\delta p(\nu) = 0.
\end{equation}
Because
\begin{equation}
\delta P(n) = \delta \left[\int_{\nu_0}^{\nu_{max}} \de \nu\, p(\nu)\, p(n|\nu)\right ] = \int_{\nu_0}^{\nu_{max}} \de \nu\, p(n|\nu)\, \delta p(\nu),
\end{equation}
we have
\begin{equation}
\delta \widetilde I = - \int_{\nu_0}^{\nu_{max}} \de \nu \sum_{n=0}^{+\infty} p(n|\nu)\, \left(\log \, P(n) + 1\right )\, \delta p(\nu) - \int_{\nu_0}^{\nu_{max}} \de \nu\, h(\nu)\, \delta p(\nu) + \lambda \int_{\nu_0}^{\nu_{max}} \de \nu\, \delta p(\nu) = 0
\end{equation}
which leads to
\begin{equation}
-\sum_{n=0}^{+\infty} p(n|\nu)\, \left(\log P(n) + 1\right)-h(\nu)+\lambda = 0.
\end{equation}
Absorbing $ - \sum_{n=0}^{+\infty} p(n|\nu) = -1$ into $\lambda$, multiplying by $p(\nu)$, and integrating over $\nu$, we have $I + \lambda = 0$. As a result,
\begin{equation} 
I = -\lambda = -\sum_n p(n|\nu)\log \, P(n) -h(\nu) = \sum_n p(n|\nu)\log \frac{p(n|\nu)}{P(n)} = i(\nu),\ \ \text{for\ } \nu \in [\nu_0, \nu_{max}]
\end{equation}
which means the density of mutual information $i(\nu)$ is a constant for all firing rates $\nu$. One can still define the density function with stimulus $s$ as  Eq.~\ref{eq:i_s_defn} and the $i_s(s)$ is still a constant when the mutual information is optimized.
\\
\\
In summary, we have shown that the density of mutual information $i(\nu)$ is a constant for all possible firing rates, independent of whether the activation function is discrete or continuous. We note that this result has also been proven in previous work using a different approach based on the convexity of mutual information \cite{Smith1971, Shamai1990}.

%%%%%%%%%%%%%%%%%%%%%%%%%%%%%%%%%%%%%%%%%%%%%%%%%%%%%%%%%%%%
\setcounter{equation}{0}

\section{The optimal activation functions of a population of neurons are discrete\label{sec:discrete_tuning_curves}}

To prove that the optimal activation functions are discrete, we first need to prove that when the mutual information of a population of $N$ neurons is maximized, the density of mutual information $\widetilde i(\nu_1)$ that we defined in the main text is a constant and equals to the maximal mutual information $I_N^{\max}$ (Eq. 35). Consistent with the main text, we denote $p(n_i|\nu_i)$ by $L(n_i, \nu_i T)$ from now on.
\\
\\
According to the definition in the main text (Eq. 34), we have
\begin{equation}
\begin{aligned}
\widetilde i(\nu_1) &= \sum_{n_1} p(n_1|\nu_1)\log \frac{p(n_1|\nu_1)}{P(n)}+ p(n_1=0|\nu_1)I_{N-1}^{\max} \\
&= \sum_{n_1} L(n_1, \nu_1 T)\log \frac{L(n_1, \nu_1 T)}{P(n)}+ L(0, \nu_1 T)I_{N-1}^{\max}
\end{aligned}
\end{equation}
and we can see that when neurons $2, ..., N$ are all optimized,
\begin{equation}
    \int \de \nu_1\,i(\nu_1)\,p(\nu_1) = I_N  = I(F_1) + P_1(0)I_{N-1}^{\max}.
\end{equation}
\\
Similar as in the last section, we define $\widetilde I_N = I_N + \lambda \left(\int_{0}^{\nu_{max}} p(\nu_1)\text d\nu_1 - 1 \right)$, and write
\begin{equation}
\begin{aligned}
\widetilde I_N =& -\sum_{n_1=0}^{+\infty}P(n_1)\log P(n_1)+\sum_{n_1=0}^{+\infty}\int_{0}^{\nu_{max}} \de \nu_1\, L(n_1,\nu_1 T)\, p(\nu_1)\log L(n_1,\nu_1 T) \\
&+ I_{N-1}^{\max}\int_{0}^{\nu_{max}} \de \nu_1\, p(\nu_1)L(0, \nu_1T) + \lambda \left(\int_{0}^{\nu_{max}} \de \nu_1\, p(\nu_1) -1\right ).
\end{aligned}
\end{equation}
When optimized,
\begin{equation}
\delta \widetilde I_N = -\sum_{n_1=0}^{+\infty}\left (\log \, P(n_1)+1 \right)\, \delta P(n_1)+\int_{0}^{\nu_{max}} \de \nu_1 \, \left(I_{N-1}^{\max}L(0, \nu_1T) - h(\nu_1)\right)\, \delta p(\nu_1) + \lambda \int_{0}^{\nu_{max}} \de \nu_1 \, \delta p(\nu_1) = 0.
\end{equation}
Because
\begin{equation}
\delta P(n_1) = \delta \left[\int_0^{\nu_{max}} \de \nu_1\, p(\nu_1)\, L(n_1,\nu_1 T)\right ] = \int_0^{\nu_{max}} \de \nu_1\, L(n_1,\nu_1 T)\, \delta p(\nu_1),
\end{equation}
we have
\begin{equation}
\begin{aligned}
\delta \widetilde I_N =& - \int_0^{\nu_{max}} \de \nu_1 \sum_{n_1=0}^{+\infty} L(n_1,\nu_1 T)\, \left(\log \, P(n_1) + 1\right )\, \delta p(\nu_1) \\
&+ \int_0^{\nu_{max}} \de \nu_1 \, \left(I_{N-1}^{\max}\,L(0, \nu_1T) - h(\nu_1)\right)\, \delta p(\nu_1) + \lambda \int_0^{\nu_{max}} \de \nu_1\,\delta p(\nu_1) = 0
\end{aligned}
\end{equation}
which leads to
\begin{equation}
-\sum_{n_1=0}^{+\infty} L(n_1,\nu_1 T)\, \left(\log P(n_1) + 1\right)+I_{N-1}^{\max}\,L(0, \nu_1T)-h(\nu_1)+\lambda = 0.
\end{equation}
Absorbing $ - \sum_{n_1=0}^{+\infty} L(n_1,\nu_1 T) = -1$ into $\lambda$, multiplying by $p(\nu_1)$, and integrating over $\nu_1$, we have $I_N + \lambda = 0$. As a result,
\begin{equation} 
\begin{aligned}
I_N^{\max} = -\lambda &= -\sum_{n_1} L(n_1,\nu_1 T)\log \, P(n_1) + I_{N-1}^{\max}\, L(0, \nu_1T) -h(\nu_1) \\
&= \sum_{n_1} L(n_1,\nu_1 T)\log \frac{L(n_1,\nu_1 T)}{P(n_1)} + I_{N-1}^{\max}\, L(0, \nu_1T) = \widetilde i(\nu_1),\ \ \text{for\ } \nu_1 \in [0, \nu_{max}].
\end{aligned}
\label{eq:max_info_and_density_info}
\end{equation}
Second, we need to prove that the above result (Eq.~\ref{eq:max_info_and_density_info}) will lead to a contradiction if the optimal activation function $F_1$ is continuous. From the discussion in the main text, this is equivalent to finding a paradox in
\begin{equation}
\widetilde i(\nu_1) = \sum_{n_1=0}^{+\infty} L(n_1,\nu_1 T)\log \frac{L(n_1, \nu_1 T)}{P(n_1)} + I_{N-1}^{\max}\, L(0, \nu_1T) = I_N^{\max} = \text{const}.
\label{eq:info_density_const_N}
\end{equation}
Following the same procedure as in the main text, we can prove that if we write the Maclaurin series
\begin{equation}
L(n_1, \nu_1 T) = \sum_{k=1}^{+\infty} a_{n_1, k}(\nu_1 T)^k
\end{equation}
for any $n_1 \geq 1$, the sum of the coefficients of $\log (\nu T)$ terms in the $m^{th}$ derivative of $\widetilde i(\nu_1)$ is then $\sum_{n_1=1}^{+\infty}a_{n_1, m}\,j(n_1)$, where $j(n_1)$ is the minimal index of $k$ that makes $a_{n_1, k} > 0$. This follows the same formalism as Eq. 22 in the main text, because the additional term here, $I_{N-1}^{\max}\, L(0, \nu_1T)$, does not contribute to $\log (\nu T)$ terms when it is written as a Maclaurin series.
\\
\\
If Eq.~\ref{eq:info_density_const_N} were correct, all the derivatives of $\widetilde i(\nu_1)$ would be 0, and we would have 
\begin{equation}
    \sum_{n_1=1}^{+\infty}a_{n_1, m}\,j(n_1) = 0 \text{ for any } m \geq 1
\end{equation}
because $\log (\nu_1 T)$ diverges as $\nu_1 T \rightarrow 0$. Again, similar as in the main text, we could show that in this case,
\begin{equation}
L(n_1 = 0, \nu_1 T) = 1 \text{ for any } \nu_1,
\end{equation}
which cannot be true. Therefore, we have proved that the optimal $F_1$ being continuous will lead to a paradox. Therefore, in a population of $N$ neurons, given that neurons $2, ..., N$ are all optimized, the optimal activation function of neuron 1 will be discrete.

%%%%%%%%%%%%%%%%%%%%%%%%%%%%%%%%%%%%%%%%%%%%%%%%%%%%%%%%%%%%
\setcounter{equation}{0}

\section{The number of steps in the optimal activation functions increases as a function of the maximal firing rate constraint \label{sec:increasing_steps}}

\begin{figure}
    \centering
    \includegraphics[width=17.2cm]{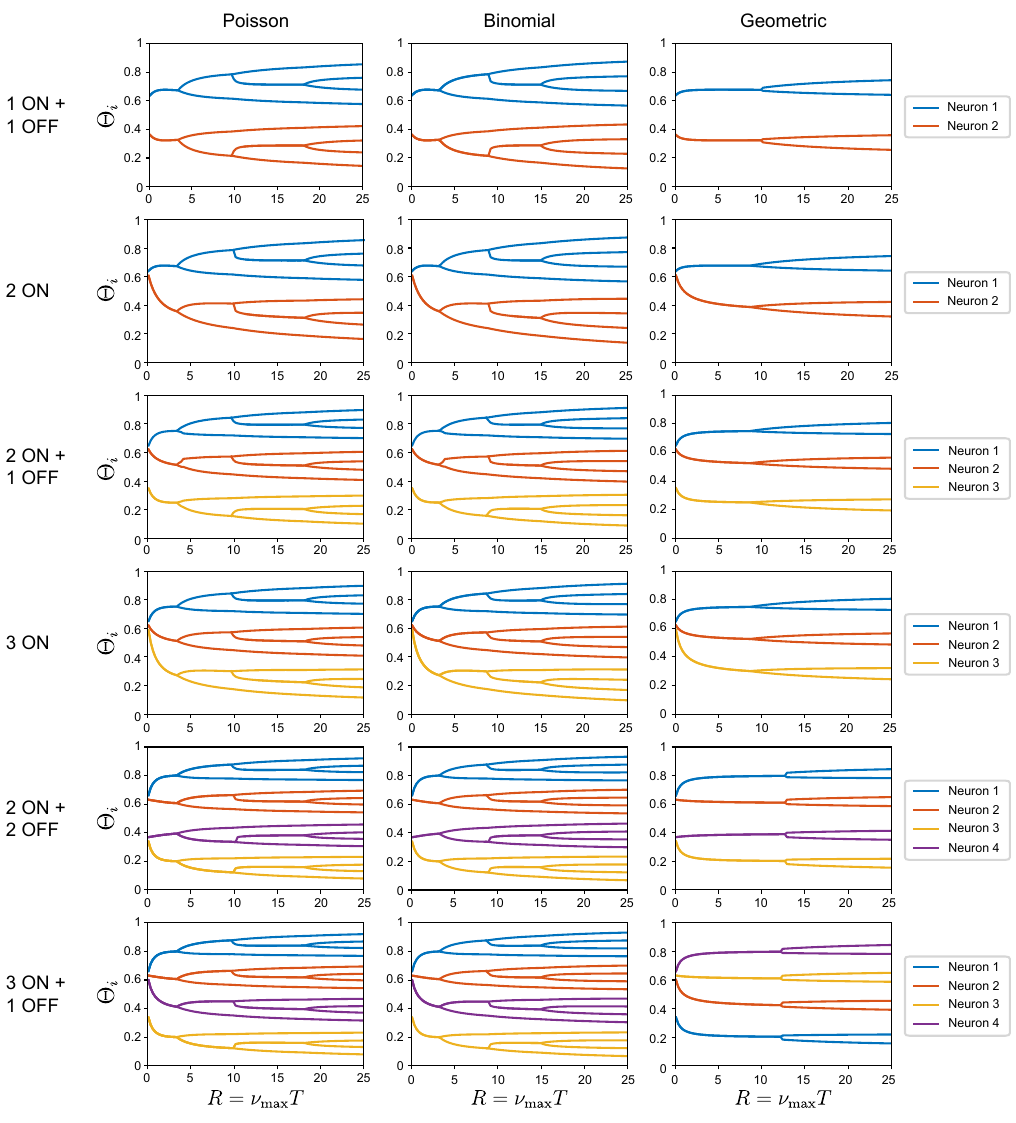}
    \caption{Optimal thresholds in different neuronal populations (1 ON + 1 OFF, 2 ON, 2 ON + 1 OFF, 3 ON, 2 ON + 2 OFF, and 3 ON + 1 OFF) with different noise generation functions (Poisson, Binomial, and Geometric distribution).}
    \label{fig:numerics}
\end{figure}

Here, we perform extensive numerical calculations on neuronal populations with up to four neurons and any ON-OFF mixture to demonstrate that as the maximal firing rate constraint $\nu_{\max}$ increases, the number of steps in the optimal activation functions increases. We calculated the optimal thresholds numerically using three different noise generation functions $L(n, \nu T)$ (Fig.~\ref{fig:numerics}): 

\noindent (1) Poisson distribution 
\begin{equation}
    L(n, \nu T) = \frac{(\nu T)^n}{n!} \exp(-\nu T),
    \label{eq:Poisson_supp}
\end{equation}

\noindent (2) Binomial distribution
\begin{equation}
    L(n, \nu T) = \tbinom{N}{n}p^n(1-p)^{N-n},
    \label{eq:Binomial_supp}
\end{equation}
where $N = 30$ and $p = \nu T/N$, and 

\noindent (3) Geometric distribution
\begin{equation}
    L(n, \nu T) =  p^n(1-p)
    \label{eq:Geometric_supp}
\end{equation}
where $p = \nu T/(1+\nu T)$. 

In detail, we calculated the mutual information as a function of the firing thresholds and intermediate firing rates, based on Eq. 5. These parameters were initialized randomly, and then optimized using SLSQP method \cite{Kraft1988}. All code was written in Python 2.7. A sample is publicly available at \href{https://zenodo.org/record/8083056}{https://zenodo.org/record/8083056}.

We find that the number of thresholds increases as the maximal firing rate constraint $\nu_{\max}$ increases. Moreover, for all neurons in the same population, the threshold splitting occurs at the same firing rate, meaning that every neuron in the population has an optimal discrete activation function with the same number of steps. Hence, the optimal neuronal population consists of exclusively binary neurons, or exclusively ternary neurons, or exclusively quaternary neurons, etc. But it can never be a mixture of neurons with different numbers of steps, e.g., binary and ternary.

\section{Population coding of binary neurons with any noise generation function \label{sec:binary_neurons}}

\setcounter{equation}{0}

From now on, we denote the probability function of spike generation as $L(n_i,r_i)$, where $n_i$ is the spike count of neuron $i$, and $r_i$ is the expected value of $n_i$. When the firing rate of neuron $i$ is $\nu_i$ and the time window is $T$, we have $r_i = \nu_i T$. Consistent with the main text, we assume the neurons do not have a spontaneous firing rate, i.e., $\nu_0 = 0$. $R = \nu_{\max}T$ is the maximal value of  any $r_i$.

For a binary neuron, we define the interval of stimulus space partitioned by its threshold as $u_i = \text{Prob}(\nu_i = \nu_{max})$, which is the same as Eq. 38 in the main text (Fig. 2B). The mutual information between stimuli and spikes can be formulated as
\begin{equation}
\begin{aligned}
I_1 &= g(u_1) = \sum_n \int_s \de s\,p(s)\, p(n|s)\log \frac{P(n|s)}{P(n)} \\
&= -(1-u_1)\log P(0) + u_1L(0, R)\log \frac{L(0,R)}{P(0)} + u_{1}\sum_{n=1}^{+\infty}L(n,R)\log \frac{L(n,R)}{u_1L(n,R)}\\
&= -P(0)\log P(0) + u_1L(0, R)\log L(0, R) - u_1\log u_1\sum_{n=1}^{+\infty}L(n,R).
\label{eq:I1_binary}
\end{aligned}
\end{equation}
\noindent Define $q = L(0, R) = 1- \sum_{n=1}^{+\infty}L(n, R)$, we have
\begin{equation}
P(0) = 1 - u_1 + u_1q
\end{equation}
\begin{equation}
I_1 = -P(0)\log P(0) + u_1q\log q - u_1(1-q)\log u_1.
\end{equation}
Here, all the nonzero spike counts have been merged as in previous work with Poisson spike statistics \cite{Gjorgjieva2019}. It is equivalent to only having a firing state $n \neq 0$ and a non-firing state $n = 0$. The only difference from Poisson spike statistics is the exact formulation of the function $L$ and the value of $q$. This similarity allows us to use some of the results derived in previous literature \cite{Gjorgjieva2019}. For example, given a population of $N$ neurons, its mutual information can be written as
\begin{equation}
I_N = g(u_1) + (1-u_1(1-q)) \left [g(u_2^{(1)})+...+\left(1-u_{N-1}^{(N-2)}(1-q)\right )g(u_N^{(N-1)}) \right ]
\label{eq:I_decompose_binary}
\end{equation}
\noindent where $u_i^{(j)}$ means the revised probability of $u_i$ after knowing that all neurons $1, ..., j\, (j < i)$ did not spike, and
$g(u) = -(1 - u + uq)\text{log}(1 - u + uq)  + uq\log q - u(1-q)\log u$. The index of neurons follow Eq. 37 in the main text (Fig. 2B).

One can show that $u_i^{(j)}$ follows the rule below \cite{Gjorgjieva2019}:
\begin{equation}
u_i^{(j)} = \frac{u_i^{(j-1)} - \left(1- P_j^{(j-1)}(0)\right)}{P_j^{(j-1)}(0)} = \frac{u_i^{(j-1)} - u_j^{(j-1)}(1-q)}{1-u_j^{(j-1)}(1-q)}
\label{eq:revised_u_binary_ON}
\end{equation}
if neuron $i$ and neuron $j$ are both ON, or both OFF, and
\begin{equation}
u_i^{(j)} = \frac{u_i^{(j-1)}}{P_j^{(j-1)}(0)} = \frac{u_i^{(j-1)}}{1-u_j^{(j-1)}(1-q)}
\label{eq:revised_u_binary_OFF}
\end{equation}
if neuron $i$ is OFF but neuron $j$ is ON. Here $P_j^{(k)}(0)$ is the probability that neuron $j$ does not fire after knowing that none of the neurons $1, ..., k\,(k < j)$ spikes, we also have
\begin{equation}
P_j^{(k)}(0) = 1 - u_j^{(k)}(1-q).
\end{equation}
Taking derivatives of $I_N$ with respect to $u_N^{(N-1)}$, $u_{N-1}^{(N-2)}$, ..., and $u_1$ yields \cite{Gjorgjieva2019}
\begin{equation}
    u_i^{(i-1)} = \frac{1}{(N-i+1)(1-q)+q^{-q/(1-q)}}.
    \label{eq:u_N-k}
\end{equation}
Using Eq.~\ref{eq:u_N-k}, Eq.~\ref{eq:revised_u_binary_ON}, and Eq.~\ref{eq:revised_u_binary_OFF} iteratively, we can then derive the optimal solution
\begin{equation}
u_i = \frac{1+(i-1)(1-q)}{N(1-q)+q^{-q/(1-q)}}
\end{equation}
\noindent for ON neurons and
\begin{equation}
u_i = \frac{1+(m-i+1)(1-q)}{N(1-q)+q^{-q/(1-q)}}
\end{equation}
\noindent for OFF neurons.
With the definition in Eq. 44, we instantly get
\begin{equation}
\begin{aligned}
&p_1 = p_{m+1} = \frac{1}{N(1-q)+q^{-q/(1-q)}} \overset{\text {def}}{=} p_{\text {edge}} \\
&p_2 = ... = p_m = p_{m+2} = ... = p_N = \frac{1-q}{N(1-q)+q^{-q/(1-q)}} = \overset{\text {def}}{=} p \\
&p = (1-q)p_{\text {edge}}
\label{eq:binary_p}
\end{aligned}
\end{equation}
which is summarized in Eq. 45 of the main text. In this case the maximal mutual information
\begin{equation}
I_N^{\max} = \text{log} \left(1 + N(1-q)q^{q/(1-q)} \right) = -\log P(\vec 0)
\label{I_N_binary}
\end{equation}
and we also have
\begin{equation}
I_N^{\max} = -\log \left(1-Np \right).
\end{equation}
Using Eq.~\ref{I_N_binary} and comparing $I_1^{\max}$ and $I_N^{\max}$, we have
\begin{equation}
I_N^{\max} = \log \left[N(\exp({I_1^{\max}})-1)+1 \right].
\end{equation}
We can also calculate the overall mean firing rate ($\bar \nu$) of a population given the optimal thresholds of Eq.~\ref{eq:binary_p}. If the stimulus $s$ is higher than $\theta_1$ (the threshold of the highest ON neuron), all the $m$ ON neurons fire at $\nu_{\max}$ together. If $s \in (\theta_2, \theta_1)$, $m-1$ ON neurons (neuron $2, ..., m$) fire at $\nu_{\max}$, and so on so forth. Following this idea, we can write
\begin{equation}
\begin{aligned}
    \bar \nu &= \frac{\nu_{\max}}{N}\left[ \pe m + p \sum_{i=1}^{m-1}(m-i) + \pe (N-m) + p\sum_{i=1}^{N-m-1}(N-m-i)\right] \\
    &= \frac{\nu_{\max}}{N}\left[ \pe N + p\, \frac{m(m-1)}{2} + p\, \frac{(N-m)(N-m-1)}{2} \right] \\
    &= \left[\pe + \frac{N-1}{2}\,p + \frac{m}{N}\,(m-N)\,p \right]\nu_{\max},
    \label{eq:avg_fr_binary_supp}
\end{aligned}
\end{equation}
which produces Eq. 82 in the main text.

\setcounter{equation}{0}

\section{Population coding of binary neurons with heterogenous maximal firing rates \label{sec:binary_neurons_hetero}}

\subsection{Maximal mutual information and optimal firing thresholds}
As mentioned in the main text, we define $\nu_{\max, i}$ as the maximal firing rate of neuron $i$, and then
\begin{equation}
    R_i = \nu_{\max,i}\,T, \quad \quad \quad q_i = L(0, R_i),
\end{equation}
and
\begin{equation}
 u_i = \text{Prob}(\nu_i = \nu_{\max,i}) = 
\begin{cases}
\int_{\theta_i}^{+\infty} \de s \, p(s), \, \text{for ON}\\
\int_{-\infty}^{\theta_i} \de s \, p(s), \, \text{for OFF.}
\end{cases}
\end{equation}
Similar to Eq.~\ref{eq:I_decompose_binary}, the mutual information of a population of $N$ neurons can be decomposed into $N$ terms,
\begin{equation}
    I_N = g_1(u_1) + (1-u_1(1-q_1)) \left [g_2(u_2^{(1)})+...+\left(1-u_{N-1}^{(N-2)}(1-q_{N-1})\right )g_N(u_N^{(N-1)}) \right ]
    \label{eq:I_decompose_binary_hetero}
\end{equation}
where $u_i^{(j)}$ means the revised probability of $u_i$ after knowing that all neurons $1, ..., j (j < i)$ did not spike, and $g_i(u)$ denotes the information encoded by neuron $i$. With binary neurons and heterogeneous maximal firing rates
\begin{equation}
    g_i(u) = -(1-u+uq_i)\log (1-u+uq_i)+uq_i\log q_i - u(1-q_i)\log u.
\end{equation}
Also similarly to the case with identical maximal firing rates for all cells (Eq.~\ref{eq:revised_u_binary_ON} and Eq.~\ref{eq:revised_u_binary_OFF}), $u_i^{(j)}$ relates to $u_i^{(j-1)}$ with
\begin{equation}
u_i^{(j)} = \frac{u_i^{(j-1)} - \left(1- P_j^{(j-1)}(0)\right)}{P_j^{(j-1)}(0)} = \frac{u_i^{(j-1)} - u_j^{(j-1)}(1-q_j)}{1-u_j^{(j-1)}(1-q_j)}
\label{eq:revised_u_binary_ON_hetero}
\end{equation}
if neuron $i$ and neuron $j$ are both ON, or both OFF, and
\begin{equation}
u_i^{(j)} = \frac{u_i^{(j-1)}}{P_j^{(j-1)}(0)} = \frac{u_i^{(j-1)}}{1-u_j^{(j-1)}(1-q_j)}
\label{eq:revised_u_binary_OFF_hetero}
\end{equation}
if neuron $i$ is OFF but neuron $j$ is ON. $P_j^{(k)}(0)$ is the probability that neuron $j$ does not fire after knowing that none of the neurons $1, ..., k\,(k < j)$ spikes, and can be calculated as
\begin{equation}
P_j^{(k)}(0) = 1 - u_j^{(k)}(1-q_j).
\end{equation}
Taking derivatives of $I_N$ with respect to $u_N^{(N-1)}$, $u_{N-1}^{(N-2)}$, ..., and $u_1$ yields
\begin{equation}
    u_i^{(i-1)} = \frac{q_i^{q_i/(1-q_i)}}{1+\sum_{j=i}^N (1-q_j)q_j^{q_j/(1-q_j)}}.
    \label{eq:u_i_i-1_hetero}
\end{equation}
Using Eq.~\ref{eq:revised_u_binary_ON_hetero} and Eq.~\ref{eq:revised_u_binary_OFF_hetero} iteratively, we can derive the optimal solution
\begin{equation}
    u_i = \frac{q_i^{q_i/(1-q_i)}+\sum_{j=1}^{i-1}(1-q_j)q_j^{q_j/(1-q_j)}}{1+\sum_{j=1}^N (1-q_j)q_j^{q_j/(1-q_j)}}
    \label{eq:u_i_ON_hetero}
\end{equation}
for ON neurons and
\begin{equation}
    u_i = \frac{q_i^{q_i/(1-q_i)}+\sum_{j=m+1}^{i-1}(1-q_j)q_j^{q_j/(1-q_j)}}{1+\sum_{j=1}^N (1-q_j)q_j^{q_j/(1-q_j)}}.
    \label{eq:u_i_OFF_hetero}
\end{equation}
for OFF neurons. Also, inserting Eq.~\ref{eq:u_i_i-1_hetero} into Eq.~\ref{eq:I_decompose_binary_hetero}, we can calculate the maximal mutual information as
\begin{equation}
    I_N = \log \left[1+\sum_{i=1}^N (1-q_i)q_i^{q_i/(1-q_i)}\right]
    \label{eq:IN_max_hetero}
\end{equation}
With the definition of $p_i$ (Eq. 54), we can calculate the cumulative stimulus intervals partitioned by firing thresholds as
\begin{equation}
    \begin{aligned}
    &p_1 = \frac{q_1^{q_1/(1-q_1)}}{1+\sum_{j=1}^N(1-q_j)\,q_j^{q_j/(1-q_j)}} = e^{-I_N}\,q_1^{q_1/(1-q_1)}\\
    &p_{m+1} = \frac{q_{m+1}^{q_{m+1}/(1-q_{m+1})}}{1+\sum_{j=1}^N(1-q_j)\,q_j^{q_j/(1-q_j)}} = e^{-I_N}\,q_{m+1}^{q_{m+1}/(1-q_{m+1})}\\
    &p_i = \frac{q_i^{q_i/(1-q_i)} - q_{i-1}^{1/(1-q_{i-1})}}{1+\sum_{j=1}^N(1-q_j)\,q_j^{q_j/(1-q_j)}} = e^{-I_N}\left[q_i^{q_i/(1-q_i)} - q_{i-1}^{1/(1-q_{i-1})}\right], \quad i \neq 1 \text{ and } i \neq m
    \end{aligned}
    \label{eq:binary_p_hetero}
\end{equation}
which produces Eq. 55 in the main text.
Next, we ask how the optimal stimulus intervals $\lbrace p_i \rbrace$ depend on the noise levels $q_i$. To proceed, we define 
\begin{equation}
    f_1(x) = x^{x/(1-x)}, \quad \quad \quad f_2(x) = x^{1/(1-x)}, \quad \quad x \in (0,1)
\end{equation}
which are the two functions appearing in Eq.~\ref{eq:binary_p_hetero} above. One can prove that
\begin{equation}
\begin{aligned}
    \frac{\de \log f_1(x)}{\de x} &= \frac{1-x+\log x}{(1-x)^2} < 0 \\
    \frac{\de \log f_2(x)}{\de x} &= \frac{1-x+x\log x}{(1-x)^2x} > 0.
    \label{eq:d_f1_f2}
\end{aligned}
\end{equation}
Therefore, given the fixed amount of mutual information $I_N$, $p_1$ decreases with $q_1$, $p_{m+1}$ decreases with $q_{m+1}$, and any other $p_i$ decreases with both $q_i$ and $q_{i-1}$. When $q_i \rightarrow 1$ and  $q_{i-1} \rightarrow 1$, $p_i$ approaches its lower limit 0. In contrast, when $q_i \rightarrow 0$ and  $q_{i-1} \rightarrow 0$, $p_i$ approaches its upper limit $e^{-I_N}$.

\subsection{Mean firing rate \label{sec:binary_hetero_avg_fr}}
The mean firing rate can be calculated as
\begin{equation}
\begin{aligned}
    \bar{\nu} &= \frac{1}{N} \sum_{i=1}^N \nu_{\max_i}\, u_i \\
    &= \frac{e^{-I_N}}{NT} \left[-\sum_{i=1}^N q_i^{q_i/(1-q_i)}\log q_i - \sum_{i=1}^m \log q_i \sum_{j=1}^{i-1}(1-q_j)q_j^{q_j/(1-q_j)} - \sum_{i=m+1}^N \log q_i \sum_{j=m+1}^{i-1}(1-q_j)q_j^{q_j/(1-q_j)}\right]
\end{aligned}
\end{equation}
Here, we prove that (1) homogeneous populations of ON or OFF neurons generate the highest mean firing rate $\bar{\nu}$; (2) When the lowest $\bar{\nu}$ is reached, Eq. 85 and Eq. 86 must be valid. To proceed, we denote 
\begin{equation}
    A_i = - \log q_i, \quad B_i = (1-q_i)\, q_i^{q_i/(1-q_i)}. 
\end{equation}
Note that for any $i$, $A_i > 0$ and $B_i > 0$.

To prove the first argument, we rewrite the mean firing rate as a function of $m$, the number of ON neurons in the population:
\begin{equation}
    \bar{\nu}(m) = \frac{e^{-I_N}}{NT} \left[\sum_{i=1}^N q_i^{q_i/(1-q_i)}A_i + \sum_{i=1}^m A_i \sum_{j=1}^{i-1}B_j + \sum_{i=m+1}^N A_i \sum_{j=m+1}^{i-1}B_j\right].
    \label{eq:nu_A_B}
\end{equation}
One can then obtain
\begin{equation}
\begin{aligned}
    \bar{\nu}(m = N) - \bar{\nu}(m) &= \frac{e^{-I_N}}{NT} \left[\sum_{i=1}^N A_i \sum_{j=1}^{i-1}B_j - \sum_{i=1}^m A_i \sum_{j=1}^{i-1}B_j - \sum_{i=m+1}^N A_i \sum_{j=m+1}^{i-1}B_j\right] \\
    &= \frac{e^{-I_N}}{NT} \left[\sum_{i=m+1}^N A_i \sum_{j=1}^{i-1}B_j - \sum_{i=m+1}^N A_i \sum_{j=m+1}^{i-1}B_j\right] \\
    &= \frac{e^{-I_N}}{NT}\sum_{i=m+1}^N A_i \sum_{j = 1}^m B_j \geq 0,
\end{aligned}
\end{equation}
which means a homogeneous ON population always generates higher mean firing rate than a mixed population with the same maximal firing rates $\lbrace \nu_{\max, i} \rbrace$. Similarly, one can also find $\bar{\nu}(m = 0) - \bar{\nu}(m) \geq 0$. Therefore, we have proved the first proposition above, that homogeneous populations of ON or OFF neuron generate the highest mean firing rate $\bar{\nu}$.

To prove Eq. 85, we assume for neuron $t$ and $t+1$ ($1 \leq t < m$), $q_t > q_{t+1}$, which results in $A_t < A_{t+1}$ and $B_t < B_{t+1}$. Then we swap neuron $t$ and $t+1$ and show that $\bar{\nu}$ becomes lower. The change of $\bar{\nu}$ due to the swap can be expressed as
\begin{equation}
    \begin{aligned}
    \Delta \bar{\nu} &= \frac{e^{-I_N}}{NT} \left[ \sum_{i = 1}^{t-1}A_i\sum_{j=1}^{i-1}B_j + A_{t+1}\sum_{j=1}^{t-1}B_j +A_t\left(\sum_{j=1}^{t-1} B_j +B_{t+1} \right)  + \sum_{i=t+2}^m A_i\sum_{j=1}^{i-1}B_j - \sum_{i=1}^N A_i\sum_{j=1}^{i-1}B_j \right] \\
    &= \frac{e^{-I_N}}{NT} \left[ (A_{t+1} - A_t)\sum_{j=1}^{t-1}B_j + (A_t - A_{t+1})\sum_{j=1}^t B_j +  (B_{t+1} -B_t)A_t \right]\\
    &= \frac{e^{-I_N}}{NT} \left(B_{t+1}A_t - A_{t+1}B_t\right).
    \end{aligned}
\end{equation}
To compare $B_{t+1}A_t$ and $A_{t+1}B_t$, we calculate the derivative $\de(A_i/B_i)/\de q_i$ as follows:
\begin{equation}
    \frac{\de}{\de q_i}\left(\frac{A_i}{B_i}\right) = B_i\left[\left(\frac{\log q_i}{1-q_i}\right)^2 - \frac{1}{q_i}\right],
\end{equation}
where we have used Eq.~\ref{eq:d_f1_f2} to acquire $\de B_i/\de q_i = B_i\log q_i/(1-q_i)^2$. Then we define
\begin{equation}
    f_3(x) = \log x - \frac{x-1}{\sqrt x}
\end{equation}
and show that $f_3'(x) \leq 0$. This leads to
\begin{equation}
    \frac{\de}{\de q_i}\left(\frac{A_i}{B_i}\right) = B_i\left[\left(\frac{\log q_i}{1-q_i}\right)^2 - \frac{1}{q_i}\right] < 0,
\end{equation}
and then
\begin{equation}
    \frac{A_t}{B_t} < \frac{A_{t+1}}{B_{t+1}}.
\end{equation}
Therefore, if $q_t > q_{t+1}$, swapping neuron $t$ and $t+1$ will reduce the mean firing rate $\bar{\nu}$. This implies when the mean firing rate $\bar{\nu}$ is minimized, there must be 
\begin{equation}
    q_1 \leq ... \leq q_m.
\end{equation}
In the same way we can prove that there must also be
\begin{equation}
    q_{m+1} \leq ... \leq q_N,
\end{equation}
which is the equivalency for OFF neurons. These two equations prove Eq. 85 in the main text.

Finally, to prove Eq. 86, we calculate the difference $\bar \nu(m) - \bar \nu(m-1)$, which should converge to 0 when $\bar \nu(m)$ approaches the minimum. According to Eq.~\ref{eq:nu_A_B},
\begin{equation}
    \bar \nu(m) - \bar \nu(m-1) = A_m \sum_{j=1}^{m-1}B_j - B_m \sum_{j=m+1}^N A_i.
\end{equation}
Then $\bar \nu(m) - \bar \nu(m-1) \rightarrow 0$ gives rise to Eq. 86 in the main text.

\setcounter{equation}{0}

\section{Population coding of ternary neurons with any noise generation function \label{sec:ternary_neurons}}

\subsection{Maximal mutual information of a population of ternary neurons \label{sec:ternary_MI_supp}}
Following Eq. 29, we decompose the mutual information encoded by $N$ neurons as
\begin{equation}
I_N\left(\vec {u_1}, ..., \vec {u_N}\right) = I_m\left(\vec {u_1}, ..., \vec {u_m}\right) + Q_mI_{N-m}\left(\vec {u_{m+1}^{(m)}}, ..., \vec {u_N^{(m)}}\right).
\label{decompose_mutual_info}
\end{equation}
Here, the cumulative
stimulus intervals $\vec u_i$ are written as vectors because the activation functions are ternary. $Q_m$ denotes the probability that none of the neurons $1, 2, ..., m$ fires. 
%Note that there is a superscript $(m)$ for every neuron in the second term. This 
As before, we define the superscript $(m)$ to denote the  `revised' stimulus intervals assuming that the $m$ ON neurons do not fire. 

To generalize the definitions of these revised thresholds for $N$ ternary activation functions, we use $u_{i1}^{(j,N)}$ and $u_{i2}^{(j,N)}$ to denote the cumulative stimulus intervals given the condition that none of the neurons $1, ..., j \, (j < i)$ fires, and 
\begin{equation}
P_i^{(j,N)}(0)=1-u_{i1}^{(j,N)}(1-L(0,f_iR))-u_{i2}^{(j,N)}(1-L(0,R))
\label{eq:Pi0}
\end{equation}
to denote the probability that neuron $i$ does not
fire, when none of the neurons $1, ..., j \, (j < i)$ fires.
We can decompose the mutual information encoded by a population of $N$ neurons into $N$ single terms, each of which contains the mutual information encoded by one neuron. This allows us to calculate the mutual information and optimize the threshold in a recursive way, first for one neuron, then for two neurons and then generalize for any $N$ neurons (compare to Eq. 32), 
\begin{equation}
I_N = g(\vec {u_1}) + P_1(0)\left \{ g\left(\vec {u_2^{(1)}}\right)+ P_2^{(1, N)}(0) \left[g\left(\vec {u_3^{(2)}}\right) + ...+P_{N-1}^{(N-2,N)}(0)\,g\left(\vec {u_N^{(N-1)}}\right)\right]\right \}.
\label{recursive_calculation}
\end{equation}
Note that similar to the case of binary neurons (Eq.~\ref{eq:I1_binary} and Eq.~\ref{eq:I_decompose_binary}), the function $g$ here still denotes the mutual information of a single neuron, despite the difference that it takes in a vector instead of a scalar.

The mutual information of one single neuron can be written as 
%a function of $u_{11}$ and $u_{12}$,
%\begin{equation}
%I_1 = g(u_{11}, u_{12}).
%\label{I1_g_u1_u2}
%\end{equation}
%
%The complete expression of the function is
\begin{equation}
\begin{aligned}
I_1 &= g(u_{11}, u_{12}) =\sum_n \int_s \de s\,p(s)p(n|s)\log \frac{P(n|s)}{P(n)} \\
&=-(1-u_{11}-u_{12})\log P(0)+u_{11}L(0,fR)\log \frac{L(0,fR)}{P(0)}+u_{12}L(0,R)\log \frac{L(0,R)}{P(0)}\\
&+u_{11}\sum_{n=1}^{+\infty}L(n,fR)\log \frac{L(n,fR)}{u_{11}L(n,fR)+u_{12}L(n,R)} + u_{12}\sum_{n=1}^{+\infty}L(n,R)\log \frac{L(n,R)}{u_{11}L(n,fR)+u_{12}L(n,R)}\\
&=-P(0)\log P(0)+u_{11}L(0,fR)\log L(0,fR)+u_{12}L(0,R)\log L(0,R)\\
&+u_{11}\sum_{n=1}^{+\infty}L(n,fR)\log \frac{L(n,fR)}{u_{11}L(n,fR)+u_{12}L(n,R)}+u_{12}\sum_{n=1}^{+\infty}L(n,R)\log \frac{L(n,R)}{u_{11}L(n,fR)+u_{12}L(n,R)}.
\end{aligned}
\label{I1_ternary_neuron}
\end{equation}
Its derivatives are
\begin{equation}
\frac{\partial g(u_{11},u_{12})}{\partial u_{11}} = \frac{\partial P(0)}{\partial u_{11}}\left[-1-\log P(0)\right]+L(0, fR)\log L(0, fR)+\sum_{n=1}^{+\infty}L(n,fR) \left(\log \frac{L(n,fR)}{u_{11}L(n,fR)+u_{12}L(n,R)}-1\right)
\label{eq:derivative_g_u1}
\end{equation}
\begin{equation}
\frac{\partial g(u_{11},u_{12})}{\partial u_{12}} = \frac{\partial P(0)}{\partial u_{12}}\left[-1-\log P(0)\right]+L(0, R)\log L(0, R)+\sum_{n=1}^{+\infty}L(n,R)\left(\log \frac{L(n,R)}{u_{11}L(n,fR)+u_{12}L(n,R)}-1\right)
\label{eq:derivative_g_u2}
\end{equation}
and
\begin{equation}
\begin{aligned}
\frac{\partial g(u_{11},u_{12})}{\partial f}=&-u_{11}\frac{\partial L(0,fR)}{\partial f}\log P(0)+u_{11}\sum_{n=0}^{+\infty}\log L(n,fR)\frac{\partial L(n,fR)}{\partial f}\\
&-\sum_{n=1}^{+\infty}\log [u_{11}L(n,fR)+u_{12}L(n,R)]u_{11}\frac{\partial L(n,fR)}{\partial f}
\label{eq:derivative_g_f}.
\end{aligned}
\end{equation}
Note that
\begin{equation}
P(0) = 1- u_{11}\left[1-L(0,fR)\right] - u_{12}\left[1-L(0,R)\right],
\end{equation}
\begin{equation}
\frac{\partial P(0)}{\partial u_{11}} = L(0,fR) - 1, \, \, \, \, \, \frac{\partial P(0)}{\partial u_{12}} = L(0,R) - 1
\end{equation}
we can rewrite Eq.~\ref{eq:derivative_g_u1} and Eq.~\ref{eq:derivative_g_u2} into
\begin{equation}
\frac{\partial g(u_{11},u_{12})}{\partial u_{11}} =\left[1-L(0,fR)\right]\log P(0)+L(0,fR)\log L(0,fR)
+\sum_{n=1}^{+\infty}L(n,fR)\log \frac{L(n,fR)}{u_{11}L(n,fR)+u_{12}L(n,R)}
\end{equation}
and
\begin{equation}
\frac{\partial g(u_{11},u_{12})}{\partial u_{12}} =\left[1-L(0,R)\right]\log P(0)+L(0,R)\log L(0,R)
+\sum_{n=1}^{+\infty}L(n,R)\log \frac{L(n,R)}{u_{11}L(n,fR)+u_{12}L(n,R)}.
\end{equation}
\noindent Combining these two derivatives, and comparing to Eq.~\ref{I1_ternary_neuron}, we can find
\begin{equation}
g(u_{11},u_{12})=u_{11} \frac{\partial g}{\partial u_{11}}+u_{12} \frac{\partial g}{\partial u_{12}}-\log P_1(0).
\label{g_u1_u2}
\end{equation}
When $g(u_{11},u_{12})$ is at its maximal value, $\partial g/\partial u_{11}=0$, $\partial g/\partial u_{12}=0$, then
\begin{equation}
I_1^{\max} = g(u_{11}^*,u_{12}^*)=-\log P_1(0).
\end{equation}

We then examine the mutual information of a population of two neurons,
\begin{equation}
I_2 = g(u_{11}, u_{12}) + P_1(0)\,g\left(u_{21}^{(1)}, u_{22}^{(1)}\right).
\end{equation}
Maximizing $g\left(u_{21}^{(1)}, u_{22}^{(1)}\right)$ gives
\begin{equation}
I_2=g(u_{11},u_{12})-P_1(0)\log P_2^{(1)}(0).
\label{eq:I2_I1}
\end{equation}
Note that $P_1(0)=1-u_{11}(1-L(0, fR))-u_{22}(1-L(0,R))$. When $I_2$ is at its maximum, we have $\partial I_2/\partial u_{11} = 0$ and $\partial I_2/\partial u_{12} = 0$, which leads to 
\begin{equation}
I_2^{\max}=-\log P_2^{(1)}(0)-\log P_1(0).
\label{I_2_max}
\end{equation}

Next, we use mathematical induction to prove that for arbitrary $N$ (Eq.~\ref{I_N_max}), there is
\begin{equation}
I_N^{\max}=-\sum_{j=1}^N\log P_j^{(j-1)}(0).
\end{equation}
Assume $I_m^{\max}=-\sum_{j=1}^{m}\log P_j^{(j-1)}(0)$, we have
\begin{equation}
I_{m+1}=g(u_{11},u_{12})-P_1(0)\sum_{j=2}^{m+1}\log P_j^{(j-1)}(0).
\end{equation}
\noindent When optimized,
\begin{equation}
\nonumber
\frac{\partial I_{m+1}}{\partial u_{11}}=\frac{\partial g}{\partial u_{11}}+\left[1-L(0,fR)\right]\sum_{j=2}^{m+1}\log P_j^{(j-1)}(0)=0
\end{equation}
\begin{equation}
\frac{\partial I_{m+1}}{\partial u_{12}}=\frac{\partial g}{\partial u_{12}}+\left[1-L(0,R)\right]\sum_{j=2}^{m+1}\log P_j^{(j-1)}(0)=0
\end{equation}
\begin{equation}
\begin{aligned}
I_{m+1}^{\max}&=u_{11}\frac{\partial g}{\partial u_{11}}+u_{12}\frac{\partial g}{\partial u_{12}}-\log P_1(0)-P_1(0)\sum_{j=2}^{m+1}\log P_j^{j-1}(0) \\
&=-\left[u_{11}(1-L(0,fR))+u_{12}(1-L(0,R))\right]\sum_{j=2}^{m+1}\log P_j^{(j-1)}(0)-\log P_1(0)-P_1(0)\sum_{j=2}^{m+1}\log P_j^{(j-1)}(0) \\
&=-\sum_{j=1}^{m+1}\log P_j^{(j-1)}(0).
\end{aligned}
\end{equation}
\noindent Hence we have verified that if $I_m^{\max}=-\sum_{j=1}^{m}\log P_j^{(j-1)}(0)$, we can show $I_{m+1}^{\max}=-\sum_{j=1}^{m+1}\log P_j^{(j-1)}(0)$. This means we can use mathematical induction to generalize the mutual information from $N = 2$ to arbitrary number of neurons $N$, and obtain
\begin{equation}
I_N^{\max}=-\sum_{j=1}^{N}\log P_j^{(j-1)}(0) = - \text{log}P(\vec 0).
\label{I_N_max}
\end{equation}

%%%%%%%%%%%

\subsection{Revised probability of ternary neurons}
\noindent The mutual information of $N$ neurons can be formulated by Eq.~\ref{recursive_calculation}, i.e.,
\begin{equation}
\nonumber
I_N = g(\vec {u_1}) + P_1(0)\left \{ g\left(\vec {u_2^{(1)}}\right)+ P_2^{(1, N)}(0) \left[g\left(\vec {u_3^{(2)}}\right) + ...+P_{N-1}^{(N-2,N)}(0)g\left(\vec {u_N^{(N-1)}}\right)\right]\right \}.
\end{equation}
Assuming we already know that neuron $1, 2, ..., j$ do not fire, we can then derive that for ON neurons, if $j+1 < i$,
\begin{equation}
    u_{i1}^{(j+1,N)} =\int_{\theta_{i1}^{(j,N)}}^{\theta_{i2}^{(j,N)}} p(s|n_{j+1}= 0)\, \text ds =\int_{\theta_{i1}^{(j,N)}}^{\theta_{i2}^{(j,N)}} \frac{p(s)\, p(n_{j+1}=0|s)}{p(n_{j+1} = 0)}\, \text ds.
\label{eq:u1_revised_1}
\end{equation}
Since we already know that neuron $1, 2, ..., j$ do not fire, here we have $p(n_{j+1} = 0) = P_{j+1}^{(j,N)}(0)$. Also, given $j+1 < i$, stimulus $s$ between $\theta_{i1}^{(j,N)}$ and $\theta_{i2}^{(j,N)}$ is too low to trigger a nonzero firing rate of neuron $j+1$. Therefore, within the interval of the integral of Eq.~\ref{eq:u1_revised_1}, $p(n_{j+1}=0|s) = 1$. Eq.~\ref{eq:u1_revised_1} then becomes
\begin{equation}
    u_{i1}^{(j+1,N)} = \frac{1}{P_{j+1}^{(j,N)}(0)} \int_{\theta_{i1}^{(j,N)}}^{\theta_{i2}^{(j,N)}} p(s) \text ds = \frac{u_{i1}^{(j,N)}}{P_{j+1}^{(j,N)}(0)}.
\label{eq:u1_revised}
\end{equation}
Similarly, we have
\begin{equation}
    u_{i2}^{(j+1,N)} =\int_{\theta_{i2}^{(j,N)}}^{+\infty} p(s|n_{j+1}= 0)\, \text ds
    = \int_{\theta_{i2}^{(j,N)}}^{+\infty} \frac{p(s)\, p(n_{j+1}=0|s)}{p(n_{j+1} = 0)}\, \text ds.
\label{eq:u2_revised_1}
\end{equation}
Same as Eq.~\ref{eq:u1_revised_1}, we have $p(n_{j+1} = 0) = P_{j+1}^{(j,N)}(0)$. Also because $p(n_{j+1}=0|s) = 1-p(n_{j+1} \neq 0|s)$, we can rewrite Eq.~\ref{eq:u2_revised_1} as
\begin{equation}
    u_{i2}^{(j+1,N)} = \frac{1}{P_{j+1}^{(j,N)}(0)} \int_{\theta_{i2}^{(j,N)}}^{+\infty} p(s) \left[1 - p(n_{j+1} \neq 0|s) \right]\, \text ds.
    \label{eq:u2_revised_2}
\end{equation}
Because stimulus $s$ lower than $\theta_{i2}^{(j,N)}$ cannot trigger a nonzero firing rate of neuron $j+1$, we have
\begin{equation}
    \int_{\theta_{i2}^{(j,N)}}^{+\infty} p(s)\, p(n_{j+1} \neq 0|s)\, \text ds = 1 - P_{j+1}^{(j,N)}(0).
\end{equation}
Substituting back into Eq.~\ref{eq:u2_revised_2} gives rise to
\begin{equation}
    u_{i2}^{(j+1,N)} = \frac{1}{P_{j+1}^{(j,N)}(0)}\left[u_{i2}^{(j,N)}-\left(1-P_{j+1}^{(j,N)}(0)\right)\right].
    \label{eq:u2_revised}
\end{equation}
Notably, Eq.~\ref{eq:u1_revised} and Eq.~\ref{eq:u2_revised} have similar formulation compared to Eq.~\ref{eq:revised_u_binary_ON} and Eq.~\ref{eq:revised_u_binary_OFF}. 
%Eq.~\ref{u1_u2_opt_supp}, Eq.~\ref{recursive_calculation_equality_supp}, Eq.~\ref{eq:u1_revised_supp}, and Eq.~\ref{eq:u2_revised_supp} produce Eq.~\ref{u1_u2_opt} - \ref{eq:u2_revised} in the main text.

When $I_N$ is maximized,
\begin{equation}
\frac{\partial I_N}{\partial \vec {u_N^{(N-1)}}} = \left( \prod_{i=1}^{N-1} P_i^{(i-1, N)}(0) \right) \frac{\partial g\left(\vec {u_N^{(N-1)}}\right)}{\partial \vec {u_N^{(N-1)}}}= 0.
\end{equation}
Therefore, optimizing $u_N^{(N-1)}$ is equivalent to maximizing the function $g(\cdot)$, which is the mutual information of a single neuron. Denoting the optimal values using an asterisk, we have
\begin{equation}
u_{N,1}^{(N-1,N)}|_*=u_1^*, \ u_{N,2}^{(N-1,N)}|_* =u_2^*,
\label{u1_u2_opt}
\end{equation}
where $u_1^*$ and $u_2^*$ correspond to the $(u_11, u_12)$ that maximizes $g(\vec u_1) = g(u_{11}, u_{12})$. Also, we can see that 
\begin{equation}
g \left( \vec {u_2^{(1)}}\right)+ P_2^{(1, N)}(0)\Bigg[g\left(\vec {u_3^{(2)}}\right) + ...+P_{N-1}^{(N-2,N)}(0)g\left(\vec {u_N^{(N-1)}}\right)\Bigg]
\end{equation}
has the exactly same formulation as $I_{N-1}$,
which means
\begin{equation}
u_{i,1}^{(j,N)}|_*=u_{i+1,1}^{(j+1,N+1)}|_*,\ u_{i,2}^{(j,N)}|_*=u_{i+1,2}^{(j+1,N+1)}|_*.
\label{recursive_calculation_equality}
\end{equation}

\subsection{A lemma that connects two adjacent neurons \label{sec:lemma}}
Next, we seek to find the optimal thresholds by deriving the relationship among $\lbrace u_{i1} \rbrace$ and $\lbrace u_{i2} \rbrace$. We start with proving the following lemma that links two adjacent neurons, i.e., $(u_{i1}, u_{i2}, f_i)$ and $(u_{i+1, 1}, u_{i+1, 2}, f_{i+1})$.
\\
\\
\textbf{Lemma:} For any $N$ neurons, when $I_N$ is optimized, 
\begin{equation}
u_{i1}^{(i-1,N)}=P_i^{(i-1,N)}(0)u_{i+1,1}^{(i,N)},\ \ \  u_{i2}^{(i-1,N)}=P_i^{(i-1,N)}(0)u_{i+1,2}^{(i,N)},\ \ \ f_i = f_{i+1}.
\label{lemma}
\end{equation}
%Since we assumed that there is no spontaneous firing rate, the mutual information can be decomposed as
\textbf{Remark:}
\begin{equation} 
I=g(u_{11},u_{12})+P_1^{(0,N)}(0)\left[g\left(u_{21}^{(1,N)},u_{22}^{(1,N)}\right)+ ... +P_{N-1}^{(N-2,N)}(0)g\left(u_{N1}^{(N-1,N)},u_{N2}^{(N-1,N)}\right)\right].
\label{eq:remark_1_1}
\end{equation}
When all the $u_{i1}, u_{i2}$ and $f_i$ are optimized, according to the previous subsection,
\begin{equation}
\begin{aligned}
&g\left(u_{i1}^{(i-1,N)},u_{i2}^{(i-1,N)}\right)+P_i^{(i-1,N)}(0)\left[g\left(u_{i+1,1}^{(i,N)},u_{i+1,2}^{(i,N)}\right)+...+P_{N-1}^{(N-2,N)}(0)g\left(u_{N1}^{(N-1,N)}, u_{N2}^{(N-1,N)}\right)\right] \\
&=-\sum_{j=i}^N \log P_j^{(j-1,N)}(0).
\label{eq:remark_1_2}
\end{aligned}
\end{equation}
This means when $\left( u_{i+1,1}^{(i,N)}, u_{i+1,2}^{(i,N)}, f_{i+1} \right)$ is optimized,
\begin{equation}
\frac{\partial I_N}{\partial u_{i+1,1}^{(i,N)}} = P_i(0) \left \{ \frac{\partial g\left(u_{i+1,1}^{(i,N)},u_{i+1,2}^{(i,N)}\right)}{\partial u_{i+1,1}^{(i,N)}}+\left[1-L(0,f_{i+1}R)\right]\sum_{j=i+2}^N\log P_j^{(j-1,N)}(0)  \right \} = 0
\label{derivative_u_i+1}
\end{equation}

\begin{equation}
\frac{\partial I_N}{\partial f_{i+1}}=P_i(0) \left \{ \frac{\partial g\left(u_{i+1,1}^{(i,N)},u_{i+1,2}^{(i,N)}\right)}{\partial f_{i+1}}-u_{i+1,1}^{(i,N)}\frac{\partial L(0, f_{i+1}R)}{\partial f_{i+1}}\sum_{j=i+2}^N\log P_j^{(j-1,N)}(0)  \right \} = 0.
\label{derivative_f_i+1}
\end{equation}
Denote 
\begin{equation}
\hat{u}_{i1}=P_i^{(i-1,N)}(0)u_{i+1,1}^{(i,N)},\ \ \ \hat{u}_{i2}=P_i^{(i-1,N)}(0)u_{i+1,2}^{(i,N)},\ \ \ \hat{f}_i = f_{i+1},
\label{eq:define_u1_u2_f_star}
\end{equation}
    so that with given $\hat{u}_{i1}, \hat{u}_{i2}, \hat{f}_i$,
\begin{equation}
P_i^{(i-1,N)}(0) = 1-\hat{u}_{i1}[1-L(0,f_iR)]-\hat{u}_{i2}[1-L(0,R)]
\end{equation}
\begin{equation}
\hat{u}_{i1}=\frac{u_{i+1,1}^{(i,N)}}{1+u_{i+1,1}^{(i,N)}[1-L(0,f_iR)]+u_{i+1,2}^{(i,N)}[1-L(0,R)]}
\end{equation}
\begin{equation}
\hat{u}_{i2}=\frac{u_{i+1,2}^{(i,N)}}{1+u_{i+1,1}^{(i,N)}[1-L(0,f_iR)]+u_{i+1,2}^{(i,N)}[1-L(0,R)]}.
\end{equation}
We need to prove
\begin{equation}
\frac{\partial I_N}{\partial u_{i1}^{(i-1,N)}}|_{\hat{u}_{i1},\hat{u}_{i2},\hat{f}_i}=0,\ \ \ \frac{\partial I_N}{\partial u_{i2}^{(i-1,N)}}|_{\hat{u}_{i1},\hat{u}_{i2},\hat{f}_i}=0,\ \ \ \frac{\partial I_N}{\partial f_i}|_{\hat{u}_{i1},\hat{u}_{i2},\hat{f}_i}=0
\end{equation}
\noindent i.e.~the combination of $\hat{u}_{i1},\hat{u}_{i2}$, and $\hat{f}_i$ is optimal.
\\
\\
$\textbf{Proof:}$
\begin{equation}
\frac{\partial I_N}{\partial u_{i1}^{(i-1,N)}}|_{\hat{u}_{i1}, \hat{u}_{i2}, \hat{f}_i}=P_{i-1}(0) \left \{ \frac{\partial g\left(u_{i1}^{(i-1,N)},u_{i2}^{(i-1,N)}\right)}{\partial u_{i1}^{(i-1,N)}}|_{\hat{u}_{i1}, \hat{u}_{i2}, \hat{f}_i}+[1-L(0,\hat{f}_iR)]\sum_{j=i+1}^N\log P_j^{(j-1,N)}(0)  \right \}
\end{equation}
\begin{equation}
\frac{\partial I_N}{\partial u_{i2}^{(i-1,N)}}|_{\hat{u}_{i1}, \hat{u}_{i2}, \hat{f}_i}=P_{i-1}(0) \left \{\frac{\partial g\left(u_{i1}^{(i-1,N)},u_{i2}^{(i-1,N)}\right)}{\partial u_{i2}^{(i-1,N)}}|_{\hat{u}_{i1}, \hat{u}_{i2}, \hat{f}_i}+[1-L(0,R)]\sum_{j=i+1}^N\log P_j^{(j-1,N)}(0) \right \}.
\end{equation}
According to Eq.~\ref{derivative_u_i+1},
\begin{equation}
\begin{aligned}
&\frac{1}{P_{i-1}(0)}\frac{\partial I_N}{\partial u_{i1}^{(i-1,N)}}|_{\hat{u}_{i1}, \hat{u}_{i2}, \hat{f}_i} - \frac{1}{P_i(0)}\frac{\partial I_N}{\partial u_{i+1,1}^{(i, N)}}|_{u_{i+1,1}^{(i,N)},u_{i+1,2}^{(i,N)},f_{i+1}} \\
&\xlongequal{Eq.~\ref{derivative_u_i+1}} \frac{\partial g\left(u_{i1}^{(i-1,N)},u_{i2}^{(i-1,N)}\right)}{\partial u_{i1}^{(i-1,N)}}|_{\hat{u}_{i1}, \hat{u}_{i2}, \hat{f}_i}- \frac{\partial g\left(u_{i+1,1}^{(i,N)},u_{i+1,2}^{(i,N)}\right)}{\partial u_{i+1,1}^{(i,N)}}+\left[1-L(0,\hat{f}_iR)\right]\log P_{i+1}^{(i,N)}(0)\\
&\xlongequal{Eq.~\ref{eq:derivative_g_u1}}[1-L(0,\hat{f}_iR)]\left(\log P_i^{(i-1,N)}(0)-\log P_{i+1}^{(i,N)}(0)\right)-\sum_{n=1}^{+\infty}L(n,\hat{f}_iR)\log \left[\hat{u}_{i1}L(n,fR)+\hat{u}_{i2}L(n,R)\right]+ \\
&\, \, \, \, \, \, \, \, \sum_{n=1}^{+\infty}L(n,\hat{f}_iR)\log \left[u_{i+1,1}^{(i,N)}L(n,fR)+u_{i+1,2}^{(i,N)}L(n,R)\right]+[1-L(0,\hat{f}_iR)]\log P_{i+1}^{(i,N)}(0) \\
&=[1-L(0,\hat{f}_iR)]\left(\log P_i^{(i-1,N)}(0)-\log P_{i+1}^{(i,N)}(0)\right)-\sum_{n=1}^{+\infty}L(n,\hat{f}_iR)\log P_i^{(i-1,N)}(0)+[1-L(0,\hat{f}_iR)]\log P_{i+1}^{(i,N)}(0).
\end{aligned}
\end{equation}
Because
\begin{equation}
L(0,\hat{f}_iR)+\sum_{n=1}^{+\infty}L(n,\hat{f}_iR)=1\ \ \ \text{(the sum of probabilities),}
\end{equation}
\begin{equation}
\frac{1}{P_{i-1}(0)}\frac{\partial I_N}{\partial u_{i1}^{(i-1,N)}}|_{\hat{u}_{i1}, \hat{u}_{i2}, \hat{f}_i} - \frac{1}{P_i(0)}\frac{\partial I_N}{\partial u_{i+1,1}^{(i, N)}}=0,\ \ \ \frac{\partial I_N}{\partial u_{i1}^{(i-1,N)}}|_{\hat{u}_{i1},\hat{u}_{i2},\hat{f}_i}=0.
\label{eq:derivative_u1_0}
\end{equation}
Similarly, we have
\begin{equation}
\frac{\partial I_N}{\partial u_{i2}^{(i-1,N)}}|_{\hat{u}_{i1},\hat{u}_{i2},\hat{f}_i}=0.
\label{eq:derivative_u2_0}
\end{equation}
Also,
\begin{equation}
\begin{aligned}
&\quad \quad\frac{1}{\hat{u}_{i1}P_{i-1}(0)}\frac{\partial I_N}{\partial f_i}|_{\hat{u}_{i1}, \hat{u}_{i2}, \hat{f}_i}-\frac{1}{u_{i+1,1}^{(i,N)} P_i(0)}\frac{\partial I_N}{\partial f_{i+1}}|_{u_{i+1,1}^{(i,N)},u_{i+1,2}^{(i,N)},f_{i+1}} \\
&\xlongequal{Eq.~\ref{derivative_f_i+1}}-\frac{\partial L(0,\hat{f}_iR)}{\partial \hat{f}_i}\log P_i^{(i-1, N)}(0)+\sum_{n=0}^{+\infty}\log L(n,\hat{f}_iR)\frac{\partial L(n,\hat{f}_iR)}{\partial \hat{f}_i}\\
&\quad-\sum_{n=1}^{+\infty}\log \left[\hat{u}_{i1}L(n,\hat{f}_iR)+\hat{u}_{i2}L(n,R)\right]\frac{\partial L(n,\hat{f}_iR)}{\partial \hat{f}_i} -\sum_{n=0}^{+\infty}\log L(n,f_{i+1}R)\frac{\partial L(n,f_{i+1}R)}{\partial f_{i+1}} \\
&\, \, \, \, \, \, \, \, +\sum_{n=1}^{+\infty}\log \left[u_{i+1,1}L(n,f_{i+1}R)+u_{i+1,2}L(n,R)\right]\frac{\partial L(n,f_{i+1}R)}{\partial f_{i+1}} \\
&\, \, \, \, \, \, \, \, -\left(\frac{\partial L(0,\hat{f}_iR)}{\partial \hat{f}_i}-\frac{\partial L(0,f_{i+1}R)}{\partial f_{i+1}}\right)\sum_{j=i+1}^N \log P_j^{(j-1,N)}(0).
\end{aligned}
\end{equation}
According to the definition of $\hat{f}_i, \hat{u}_{i1}$ and $\hat{u}_{i2}$ (Eq.~\ref{eq:define_u1_u2_f_star}), we have
\begin{equation}
\begin{aligned}
&\frac{1}{\hat{u}_{i1}P_{i-1}(0)}\frac{\partial I_N}{\partial f_i}|_{\hat{u}_{i1}, \hat{u}_{i2}, \hat{f}_i}-\frac{1}{u_{i+1,1}^{(i,N)} P_i(0)}\frac{\partial I_N}{\partial f_{i+1}}\\
&=-\frac{\partial L(0,\hat{f}_iR)}{\partial \hat{f}_i}\log P_i^{(i-1,N)}(0)-\log P_i^{(i-1,N)}(0)\sum_{n=1}^{+\infty}\frac{\partial L(n,\hat{f}_iR)}{\partial \hat{f}_i}\\
&=-\sum_{n=0}^{+\infty}\frac{\partial L(n,\hat{f}_iR)}{\partial \hat{f}_i}\log P_i^{(i-1,N)}(0) \\
&=-\frac{\partial}{\partial \hat{f}_i}\left[\sum_{n=0}^{+\infty}L(n,\hat{f}_iR)\right]\log P_i^{(i-1,N)}(0).
\end{aligned}
\end{equation}
Since $\sum_{n=0}^{+\infty}L(n,\hat{f}_iR)=1$ is a constant, $\frac{\partial}{\partial \hat{f}_i}\left[\sum_{n=0}^{+\infty}L(n,\hat{f}_iR)\right]=0$, then
\begin{equation}
\frac{1}{\hat{u}_{i1}P_{i-1}(0)}\frac{\partial I_N}{\partial f_i}|_{\hat{u}_{i1}, \hat{u}_{i2}, \hat{f}_i}-\frac{1}{u_{i+1,1}^{(i,N)} P_i(0)}\frac{\partial I_N}{\partial f_{i+1}} = 0
\end{equation}
which leads to
\begin{equation}
\frac{\partial I_N}{\partial f_i}|_{\hat{u}_{i1},\hat{u}_{i2},\hat{f}_i}=0.
\label{eq:derivative_f_0}
\end{equation}
Combined with Eq.~\ref{eq:derivative_u1_0} and Eq.~\ref{eq:derivative_u2_0}, the lemma has been proved. $\blacksquare$

\subsection{Optimal thresholds for a homogeneous population of only ON neurons}

We first consider a homogeneous population with only ON neurons. All the variables in this subsection are optimized, so we omit the * symbol (that we previously used to indicate optimum, e.g., Eq.~\ref{recursive_calculation_equality}) for clarity. From the lemma (Eq.~\ref{lemma}) we know that $f_i = f_{i+1}$, which we denote as $f$. For brevity, we also denote $q_1 = L(0, fR)$ and $q_2 = L(0,R)$.

Combining the lemma (Eq.~\ref{lemma}) and Eq.~\ref{eq:Pi0}
\begin{equation}
P_i^{(i-1,N)}(0) = 1 - u_{i1}^{(i-1, N)}\left(1-q_1\right) - u_{i2}^{(i-1, N)}\left(1-q_2\right),
\label{eq:P0_i_i-1}
\end{equation}
we can write
\begin{equation}
\begin{aligned}
P_i^{(i-1,N)}(0) &= \left[1 + u_{i+1,1}^{(i, N)}\left(1-q_1\right) + u_{i+1,2}^{(i, N)}\left(1-q_2\right) \right]^{-1} \\
u_{i1}^{(i-1,N)} &= \frac{ u_{i+1,1}^{(i, N)}}{1 + u_{i+1,1}^{(i, N)}\left(1-q_1\right) + u_{i+1,2}^{(i, N)}\left(1-q_2\right)} \\
u_{i2}^{(i-1,N)} &= \frac{ u_{i+1,2}^{(i, N)}}{1 + u_{i+1,1}^{(i, N)}\left(1-q_1\right) + u_{i+1,2}^{(i, N)}\left(1-q_2\right)}.
\label{eq:recursive_ui_i-1}
\end{aligned}
\end{equation}
Eq.~\ref{eq:recursive_ui_i-1} allows us to recursively calculate $\left(u_{i1}^{(i-1,N)}, u_{i2}^{(i-1,N)}\right)$ from $\left(u_{i+1,1}^{(i,N)}, u_{i+1,2}^{(i,N)}\right)$. Starting from $u_{N1}^{(N-1, N)} = u_1,\, u_{N2}^{(N-1, N)} = u_2$, we can reach
\begin{equation}
\begin{aligned}
P_i^{(i-1,N)}(0) &= \frac{1 + (N-i-1)\left[u_1\left(1-q_1\right) + u_2\left(1-q_2\right) \right]}{1 + (N-i)\left[u_1\left(1-q_1\right) + u_2\left(1-q_2\right) \right]}\\
u_{i1}^{(i-1,N)} &= \frac{u_1}{1 + (N-i)\left[u_1\left(1-q_1\right) + u_2\left(1-q_2\right) \right]} \\
u_{i2}^{(i-1,N)} &= \frac{u_2}{1 + (N-i)\left[u_1\left(1-q_1\right) + u_2\left(1-q_2\right) \right]}.
\end{aligned}
\label{eq:u_i_i-1}
\end{equation}
Using Eq.~\ref{eq:u1_revised} and Eq.~\ref{eq:u2_revised}, we can do the inverse calculation of revising the probabilities and obtain
\begin{equation}
\begin{aligned}
u_{i1}^{(j,N)} &= \frac{u_1}{1 + (N-j-1)\left[u_1\left(1-q_1\right) + u_2\left(1-q_2\right) \right]} \\
u_{i2}^{(j,N)} &= \frac{u_2+(i-j-1)\left[u_1\left(1-q_1\right) + u_2\left(1-q_2\right) \right]}{1 + (N-j-1)\left[u_1\left(1-q_1\right) + u_2\left(1-q_2\right) \right]}.
\label{eq:inverse_revise}
\end{aligned}
\end{equation}
Letting $j=0$, this becomes the non-revised stimulus intervals (Eq. 58), which is
\begin{equation}
\begin{aligned}
u_{i1} &= \frac{u_1}{1 + (N-1)\left[u_1\left(1-q_1\right) + u_2\left(1-q_2\right) \right]} \\
u_{i2} &= \frac{u_2+(i-1)\left[u_1\left(1-q_1\right) + u_2\left(1-q_2\right) \right]}{1 + (N-1)\left[u_1\left(1-q_1\right) + u_2\left(1-q_2\right) \right]}.
\label{eq:u1_u2_solution_supp}
\end{aligned}
\end{equation}
With the definition of $p_{i1}$ and $p_{i2}$ (Eq. 61), we can write
\begin{equation}
\begin{aligned}
&p_{11} = p_{21} = ... = p_{N1} = \frac{u_1}{1 + (N-1)\left[u_1\left(1-q_1\right) + u_2\left(1-q_2\right) \right]} \overset{\text {def}}{=} p_1 \\
&p_{22} = ... = p_{N2} = \frac{-u_1q_1 + u_2\left(1-q_2\right) }{1 + (N-1)\left[u_1\left(1-q_1\right) + u_2\left(1-q_2\right) \right]} \overset{\text {def}}{=} p_2 \\
&p_{12} = \frac{u_2}{1 + (N-1)\left[u_1\left(1-q_1\right) + u_2\left(1-q_2\right) \right]} \overset{\text {def}}{=} \pe \\
&p_1 q_1 + p_2 = \pe\left(1-q_2 \right).
\end{aligned}
\label{eq:ternary_p_results_ON}
\end{equation}
This derives the optimal thresholds for a homogeneous population of ternary ON cells (Fig. 5B).

\subsection{Revised probability in ON-OFF mixtures}

\noindent For ON-OFF mixtures, the revised probability (Eq.~\ref{eq:u1_revised}, Eq.~\ref{eq:u2_revised}) needs to be adjusted. We derive the equivalency of Eq.~\ref{eq:u1_revised} and Eq.~\ref{eq:u2_revised} that we derived before for ON neurons.
\\
\\
Assuming we already know that neuron $1, 2, ..., j$ do not fire, for an OFF neuron $i$, if neuron $j+1$ is also an OFF neuron, and $j+1 < i$, we have
\begin{equation}
    u_{i1}^{(j+1,N)} =\int_{\theta_{i2}^{(j,N)}}^{\theta_{i1}^{(j,N)}} p(s|n_{j+1}= 0)\, \text ds =\int_{\theta_{i2}^{(j,N)}}^{\theta_{i1}^{(j,N)}} \frac{p(s)\,p(n_{j+1}=0|s)}{p(n_{j+1} = 0)}\, \text ds.
    \label{eq:u1_revised_OFF_1}
\end{equation}
Stimulus $s$ within the interval $\left[\theta_{i2}^{(j,N)}, \theta_{i1}^{(j,N)} \right]$ cannot trigger a nonzero firing rate of neuron $j+1$. For this reason, in the integral above, $p(n_{j+1}=0|s) = 1$. Since we already assumed that neuron $1, 2, ..., j$ do not fire, $p(n_{j+1} = 0) = P_{j+1}^{(j,N)}(0)$. Then we can rewrite Eq.~\ref{eq:u1_revised_OFF_1} as
\begin{equation}
    u_{i1}^{(j+1,N)} = \frac{1}{P_{j+1}^{(j,N)}(0)} \int_{\theta_{i2}^{(j,N)}}^{\theta_{i1}^{(j,N)}} p(s)\, \text ds = \frac{u_{i1}^{(j,N)}}{P_{j+1}^{(j,N)}(0)}.
    \label{eq:u1_revised_OFF}
\end{equation}
For $u_{i2}^{(j+1,N)}$, we have
\begin{equation}
    u_{i2}^{(j+1,N)} = \int_{-\infty}^{\theta_{i2}^{(j,N)}}p(s|n_{j+1}= 0) \, \text ds = \int_{-\infty}^{\theta_{i2}^{(j,N)}}\frac{p(s)\,p(n_{j+1}= 0|s)}{p(n_{j+1}= 0)}\, \text ds\\.
    \label{eq:u2_revised_OFF_1}
\end{equation}
Similar to Eq.~\ref{eq:u2_revised_2} before, we have $p(n_{j+1}= 0) = P_{j+1}^{(j,N)}(0)$ and $p(n_{j+1}= 0|s) = 1-p(n_{j+1}\neq 0|s)$. We can rewrite Eq.~\ref{eq:u2_revised_OFF_1} as
\begin{equation}
    u_{i2}^{(j+1,N)} = \frac{1}{P_{j+1}^{(j,N)}(0)}\int_{-\infty}^{\theta_{i2}^{(j,N)}} p(s) \left[ 1-p(n_{j+1} \neq 0|s) \right]\, \text ds.
    \label{eq:u2_revised_OFF_2}
\end{equation}
Because stimulus $s$ higher than $\theta_{i2}^{(j,N)}$ cannot lead to $n_{j+1} \neq 0$, we have
\begin{equation}
    \int_{-\infty}^{\theta_{i2}^{(j,N)}} p(s)\, p(n_{j+1} \neq 0|s)\, \text ds = P_{j+1}^{(j,N)}(0).
\end{equation}
Substituting back into Eq.~\ref{eq:u2_revised_OFF_2} gives rise to
\begin{equation}
    u_{i2}^{(j+1,N)} = \frac{1}{P_{j+1}^{(j,N)}(0)}\left[u_{i2}^{(j,N)}-\left(1-P_{j+1}^{(j,N)}(0)\right)\right].
    \label{eq:u2_revised_OFF}
\end{equation}
Here we find Eq.~\ref{eq:u1_revised_OFF} and Eq.~\ref{eq:u2_revised_OFF} are the same as Eq.~\ref{eq:u1_revised} and Eq.~\ref{eq:u2_revised} we derived before. However, if neuron $j+1$ is an ON neuron, $u_{i1}^{(j+1,N)}$ will remain the same as above while for $u_{i2}^{(j+1,N)}$, we have $p(n_{j+1}=0|s) = 1$ for $s \in (-\infty, \theta_{i2}^{(j,N)}]$. As a result, Eq.~\ref{eq:u2_revised_OFF} changes into
\begin{equation}
    u_{i2}^{(j+1,N)} = \frac{u_{i2}^{(j,N)}}{P_{j+1}^{(j,N)}(0)}.
    \label{eq:u2_revised_ON_OFF}
\end{equation}

\subsection{Optimal thresholds for a population of ON and OFF cells}

For ON-OFF mixtures, the mutual information can still be written recursively as Eq.~\ref{recursive_calculation},
% i.e.
%\begin{widetext}
%\begin{equation} 
%I=g(u_{11},u_{12})+P_1^{(0,N)}(0)\left[g\left(u_{21}^{(1,N)},u_{22}^{(1,N)}\right)+ ... +P_{N-1}^{(N-2,N)}%(0)g\left(u_{N1}^{(N-1,N)},u_{N2}^{(N-1,N)}\right)\right]
%\end{equation}
%\end{widetext}
hence,  the lemma (Eq.~\ref{lemma}) and Eq.\ref{eq:P0_i_i-1}-\ref{eq:u_i_i-1} also pertain to ON-OFF mixed populations. 
If neuron $i$ and neuron $j$ are both ON neurons ($j < i \leq m$) or both OFF neurons ($i > m$ and $j \geq m$), Eq.~\ref{eq:inverse_revise} becomes
\begin{equation}
\begin{aligned}
u_{i1}^{(j,N)} &= \frac{u_1}{1 + (N-j-1)\left[u_1\left(1-q_1\right) + u_2\left(1-q_2\right) \right]} \\
u_{i2}^{(j,N)} &= \frac{u_2+(i-j-1)\left[u_1\left(1-q_1\right) + u_2\left(1-q_2\right) \right]}{1 + (N-j-1)\left[u_1\left(1-q_1\right) + u_2\left(1-q_2\right) \right]}
\end{aligned}
\end{equation}
and if  neuron $i$ is an OFF neuron but neuron $j$ is an ON neuron ($i > m,\, j \leq m$)
\begin{equation}
\begin{aligned}
u_{i1}^{(j,N)} &= \frac{u_1}{1 + (N-j-1)\left[u_1\left(1-q_1\right) + u_2\left(1-q_2\right) \right]} \\
u_{i2}^{(j,N)} &= \frac{u_2+(i-m-1)\left[u_1\left(1-q_1\right) + u_2\left(1-q_2\right) \right]}{1 + (N-j-1)\left[u_1\left(1-q_1\right) + u_2\left(1-q_2\right) \right]}.
\end{aligned}
\end{equation}
Letting $j = 0$, we get the non-revised stimulus intervals for the ON neurons (Eq. 58)
\begin{equation}
\begin{aligned}
u_{i1} &= \frac{u_1}{1 + (N-1)\left[u_1\left(1-q_1\right) + u_2\left(1-q_2\right) \right]}, \\
u_{i2} &= \frac{u_2+(i-1)\left[u_1\left(1-q_1\right) + u_2\left(1-q_2\right) \right]}{1 + (N-1)\left[u_1\left(1-q_1\right) + u_2\left(1-q_2\right) \right]},\\
& \quad i \leq m 
\end{aligned}
\end{equation}
and for the OFF neurons (Eq. 59)
\begin{equation}
u_{i2} = \frac{u_2+(i-m-1)\left[u_1\left(1-q_1\right) + u_2\left(1-q_2\right) \right]}{1 + (N-1)\left[u_1\left(1-q_1\right) + u_2\left(1-q_2\right) \right]},\quad i > m. 
\label{eq:u1_u2_solution_ON_OFF}
\end{equation} 
The cumulative stimulus intervals then become
\begin{equation}
\begin{aligned}
&p_{11} = p_{21} = ... = p_{N1} = \frac{u_1}{1 + (N-1)\left[u_1\left(1-q_1\right) + u_2\left(1-q_2\right) \right]} \overset{\text {def}}{=} p_1 \\
&p_{22} = ... = p_{m2} = p_{m+2, 2} = ... = p_{N2} = \frac{-u_1q_1 + u_2\left(1-q_2\right) }{1 + (N-1)\left[u_1\left(1-q_1\right) + u_2\left(1-q_2\right) \right]} \overset{\text {def}}{=} p_2 \\
&p_{12} = p_{m+1, 2} = \frac{u_2}{1 + (N-1)\left[u_1\left(1-q_1\right) + u_2\left(1-q_2\right) \right]} \overset{\text {def}}{=} \pe \\
&p_1 q_1 + p_2 = \pe\left(1-q_2 \right).
\end{aligned}
\label{eq:ternary_p_results_ON_OFF_supp}
\end{equation}
This derives the optimal thresholds for a mixed population of ON and OFF cells (Eq. 62, Fig. 5B).

\subsection{Mean firing rate of an optimized population \label{sec:avg_fr_ternary}}

Similar to neuronal populations of binary neurons (Eq.~\ref{eq:avg_fr_ternary_supp}), here we can still calculate the mean firing rate ($\bar \nu$) in an optimized population of ternary neurons, based on the thresholds of Eq.~\ref{eq:ternary_p_results_ON_OFF_supp}. Similar to Eq.~\ref{eq:avg_fr_ternary_supp}, calculating the mean firing rate yields
\begin{equation}
\begin{aligned}
    \bar \nu &= \frac{1}{N}\Big[ \pe m + p_2 \sum_{i=1}^{m-1}(m-i) + p_1 \sum_{i=0}^{m-1}(m-i-1+f) + \\
    &\quad \quad \quad \pe (N-m) + p_2 \sum_{i=1}^{N-m-1}(N-m-i) + p_1\sum_{i=0}^{N-m-1}(N-m-i-1+f) \Big] \\
    &= \frac{1}{N}\Big[ \pe m + (p_1 + p_2) \frac{m(m-1)}{2} + p_1\, m - p_1\, m\, (1-f)+\\
    &\quad \quad \quad \pe (N-m) + (p_1 + p_2) \frac{(N-m)(N-m-1)}{2} + p_1\, (N-m) - p_1\, (N-m)\, (1-f) \Big] \\
    &= \pe + \frac{N-1}{2}\,(p_1 + p_2) + \frac{m}{N}\,(m-N)\, (p_1 + p_2) + p_1f.
    \label{eq:avg_fr_ternary_supp}
\end{aligned}
\end{equation}
Thus we have Eq. 83 in the main text.

\subsection{The maximal mutual information grows logarithmically with the number of neurons \label{sec:IN_I1_ternary}}

Next, we show that the maximal mutual information for a population of ternary neurons also grows logarithmically with the number of neurons $N$ as for binary neurons (Eq. 41). We first derive a universal relationship between the maximal mutual information $I_N^{\max}$ and the stimulus intervals $p_1, p_2$ for all mixtures of ON and OFF neurons. Then we apply Eq.~\ref{eq:ternary_p_results_ON} and Eq.~\ref{eq:ternary_p_results_ON_OFF_supp} to reach the conclusion.

As previously shown (Eq.~\ref{I_N_max}), the maximal mutual information is
\begin{equation}
I_N^{\max} = -\log P(\vec 0).
\end{equation}
This allows us to derive the relationship between the maximal mutual information and the stimulus interval $p$. 

For a homogeneous ON neuron population we define with $p_0 = \text{Prob}(\vec \nu = \vec 0)$ the `silent' interval that is lower than all the thresholds. We also denote $q_1 = L(0, fR)$ and $q_2 = L(0, R)$ for brevity and write:
\begin{equation}
\begin{aligned}
&P(\vec 0) \\
&= p_0 + (p_1q_1 + p_2q_2) + (p_1q_1 + p_2q_2)q_2 + ... \\
&+ (p_1q_1 + p_2q_2)q_2^{N-2} + (p_1q_1 + \pe q_2)q_2^{N-1} \\
&= p_0 + (p_1 q_1 + p_2 q_2)\frac{1-q_2^{N-1}}{1-q_2}+p_1q_1q_2^{N-1}+\pe q_2^N.
\label{eq:P_0_ternary}
\end{aligned}
\end{equation}
Since all the cumulative stimulus intervals sum up to 1, we get
\begin{equation}
p_0 = 1 - (N-1)(p_1+p_2) - p_1 - \pe.
\end{equation}
Also, from Eq.~\ref{eq:ternary_p_results_ON_OFF_supp} we know
\begin{equation}
    \pe = \frac{p_1 q_1 + p_2}{1- q_2}.
\end{equation}
Substituting these two equations back into Eq.~\ref{eq:P_0_ternary}, we can derive
\begin{equation}
P(\vec 0) = 1 - N(p_1 + p_2)
\end{equation}
which leads to
\begin{equation}
I_N^{\max} = -\log \left[1 - N(p_1 + p_2)\right].
\label{eq:I_max_p1_p2_supp}
\end{equation}
Since a mixed population has the same maximal information as a homogeneous population with the same $p_1$ and $p_2$ partitioning the stimulus intervals, this ensures that the maximal mutual information grows exponentially with the number of neurons as in Eq.~\ref{eq:I_max_p1_p2_supp} independent of the ON-OFF mixture.

Using the optimal values of $p_1$ and $p_2$ (Eq.~\ref{eq:ternary_p_results_ON} and Eq.~\ref{eq:ternary_p_results_ON_OFF_supp}), we have
\begin{equation}
I_N^{\max} = -\log \frac{1 - u_1\left(1-q_1\right) - u_2\left(1-q_2\right)}{1 + (N-1)\left[u_1\left(1-q_1\right) + u_2\left(1-q_2\right) \right]}.
\end{equation}
This allows us to write the  maximal mutual information of an $N$-neuron ternary population as a function of the the maximal mutual information of a single neuron population:
\begin{equation}
I_N^{\max} = \log \left[N(\exp({I_1^{\max}})-1)+1 \right],
\end{equation}
similar to the case with binary neurons (Eq. 41).

\setcounter{equation}{0}

\section{Population coding of neurons with any shapes of activation functions and any noise generation function \label{sec:M_ary_neurons}}
Here we derive the calculations with $(M+1)$-ary neurons. Because the calculations are similar to the last section, we omit some details and only show the framework of the calculations.
%%%%%

\subsection{Maximal mutual information}
Eq.~\ref{decompose_mutual_info} is still valid for $(M+1)$-ary neurons. The only difference from before is that every $\vec {u_i}$ or $\vec {u_i^{(m)}}$ is now an $M$-elements vector. Hence, we can still decompose $I_N$ into $N$ terms (Eq.~\ref{recursive_calculation}).
Similarly, we use $u_{ik}^{(j,N)}$ to denote the cumulative stimulus interval given the condition that none of the neurons $1, ..., j\,(j<i)$ fires (see Fig. 6A and Eq. 37)
and $P_i^{(j,N)}(0)=1-\sum_{k=1}^Mu_{ik}^{(j,N)}(1-L(0,f_{ik}R))$ to denote the probability that neuron $i$ does not
fire, when none of the neurons $1, ..., j \, (j<i)$ fires. 
Similar to Eq.~\ref{u1_u2_opt} and Eq.~\ref{recursive_calculation_equality}, denoting the optimal values using an asterisk, we derive several important relationships among
the revised probabilities,
\begin{equation}
u_{Nk}^{(N-1,N)}|_*=u_k^*, \ u_{ik}^{(j,N)}|_*=u_{i+1,k}^{(j+1,N+1)}|_*.
\end{equation}
If neuron $i$ and neuron $j+1$ are both ON neurons or both OFF neurons, the revised probabilities follow
\begin{equation}
u_{ik}^{(j+1,N)} =\frac{u_{ik}^{(j,N)}}{P_{j+1}^{(j,N)}(0)}, \quad k < M
\label{eq:uk_revised}
\end{equation}

\begin{equation}
u_{iM}^{(j+1,N)}=\frac{1}{P_{j+1}^{(j,N)}(0)}\left[u_{iM}^{(j,N)}-\left(1-P_{j+1}^{(j,N)}(0)\right)\right].
\label{eq:uM_revised}
\end{equation}
If neuron $i$ is an OFF neuron but neuron $j+1$ is an ON neuron, we have

\begin{equation}
u_{ik}^{(j+1,N)} =\frac{u_{ik}^{(j,N)}}{P_{j+1}^{(j,N)}(0)}, \quad k < M; \quad u_{iM}=\frac{u_{iM}^{(j,N)}}{P_{j+1}^{(j,N)}(0)}.
\end{equation}

Similar to Eq.~\ref{I1_ternary_neuron}, the mutual information of one single neuron can
be written as
\begin{equation}
I_1 = g(u_{11},u_{12},...,u_{1M}).
\end{equation}
We can verify that consistent with Eq.~\ref{g_u1_u2},
\begin{equation}
g=\sum_{k=1}^{M}u_k\frac{\partial g}{\partial u_k}-\log P(0),
\end{equation}
which indicates that
\begin{equation}
I_1^{\max}=-\log P(0)=-\log \left[1-\sum_{k=1}^Mu_k^*\left(1-L(0, f_k^*R)\right)\right].
\end{equation}
Same as Section~\ref{sec:ternary_neurons}.2 and Eq.~\ref{I_N_max}, we can use mathematical induction to generalize the mutual information from $N=2$ to arbitrary $N$. Here, we omit it for simplicity, and give the conclusion directly as
\begin{equation}
I_N^{\max}=-\sum_{j=1}^N\log P_j^{(j-1)}(0)|_* = -\log P(\vec 0).
\end{equation}

\subsection{Optimal thresholds}
Next, we seek to find the optimal thresholds by deriving the relationship among all $u_{ik}$. We start from a lemma that links two
adjacent neurons, $u_{ik}$ and $u_{i+1,k}$, as we did for ternary neurons (Eq.~\ref{lemma}). 

\textbf{Lemma:} For any $N$ neurons, when $I_N$ is maximized,
\begin{equation}
u_{ik}^{(i-1,N)}=P_i^{(i-1,N)}(0)u_{i+1,k}^{(i,N)},\quad f_{ik} = f_{i+1, k}.
\label{lemma_M_ary}
\end{equation}
Since the proof strictly follows those steps in Section~\ref{sec:ternary_neurons}.3, we do not repeat it here.
%%%%%%
\\
\\
We first discuss a homogeneous population with only ON neurons. All the variables in subsequent equations are optimized, so we omit the * symbol for clarity. For simplicity we also denote $q_k = L(0, f_kR)$. For a homogeneous ON population, similar to Eq.~\ref{eq:u_i_i-1}, we have
\begin{equation}
\begin{aligned}
u_{ik}^{(i-1,N)} &= \frac{u_k}{1 + (N-i)\sum_{k=1}^M u_k(1-q_k)}, \quad k<M \\
P_i^{(i-1,N)}(0) &= \frac{1 + (N-i-1)\sum_{k=1}^M u_k(1-q_k)}{1 + (N-i)\sum_{k=1}^M u_k(1-q_k)}.
\end{aligned}
\end{equation}
Using Eq.~\ref{eq:uk_revised} and Eq.~\ref{eq:uM_revised}, we can do the inverse calculation of revising the probabilities and have
\begin{equation}
\begin{aligned}
u_{ik}^{(j,N)} &= \frac{u_k}{1 + (N-j-1)\sum_{k=1}^M u_k(1-q_k)}, \quad k<M \\
u_{iM}^{(j,N)} &= \frac{u_M+(i-j-1)\sum_{k=1}^M u_k(1-q_k)}{1 + (N-j-1)\sum_{k=1}^M u_k(1-q_k)}.
\end{aligned}
\end{equation}
Letting $j=0$, these two equations turn to
\begin{equation}
\begin{aligned}
u_{ik}^{(j,N)} &= \frac{u_k}{1 + (N-1)\sum_{k=1}^M u_k(1-q_k)}, \quad k<M \\
u_{iM}^{(j,N)} &= \frac{u_M+(i-1)\sum_{k=1}^M u_k(1-q_k)}{1 + (N-j-1)\sum_{k=1}^M u_k(1-q_k)}.
\end{aligned}
\end{equation}
With the definition of $p_{ik}$ (Eq. 69), we can write
\begin{equation}
\begin{aligned}
&p_{1k} = p_{2k} = ... = p_{Nk} = \frac{u_k}{1 + (N-1)\sum_{k=1}^M u_k(1-q_k)} \overset{\text {def}}{=} p_k, \quad k < M \\
&p_{2M} = ... = p_{NM} = \frac{-\sum_{k=1}^M u_k\,q_k + u_M}{1 + (N-1)\sum_{k=1}^M u_k(1-q_k)} \overset{\text {def}}{=} p_M \\
&p_{1M} = \frac{u_M}{1 + (N-1)\sum_{k=1}^M u_k(1-q_k)} \overset{\text {def}}{=} p_{\text{edge}} \\
&\sum_{k=1}^{M-1} p_k\,q_k + p_M = p_{\text{edge}}\left(1-q_M \right)
\end{aligned}
\label{eq:M_ary_p_results_ON}
\end{equation}
which is summarized in Eq. 70 and Fig. 6B.

The equivalence to ON-OFF mixture has been discussed with ternary neurons in the main text. It still holds for $(M+1)$-ary neurons, so that we can derive that the optimal thresholds in a mixed population are
\begin{equation}
\begin{aligned}
&p_{1k} = p_{2k} = ... = p_{Nk} = \frac{u_k}{1 + (N-1)\sum_{k=1}^M u_k(1-q_k)} \overset{\text {def}}{=} p_k, \quad k < M \\
&p_{2M} = ... = p_{mM} = p_{m+2, M} = ... = p_{NM} = \frac{-\sum_{k=1}^M u_k\,q_k + u_M}{1 + (N-1)\sum_{k=1}^M u_k(1-q_k)} \overset{\text {def}}{=} p_M \\
&p_{1M} = p_{m+1, M} = \frac{u_M}{1 + (N-1)\sum_{k=1}^M u_k(1-q_k)} \overset{\text {def}}{=} p_{\text{edge}} \\
&\sum_{k=1}^{M-1}p_k\,q_k + p_M = p_{\text{edge}}\left(1-q_M\right),
\end{aligned}
\label{eq:M_ary_p_results_ON_OFF}
\end{equation}
which are also summarized in Eq. 70 and Fig. 6B.

\subsection{Mean firing rate of an optimized population \label{sec:avg_fr_M_ary}}
Similar to neuronal populations of ternary neurons (Eq.~\ref{eq:avg_fr_ternary_supp}), here we can calculate the mean firing rate $\bar \nu$ in an optimized population of $(M+1)$-ary neurons, based on the thresholds in Eq.~\ref{eq:M_ary_p_results_ON_OFF}. We can write
\begin{equation}
\begin{aligned}
    \bar \nu &= \frac{1}{N} \Big[\pe m + p_M \sum_{i=1}^{m-1} (m-i) + \sum_{k = 1}^{M-1} \sum_{i=0}^{m-1} p_k\, (m-i-1+f_k) + \\
    &\quad \quad \quad \pe (N-m) + p_M \sum_{i=1}^{N-m-1} (N-m-i) + \sum_{k = 1}^{M-1} \sum_{i=0}^{N-m-1} p_k\, (N-m-i-1+f_k) \Big]\\
    &= \frac{1}{N} \Big[\pe m + \frac{1}{2}\,m\,(m-1)\sum_{k = 1}^M p_k + m \sum_{k = 1}^{M-1} p_k\, f_k + \\
    &\quad \quad \quad \pe (N-m) + \frac{1}{2}\,(N-m)\,(N-m-1)\sum_{k = 1}^M p_k + (N-m) \sum_{k = 1}^{M-1} p_k\, f_k \Big]\\
    &= \pe + \frac{1}{2}(N-1)\sum_{k = 1}^M p_k + \frac{m}{N}(m-N)\sum_{k = 1}^M p_k + \sum_{k = 1}^{M-1} p_k\, f_k,
\end{aligned}
\end{equation}
which gives rise to Eq. 84 in the main text.

\subsection{The maximal mutual information grows logarithmically with the number of neurons}

Similar to ternary neurons, here we show that the maximal mutual information of a population of $M$-ary neurons also grows logarithmically with the number of neurons $N$ (Eq. 41). Same as before, we first derive a universal relationship between the maximal mutual information $I_N^{\max}$ and the stimulus intervals $p_k\,(k = 1, ..., M)$ for all mixtures of ON and OFF neurons. Then we apply Eq.~\ref{eq:M_ary_p_results_ON} and Eq.~\ref{eq:M_ary_p_results_ON_OFF}
to reach the conclusion (Eq. 41).
Similarly, we start from
\begin{equation}
I_N^{\max} = -\log P(\vec 0)
\end{equation}
to derive the relationship between the maximal mutual information and the stimulus interval $p$. Consider a homogeneous ON neuron population, define $p_0 = \text{Prob}(\vec \nu = 0)$, i.e. the `silent' interval that is lower than all the thresholds. Also denote $q_k = L(0, f_kR)$ and $q_M = L(0, R)$ for clarity, we have
\begin{equation}
\begin{aligned}
P(\vec 0) 
&=p_0 + \sum_{k=1}^M p_k\,q_k + \left(\sum_{k=1}^M p_k\,q_k\right)q_M + ... + \left(\sum_{k=1}^M p_k\,q_k\right)q_M^{N-2} + \left(\sum_{k=1}^{M-1} p_k\,q_k + \pe \,q_M\right)q_M^{N-1} \\
&= p_0 + \left(\sum_{k=1}^M p_k\,q_k\right)\frac{1-q_M^{N-1}}{1-q_M}+\left(\sum_{k=1}^{M-1} p_k\,q_k\right)q_M^{N-1}+\pe q_M^N.
\end{aligned}
\end{equation}
Note that
\begin{equation}
\begin{aligned}
& p_0 = 1 - (N-1)\sum_{k=1}^M p_k - \sum_{k=1}^{M-1} p_k - \pe \\
& \pe = \frac{\sum_{k=1}^{M-1} p_k + p_M}{1- q_M}
\end{aligned}
\end{equation}
we can derive
\begin{equation}
P(\vec 0) = 1 - N\sum_{k=1}^M p_k
\end{equation}
which leads to
\begin{equation}
I_N^{\max} = -\log \left[1 - N\sum_{k=1}^M p_k\right].
\label{eq:I_max_p1_p2_M_ary}
\end{equation}
The ON-OFF mixed population has been shown to have the same maximal information as a homogeneous population and also the same $p_k$. Hence Eq.~\ref{eq:I_max_p1_p2_M_ary} is valid for all possible ON-OFF neuron populations. Using the optimal values of $p_k$ (Eq.~\ref{eq:M_ary_p_results_ON} and Eq.~\ref{eq:M_ary_p_results_ON_OFF}), we have
\begin{equation}
I_N^{\max} = -\log \frac{1 - \sum_{k=1}^M u_k(1-q_k)}{1 + (N-1)\sum_{k=1}^M u_k(1-q_k)}.
\end{equation}
This allows us to relate the maximal mutual information of neuron populations of different sizes. Specifically,
\begin{equation}
I_N^{\max} = \log \left[N(\exp({I_1^{\max}})-1)+1 \right].
\end{equation}

\setcounter{equation}{0}

\section{Population coding of neurons with activation functions of any shape and heterogenous maximal firing rates \label{sec:any_neurons_hetero}}

Here we provide the calculations of the optimal thresholds of neurons with any shape of activation functions and heterogeneous maximal firing rates. Similar to before, here we still start from ternary neurons, which is the simplest case beyond binary neurons. It is straightforward that Eq.~\ref{recursive_calculation} and Eq.~\ref{I_N_max} still hold, i.e., for heterogeneous maximal firing rates across the cells, we still have
\begin{equation}
I_N = g(\vec {u_1}) + P_1(0)\left \{ g\left(\vec {u_2^{(1)}}\right)+ P_2^{(1, N)}(0) \left[g\left(\vec {u_3^{(2)}}\right) + ...+P_{N-1}^{(N-2,N)}(0)g\left(\vec {u_N^{(N-1)}}\right)\right]\right \}.
\end{equation}
and
\begin{equation}
I_N^{\max}=-\sum_{j=1}^{N}\log P_j^{(j-1)}(0) = - \text{log}P(\vec 0).
\end{equation}
In addition, the equations deriving revised probabilities (Eq.~\ref{eq:u1_revised_OFF_1}, Eq.~\ref{eq:u2_revised_OFF_1}, and Eq.~\ref{eq:u2_revised_OFF_2}) are still valid. However, the lemma (Eq.~\ref{lemma}) becomes invalid due to the heterogeneous maximal firing rates across the cells. Therefore, we need to update the lemma.

\subsection{Updated lemma that connects a neuron in a population to a single neuron with the same maximal firing rate}

\textbf{Lemma 2:} For any $N$ (ternary) neurons, when $I_N$ is optimized, 
\begin{equation}
    \begin{aligned}
        u_{i1}^{(i-1, N)} &= \frac{\prod_{j=i}^N P_j^{(j-1)}(0, R_j)}{P^*(0, R_i)}\,u_1^*(R_i) \\
        u_{i2}^{(i-1, N)} &= \frac{\prod_{j=i}^N P_j^{(j-1)}(0, R_j)}{P^*(0, R_i)}\,u_2^*(R_i) \\
        f_i &= f^*(R_i)
    \end{aligned}
    \label{lemma2}
\end{equation}
where $R_i = \nu_{\max,i}T$ is the maximal expected spike count of neuron $i$ in the population of $N$ neurons. $u_1^*(R_i)$, $u_2^*(R_i)$, and $f^*(R_i)$ are defined in the main text as the optimal thresholds and intermediate firing rate of a single neuron with the same maximal firing rate as $\nu_{\max,i}$.
\\
\\
\textbf{Remark:} According to the definition of $u_1^*(R_i)$, $u_2^*(R_i)$, and $f^*(R_i)$, using Eqs.~\ref{eq:derivative_g_u1}-\ref{eq:derivative_g_f}, we can obtain
\begin{equation}
\begin{aligned}
\frac{\partial g\left(u_1^*(R_i),u_2^*(R_i)\right)}{\partial u_1^*(R_i)} &= \frac{\partial P^*(0, R_i)}{\partial u_1^*(R_i)}\left[-1-\log P^*(0, R_i)\right]+L(0, f^*(R_i)R_i)\log L(0, f^*(R_i)R_i) \\
&+\sum_{n=1}^{+\infty}L(n,f^*(R_i)R_i) \left(\log \frac{L(n,f^*(R_i)R_i)}{u_1^*(R_i)L(n,f^*(R_i)R_i)+u_2^*(R_i)L(n,R_i)}-1\right) = 0,
\end{aligned}
\label{eq:derivative_g_u1_hetero}
\end{equation}
\begin{equation}
\begin{aligned}
\frac{\partial g\left(u_1^*(R_i),u_2^*(R_i)\right)}{\partial u_2^*(R_i)} &= \frac{\partial P^*(0,R_i)}{\partial u_2^*(R_i)}\left[-1-\log P^*(0,R_i)\right]+L(0, R_i)\log L(0, R_i)\\
&+\sum_{n=1}^{+\infty}L(n,R_i)\left(\log \frac{L(n,R_i)}{u_1^*(R_i)L(n,f^*(R_i)R_i)+u_2^*(R_i)L(n,R_i)}-1\right) = 0,
\end{aligned}
\label{eq:derivative_g_u2_hetero}
\end{equation}
and
\begin{equation}
\begin{aligned}
\frac{\partial g\left(u_1^*(R_i),u_2^*(R_i)\right)}{\partial f^*(R_i)}=&-u_1^*(R_i)\frac{\partial L(0,f^*(R_i)R_i)}{\partial f^*(R_i)}\log P^*(0,R_i)+u_1^*(R_i)\sum_{n=0}^{+\infty}\log L(n,f^*(R_i)R_i)\frac{\partial L(n,f^*(R_i)R_i)}{\partial f^*(R_i)}\\
&-\sum_{n=1}^{+\infty}\log [u_1^*(R_i)L(n,f^*(R_i)R_i)+u_2^*(R_i)L(n,R_i)]u_1^*(R_i)\frac{\partial L(n,f^*(R_i)R_i)}{\partial f^*(R_i)} = 0.
\label{eq:derivative_g_f_hetero}
\end{aligned}
\end{equation}
Denote 
\begin{equation}
    \begin{aligned}
        \hat{u_{i1}} &= \frac{\prod_{j=i}^N P_j^{(j-1)}(0, R_j)}{P^*(0, R_i)}\,u_1^*(R_i), \\
        \hat{u_{i2}} &= \frac{\prod_{j=i}^N P_j^{(j-1)}(0, R_j)}{P^*(0, R_i)}\,u_2^*(R_i), \\
        \hat{f_i} &= f^*(R_i),
    \end{aligned}
    \label{eq:define_u1_u2_f_star_hetero}
\end{equation}
we need to prove 
\begin{equation}
\frac{\partial I_N}{\partial u_{i1}^{(i-1,N)}}|_{\hat{u}_{i1},\hat{u}_{i2},\hat{f}_i}=0,\ \ \ \frac{\partial I_N}{\partial u_{i2}^{(i-1,N)}}|_{\hat{u}_{i1},\hat{u}_{i2},\hat{f}_i}=0,\ \ \ \frac{\partial I_N}{\partial f_i}|_{\hat{u}_{i1},\hat{u}_{i2},\hat{f}_i}=0
\end{equation}
\noindent i.e.~the combination of $\hat{u}_{i1},\hat{u}_{i2}$, and $\hat{f}_i$ is optimal.
\\
\\
$\textbf{Proof:}$
\begin{equation}
\frac{\partial I_N}{\partial u_{i,1}^{(i-1,N)}}|_{\hat{u}_{i1},\hat{u}_{i2},\hat{f}_i} = P_{i-1}(0) \left \{ \frac{\partial g\left(u_{i1}^{(i-1,N)},u_{i2}^{(i-1,N)}\right)}{\partial u_{i1}^{(i-1,N)}}|_{\hat{u}_{i1},\hat{u}_{i2},\hat{f}_i}+\left[1-L(0,f_iR_i)\right]\sum_{j=i+1}^N\log P_j^{(j-1,N)}(0)  \right \}.
\label{derivative_u1_hetero}
\end{equation}
\begin{equation}
\frac{\partial I_N}{\partial u_{i2}^{(i-1,N)}}|_{\hat{u}_{i1}, \hat{u}_{i2}, \hat{f}_i}=P_{i-1}(0) \left \{\frac{\partial g\left(u_{i1}^{(i-1,N)},u_{i2}^{(i-1,N)}\right)}{\partial u_{i2}^{(i-1,N)}}|_{\hat{u}_{i1}, \hat{u}_{i2}, \hat{f}_i}+[1-L(0,R_i)]\sum_{j=i+1}^N\log P_j^{(j-1,N)}(0) \right \}.
\end{equation}
\begin{equation}
\frac{\partial I_N}{\partial f_i}|_{\hat{u}_{i1}, \hat{u}_{i2}, \hat{f}_i}=P_{i-1}(0) \left \{ \frac{\partial g\left(u_{i1}^{(i-1,N)},u_{i2}^{(i-1,N)}\right)}{\partial f_i}|_{\hat{u}_{i1}, \hat{u}_{i2}, \hat{f}_i}-u_{i1}^{(i-1,N)}\frac{\partial L(0, f_iR_i)}{\partial f_i}\sum_{j=i+1}^N\log P_j^{(j-1,N)}(0)  \right \}.
\end{equation}
According to Eq.~\ref{eq:derivative_g_u1_hetero},
\begin{equation}
\begin{aligned}
&\frac{1}{P_{i-1}(0)}\frac{\partial I_N}{\partial u_{i1}^{(i-1,N)}}|_{\hat{u}_{i1}, \hat{u}_{i2}, \hat{f}_i} - \frac{\partial g\left(u_1^*(R_i),u_2^*(R_i)\right)}{\partial u_1^*(R_i)} \\
&\xlongequal{Eq.~\ref{eq:derivative_g_u1_hetero}} \frac{\partial g\left(u_{i1}^{(i-1,N)},u_{i2}^{(i-1,N)}\right)}{\partial u_{i1}^{(i-1,N)}}|_{\hat{u}_{i1}, \hat{u}_{i2}, \hat{f}_i}- \frac{\partial g\left(u_1^*(R_i),u_2^*(R_i)\right)}{\partial u_1^*(R_i)}+\left[1-L(0,\hat{f}_iR_i)\right]\sum_{j=i+1}^N\log P_j^{(j-1,N)}(0)\\
&\xlongequal{Eq.~\ref{eq:derivative_g_u1}}[1-L(0,\hat{f}_iR_i)]\left(\log P_i^{(i-1,N)}(0)-\log P_{i+1}^{(i,N)}(0)\right)-\sum_{n=1}^{+\infty}L(n,\hat{f}_iR_i)\log \left[\hat{u}_{i1}L(n,\hat{f}_iR_i))+\hat{u}_{i2}L(n,R_i))\right] \\
&\quad \quad
+\sum_{n=1}^{+\infty}L(n,f^*(R_i)R_i)\log \left[u_1^*(R_i)L(n,f^*(R_i)R_i)+u_2^*(R_i)L(n,R_i)\right]\\&
\quad \quad +\left[1-L(0,f^*(R_i)R_i)\right]\sum_{j=i+1}^N\log P_j^{(j-1,N)}(0) \\
&=[1-L(0,\hat{f}_iR_i)]\left(\log P_i^{(i-1,N)}(0)-\log P^*(0, R_i)\right)-\sum_{n=1}^{+\infty}L(n,\hat{f}_iR_i)\log \frac{\prod_{j=i}^N P_j^{(j-1)}(0, R_j)}{P^*(0, R_i)}\\
&\quad +\left[1-L(0,\hat{f}_iR_i)\right]\sum_{j=i+1}^N\log P_j^{(j-1,N)}(0).
\end{aligned}
\end{equation}
Because
\begin{equation}
L(0,\hat{f}_iR_i)+\sum_{n=1}^{+\infty}L(n,\hat{f}_iR_i)=1\ \ \ \text{(the sum of probabilities),}
\end{equation}
\begin{equation}
\frac{1}{P_{i-1}(0)}\frac{\partial I_N}{\partial u_{i1}^{(i-1,N)}}|_{\hat{u}_{i1}, \hat{u}_{i2}, \hat{f}_i} - \frac{\partial g\left(u_1^*(R_i),u_2^*(R_i)\right)}{\partial u_1^*(R_i)} =0,\ \ \ \frac{\partial I_N}{\partial u_{i1}^{(i-1,N)}}|_{\hat{u}_{i1},\hat{u}_{i2},\hat{f}_i}=0.
\label{eq:derivative_u1_0_hetero}
\end{equation}
Similarly, we have
\begin{equation}
\frac{\partial I_N}{\partial u_{i2}^{(i-1,N)}}|_{\hat{u}_{i1},\hat{u}_{i2},\hat{f}_i}=0.
\label{eq:derivative_u2_0_hetero}
\end{equation}
Also,
\begin{equation}
\begin{aligned}
&\quad \quad\frac{1}{\hat{u}_{i1}P_{i-1}(0)}\frac{\partial I_N}{\partial f_i}|_{\hat{u}_{i1}, \hat{u}_{i2}, \hat{f}_i}-\frac{1}{u_1^*(R_i)}\frac{\partial g\left(u_1^*(R_i),u_2^*(R_i)\right)}{\partial f^*} \\
&\xlongequal{Eq.~\ref{eq:derivative_g_f_hetero}}-\frac{\partial L(0,\hat{f}_iR_i)}{\partial \hat{f}_i}\log P_i(0)+\sum_{n=0}^{+\infty}\log L(n,\hat{f}_iR_i)\frac{\partial L(n,\hat{f}_iR_i)}{\partial \hat{f}_i}\\
&\quad-\sum_{n=1}^{+\infty}\log \left[\hat{u}_{i1}L(n,\hat{f}_iR_i)+\hat{u}_{i2}L(n,R_i)\right]\frac{\partial L(n,\hat{f}_iR_i)}{\partial \hat{f}_i} - \frac{\partial L(0,\hat{f}_iR_i)}{\partial \hat{f}_i}\sum_{j=i+1}^N\log P_j^{(j-1)}(0) \\
&\, \, \, \, \, \, \, \, +\frac{\partial L(0,f^*(R_i)R_i)}{\partial f^*(R_i)}\log P^*(0,R_i)-\sum_{n=0}^{+\infty}\log L(n,f^*(R_i)R_i)\frac{\partial L(n,f^*(R_i)R_i)}{\partial f^*(R_i)} \\
&\, \, \, \, \, \, \, \, +\sum_{n=1}^{+\infty}\log \left[u_1^*(R_i)L(n,f^*(R_i)R_i)+u_2^*(R_i)L(n,R_i)\right]\frac{\partial L(n,f^*(R_i)R_i)}{\partial f^*(R_i)}.
\end{aligned}
\end{equation}
According to the definition of $\hat{f}_i, \hat{u}_{i1}$ and $\hat{u}_{i2}$ (Eq.~\ref{eq:define_u1_u2_f_star}), we have
\begin{equation}
\begin{aligned}
&\quad \frac{1}{\hat{u}_{i1}P_{i-1}(0)}\frac{\partial I_N}{\partial f_i}|_{\hat{u}_{i1}, \hat{u}_{i2}, \hat{f}_i}-\frac{1}{u_1^*(R_i)}\frac{\partial g\left(u_1^*(R_i),u_2^*(R_i)\right)}{\partial f^*}\\
&=\frac{\partial L(0,\hat{f}_iR_i)}{\partial \hat{f}_i}\left(\log P^*(0,R_i) - \log P_i^{(i-1,N)}(0) \right)-\log\frac{\prod_{j=i}^N P_j^{(j-1)}(0, R_j)}{P^*(0, R_i)}\sum_{n=1}^{+\infty}\frac{\partial L(n,\hat{f}_iR_i)}{\partial \hat{f}_i}\\
&=\sum_{n=0}^{+\infty}\frac{\partial L(n,\hat{f}_iR_i)}{\partial \hat{f}_i}\left(\log P^*(0,R_i) - \log P_i^{(i-1,N)}(0) \right) \\
&=\left(\log P^*(0,R_i) - \log P_i^{(i-1,N)}(0) \right)\frac{\partial}{\partial \hat{f}_i}\left[\sum_{n=0}^{+\infty}L(n,\hat{f}_iR_i)\right].
\end{aligned}
\end{equation}
Since $\sum_{n=0}^{+\infty}L(n,\hat{f}_iR_i)=1$ is a constant, $\frac{\partial}{\partial \hat{f}_i}\left[\sum_{n=0}^{+\infty}L(n,\hat{f}_iR_i)\right]=0$, then
\begin{equation}
\frac{1}{\hat{u}_{i1}P_{i-1}(0)}\frac{\partial I_N}{\partial f_i}|_{\hat{u}_{i1}, \hat{u}_{i2}, \hat{f}_i}-\frac{1}{u_1^*(R_i)}\frac{\partial g\left(u_1^*(R_i),u_2^*(R_i)\right)}{\partial f^*} = 0
\end{equation}
which leads to
\begin{equation}
\frac{\partial I_N}{\partial f_i}|_{\hat{u}_{i1},\hat{u}_{i2},\hat{f}_i}=0.
\end{equation}
Combined with Eq.~\ref{eq:derivative_u1_0_hetero} and Eq.~\ref{eq:derivative_u2_0_hetero}, the lemma has been proved. $\blacksquare$

\begin{figure}
    \centering
    \includegraphics[width=\textwidth]{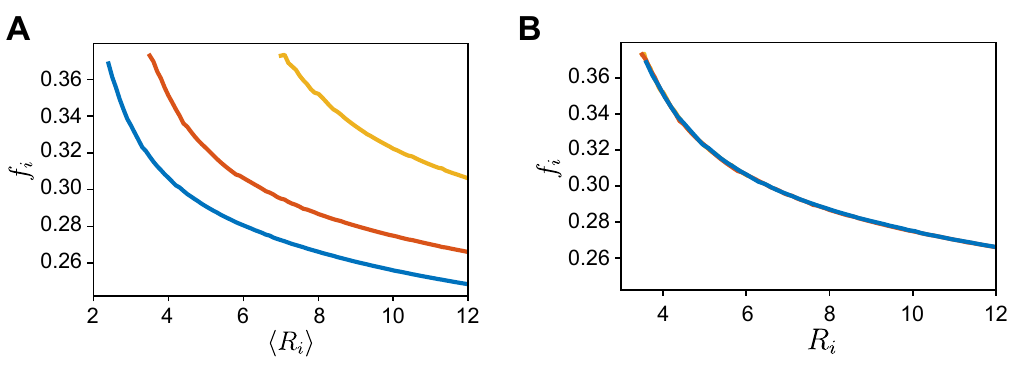}
    \caption{\textbf{Optimal intermediate firing levels in a population with heterogeneous maximal firing rate across the cells.} The population consists of $N=3$ ternary ON neurons, with $R_1:R_2:R_3 = 1:2:3$. 
    \textbf{A.} Optimal intermediate firing levels $f_i$ as a function of $\langle R_i \rangle = (R_1+R_2+R_3)/3$. \textbf{B.} Optimal intermediate firing levels $f_i$ as a function of $R_i$ of individual neurons.}
    \label{fig:any_neurons_hetero}
\end{figure}

\begin{figure}
    \centering
    \includegraphics[width=\textwidth]{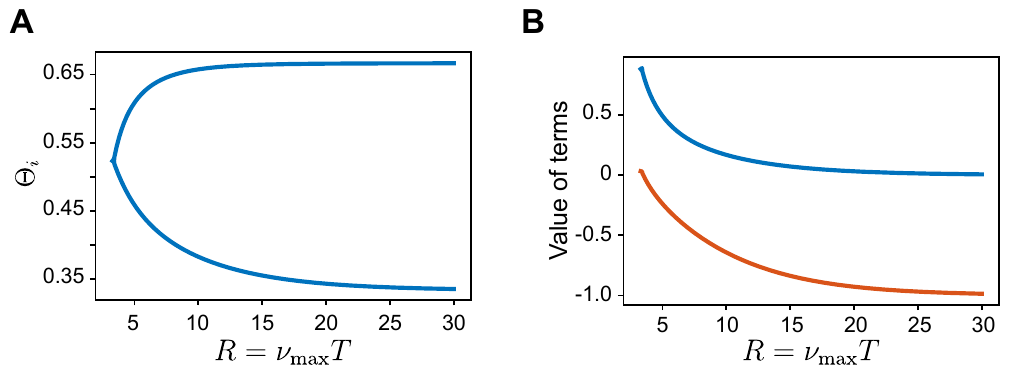}
    \caption{\textbf{Numerical calculations for a population of ternary neurons with heterogeneous maximal firing rates.}
    \textbf{A.} Optimal thresholds $(u_1, u_2)$ of a single ternary neuron. \textbf{B.} The values of two terms in Eq. 78. Blue: $\left(u_2^*(R)-u_1^*(R)\right)/P^*(0, R)$. Orange: $\left[u_2^*(R)\,q_2^*(R)-u_1^*(R)\left(1-q_1^*(R)\right)\right]/P^*(0, R)$.
    \label{fig:numerics_hetero}}
\end{figure}

\subsection{Optimal thresholds for a population with heterogeneous maximal firing rates}

Using Eq.~\ref{eq:u1_revised_OFF}, Eq.~\ref{eq:u2_revised_OFF}, and Eq.~\ref{eq:u2_revised_ON_OFF}, we can do the inverse calculation of revising the probabilities and obtain
\begin{equation}
    \begin{aligned}
        u_{i1}^{(j,N)} &= \frac{\prod_{k=j+1}^N P_k^{(k-1)}(0, R_k)}{P^*(0, R_i)}\,u_1^*(R_i), \\
        u_{i2}^{(j,N)} &= \frac{\prod_{k=j+1}^N P_k^{(k-1)}(0, R_k)}{P^*(0, R_i)}\,u_2^*(R_i)+1-\prod_{k=j+1}^{i-1}P_k^{(k-1)}(0) \\
    \end{aligned}
    \label{eq:u_i_j_hetero}
\end{equation}
if neuron $i$ and neuron $j$ are both ON neurons ($j < i \leq m$) or both OFF neurons ($i > m$ and $j \geq m$), and
\begin{equation}
    \begin{aligned}
        u_{i1}^{(j,N)} &= \frac{\prod_{k=j+1}^N P_k^{(k-1)}(0, R_k)}{P^*(0, R_i)}\,u_1^*(R_i), \\
        u_{i2}^{(j,N)} &= \frac{\prod_{k=j+1}^N P_k^{(k-1)}(0, R_k)}{P^*(0, R_i)}\,u_2^*(R_i)+\prod_{k=j+1}^{m}P_k^{(k-1)}(0)-\prod_{k=j+1}^{i-1}P_k^{(k-1)}(0) \\
    \end{aligned}
    \label{eq:u_i_j_hetero_ON_OFF}
\end{equation}
if neuron $i$ is an OFF neuron but neuron $j$ is an ON neuron ($i > m,\, j \leq m$).

Letting $j=0$, this becomes the non-revised stimulus intervals (Eq. 74), which is
\begin{equation}
    \begin{aligned}
        u_{i1} &= \frac{u_1^*(R_i)}{P^*(0, R_i)}\prod_{k=1}^N P_k^{(k-1)}(0, R_k) \, \\
        u_{i2} &= \frac{u_2^*(R_i)}{P^*(0, R_i)}\prod_{k=1}^N P_k^{(k-1)}(0, R_k) + 1 - \prod_{k=1}^{i-1} P_k^{(k-1)}(0, R_k) \\
        &\quad \quad \quad \quad\text{(ON neurons, } i = 1, ..., m \text{)}\\
        u_{i2} &= \frac{u_2^*(R_i)}{P^*(0, R_i)}\prod_{k=1}^N P_k^{(k-1)}(0, R_k) + \prod_{k=1}^m P_k^{(k-1)}(0, R_k) - \prod_{k=1}^{i-1} P_k^{(k-1)}(0, R_k) \\
        &\quad \quad \quad \quad\text{(OFF neurons, } i = m+1, ..., N \text{)}\\
        f_i &= f^*(R_i).
    \end{aligned}
    \label{eq:ternary_hetero_u_pre}
\end{equation}
Note that with
\begin{equation}
    I_N = -\log P(\vec 0) = \prod_{j=1}^N P_j^{(j-1)}(0, R_j)
\end{equation}
and
\begin{equation}
\begin{aligned}
    P_i^{(i-1)}(0,R_i) &= 1-u_{i1}^{(i-1)}\left(1-q_1(R_i)\right)-u_{i2}^{(i-1)}\left(1-q_2(R_i)\right) \\
    &= 1 - \left[\frac{u_1^*(R_i)}{P(0,R_i)}\left(1-q_1(R_i)\right) + \frac{u_2^*(R_i)}{P(0,R_i)}\left(1-q_2(R_i)\right)\right]\prod_{j=i}^N P_j^{(j-1)}(0, R_j),
\end{aligned}
\end{equation}
we can obtain
\begin{equation}
    P_i^{(i-1)}(0,R_i) = \left[1+\frac{u_1^*(R_i)}{P(0,R_i)}\left(1-q_1(R_i)\right) + \frac{u_2^*(R_i)}{P(0,R_i)}\left(1-q_2(R_i)\right)\right]^{-1},
\end{equation}
and then
\begin{equation}
    \left(\prod_{j=i}^N P_j^{(j-1)}(0, R_j)\right)^{-1} = \sum_{j=i}^N \frac{1}{P(0, R_j} - (N-i).
    \label{eq:prod_sum_relation}
\end{equation}
With Eq.~\ref{eq:prod_sum_relation}, we can rewrite the optimal thresholds (Eq.~\ref{eq:ternary_hetero_u_pre}) as
\begin{equation}
    \begin{aligned}
        u_{i1} &= \frac{u_1^*(R_i)}{P^*(0, R_i)}e^{-I_N} \, \\
        u_{i2} &= \left[\frac{u_2^*(R_i)}{P^*(0, R_i)}-\sum_{j=i}^N \frac{1}{P^*(0, R_j)}+(N-i)\right]e^{-I_N} + 1 \quad \quad \text{(ON neurons, } i = 1, ..., m \text{)}\\
        u_{i2} &= \left[\frac{u_2^*(R_i)}{P^*(0, R_i)}-\sum_{j=i}^N \frac{1}{P^*(0, R_j)}-\sum_{j=1}^m \frac{1}{P^*(0, R_j)}+(N-i+m)\right]e^{-I_N} + 1 \quad\text{(OFF neurons, } i = m+1, ..., N \text{)}\\
        f_i &= f^*(R_i).
    \end{aligned}
    \label{eq:ternary_hetero_u}
\end{equation}
According to the definition of $p_{i1}$ and $p_{i2}$, we can write
\begin{equation}
    \begin{aligned}
        p_{i1} &= \frac{u_1^*(R_i)}{P^*(0, R_i)}e^{-I_N} \, \\
        p_{i2} &= \left[\frac{u_2^*(R_i)-u_1^*(R_i)}{P^*(0, R_i)}-\frac{u_2^*(R_{i-1})\,q_2^*(R_{i-1})-u_1^*(R_{i-1})\left(1-q_1^*(R_{i-1})\right)}{P^*(0, R_{i-1})}\right]e^{-I_N}\quad(i = 2, ..., m, m+2, ..., N)\\
        p_{12} &= \left[\frac{u_2^*(R_1)}{P^*(0, R_1)}-\sum_{j=1}^N \frac{1}{P^*(0, R_j)}+(N-1)\right]e^{-I_N} + 1\\
        p_{m+1,2} &= \left[\frac{u_2^*(R_{m+1})}{P^*(0, R_{m+1})}-\sum_{j=1}^N \frac{1}{P^*(0, R_j)}+(N-1)\right]e^{-I_N} + 1\\
        f_i &= f^*(R_i).
    \end{aligned}
    \label{eq:optimal_p_ternary_hetero_supp}
\end{equation}

This derives the optimal thresholds for a mixed population of ON and OFF ternary cells with heterogeneous maximal firing rates (Eq. 62, Fig. 5B). 

Here, $f_i = f^*(R_i)$ indicates that the optimal intermediate firing level of a neuron in a population only depends on the maximal firing rate of that specific neuron (Fig.~\ref{fig:any_neurons_hetero}). Consequently, the optimal activation functions of different neurons may consist of different numbers of steps depending on the maximal firing rate constraint of those neurons (Fig. 7).

We performed numerical calculations for Poisson noise to better understand this optimal thresholds structure (Fig.~\ref{fig:numerics_hetero}). We found that $u_1^*(R)$ increases with $R$ (Fig.~\ref{fig:numerics_hetero}A). Since $P^*(0, R)$ decreases with $R$, within a population, $p_{i1}$ is larger for neurons with higher $R_i$. To analyze how $p_{i2}$ depends on $R_i$ and $R_{i-1}$, we plotted the two terms in Eq.~\ref{eq:optimal_p_ternary_hetero_supp} (second line), as functions of $R$ (Fig.~\ref{fig:numerics_hetero}B). We found that both of them decrease with $R$ (Fig.~\ref{fig:numerics_hetero}B), which indicates that within a neuron population, $p_{i2}$ is smaller for higher $R_i$ and lower $R_{i-1}$.

Again, using Eq.~\ref{eq:prod_sum_relation}, the maximal mutual information can be rewritten as
\begin{equation}
    I_N = \log\left[\sum_{j=i}^N \frac{1}{P^*(0, R_j)}-(N-1)\right],
\end{equation}

\subsection{$(M+1)$-ary neurons}
Above we derived the maximal mutual information solution for ternary neurons with heterogeneous maximal firing rates. Generalizing this from ternary neurons to activation functions with any any shape is straightforward (see Section~\ref{sec:M_ary_neurons}). For $(M+1)$-ary neurons, the optimal thresholds are
\begin{equation}
    \begin{aligned}
        u_{ik} &= \frac{u_k^*(R_i)}{P^*(0, R_i)}e^{-I_N} \quad k = 1, ..., M-1 \\
        u_{iM} &= \left[\frac{u_M^*(R_i)}{P^*(0, R_i)}-\sum_{j=i}^N \frac{1}{P^*(0, R_j)}+(N-i)\right]e^{-I_N} + 1  \quad\text{(ON neurons, } i = 1, ..., m \text{)}\\
        u_{iM} &= \Bigg[\frac{u_M^*(R_i)}{P^*(0, R_i)}-\sum_{j=i}^N \frac{1}{P^*(0, R_j)}-\sum_{j=1}^m \frac{1}{P^*(0, R_j)}+(N-i+m)\Bigg]e^{-I_N} + 1\quad\text{(OFF neurons, } i = m+1, ..., N \text{)}\\
        f_{ik} &= f_k^*(R_i). \quad k = 1, ..., M-1,
    \end{aligned}
\end{equation}
the maximal mutual information is:
\begin{equation}
    I_N = \log\left[\sum_{j=i}^N \frac{1}{P^*(0, R_j)}-(N-1)\right],
\end{equation}
and the optimal stimulus intervals $p$ are given by
\begin{equation}
    \begin{aligned}
        p_{ik} &= \frac{u_1^*(R_i)}{P^*(0, R_i)}e^{-I_N} \quad k = 1, ..., M-1 \\
        p_{iM} &= \left[\frac{u_M^*(R_i)-\sum_{k=1}^{M-1}u_k^*(R_i)}{P^*(0, R_i)}-\frac{u_M^*(R_{i-1})\,q_M^*(R_{i-1})-\sum_{k=1}^{M-1}u_k^*(R_{i-1})\left(1-q_k^*(R_{i-1})\right)}{P^*(0, R_{i-1})}\right]e^{-I_N} \\
        &\quad \quad(i = 2, ..., m, m+2, ..., N)\\
        p_{1M} &= \left[\frac{u_M^*(R_1)}{P^*(0, R_1)}-\sum_{j=1}^N \frac{1}{P^*(0, R_j)}+(N-1)\right]e^{-I_N} + 1\\
        p_{m+1,M} &= \left[\frac{u_2^*(R_{m+1})}{P^*(0, R_{m+1})}-\sum_{j=1}^N \frac{1}{P^*(0, R_j)}+(N-1)\right]e^{-I_N} + 1\\
        f_{ik} &= f_k^*(R_i), \quad k = 1, ..., M-1.
    \end{aligned}
\end{equation}

%\bibliographystyle{ieeetr}
%\bibliography{reference}